\documentclass{article}

\usepackage{arxiv}
\usepackage[utf8]{inputenc}
\usepackage[T1]{fontenc}
\usepackage{hyperref}
\usepackage{url}
\usepackage{booktabs}
\usepackage{amsfonts}
\usepackage{amssymb}
\usepackage{nicefrac}
\usepackage{microtype}
\usepackage{graphicx}
\usepackage[dvipsnames]{xcolor}
\usepackage{natbib}
\usepackage{doi}
\usepackage{multirow}
\usepackage{appendix}
\usepackage{array}
\usepackage{flafter}
\usepackage{float}
\usepackage{caption}
\usepackage{subcaption}
\usepackage{amsmath}
\usepackage[ruled,vlined]{algorithm2e}

\hypersetup{
    colorlinks=true,
    linkcolor=Black,
    filecolor=RoyalPurple,
    urlcolor=RoyalPurple,
    citecolor=RoyalPurple,
}

\title{AgenticPosesRanker: An Agentic AI Framework for Physically Grounded Ranking of Protein--Ligand Docking Poses}
\author{
    Sofiene Khiari\thanks{Department of Pharmaceutical Sciences, University of Basel, Basel, Switzerland}
    \thanks{Swiss Institute of Bioinformatics, Basel, Switzerland}
    \thanks{Corresponding author: \href{mailto:research@sofk.ch}{research@sofk.ch}}
    \And
    Amr H. Mahmoud\footnotemark[1]\footnotemark[2]
    \And
    Markus A. Lill\footnotemark[1]\footnotemark[2]
}
\date{}

\begin{document}
\maketitle

\begin{abstract}

Scoring functions remain the principal bottleneck in molecular docking:
they routinely fail to rank near-native poses above decoys, and their
composite single-score design obscures the physicochemical basis of each
ranking error. We present AgenticPosesRanker, an agentic AI framework
that combines six deterministic, physically grounded analysis
tools (interaction fingerprinting, solvent-accessible burial,
conformational strain, steric-clash detection, unsatisfied-polar-atom
penalty, and chemical-identity extraction) with large-language-model
(GPT-5) chain-of-thought reasoning to evaluate and rank docking poses.
On a curated benchmark of ten protein--ligand
systems (162 poses) balanced by construction between Smina scoring-function
successes and failures, the agent achieved 50.0\% best-pose accuracy,
matching the design-fixed Smina baseline of 50.0\% and significantly
exceeding a 7.7\% uniformly random baseline ($p < 0.001$, one-sided
exact binomial test). The balanced-benchmark accuracy decomposes
symmetrically: the agent retained 80\% (4/5) of the Smina-success systems
and recovered 20\% (1/5) of the Smina-failure systems, so the aggregate
50\% reflects one regression offset by one recovery rather than any
net improvement over the Smina reference.
Decision-attribution analysis showed high alignment between the agent's
self-reported tool weights and objective metric separations of the
selected pose (median $\rho = +0.83$), consistent across correct and
incorrect outcomes, localising the performance ceiling to tool-suite
coverage rather than reasoning inconsistency.
These results establish a methodological template for evaluating agentic
AI against objective ground truth in the natural sciences and position
the framework as an interpretable curation layer for late-stage pose
refinement in structure-based drug design.

\end{abstract}

\section{Introduction}

Molecular docking is a cornerstone of structure-based drug design, yet its
utility is limited by a persistent gap between sampling and scoring:
modern docking programs generate at least one near-native pose in the vast
majority of cases, but their scoring functions rank that pose first far less
reliably. The bottleneck is not geometric search but energetic evaluation.
The Gibbs free energy of binding would, in principle, be the single correct
ranking criterion, but conventional scoring functions approximate it through
heavily simplified functional forms that collapse van~der~Waals,
electrostatic, desolvation, and torsional contributions into a small number
of terms with fixed weights, neglecting conformational entropy and solvent
reorganisation and assuming additivity where the underlying physics is
coupled~\cite{trott2010vina, bissantz2010guide, greenidge2014mmgbsa}.
Because neither the relative importance of each term nor the coupling
between terms is constant across binding sites, the resulting score is an
unreliable surrogate for the true $\Delta G$, particularly when closely
ranked poses differ in the balance of their physicochemical trade-offs.
Rescoring
approaches, consensus methods, and machine-learning models have improved
ranking accuracy, but each carries characteristic limitations: physics-based
rescorers are computationally expensive and still rely on fixed functional
forms; consensus protocols apply uniform weights across all targets;
data-driven models achieve high accuracy yet operate as opaque functions
whose predictions resist mechanistic interrogation.

A complementary strategy is to preserve each physicochemical observable as
a separate, interpretable input and delegate the integration step to an
explicit reasoning process. Large language models, when equipped with
domain-specific tools and structured evaluation guidelines, can perform
this kind of multi-criterion scientific reasoning, weighing competing
evidence, adapting emphasis to context, and articulating the rationale
behind each decision. Such agentic AI systems have been applied to
synthesis planning, reaction optimisation, and molecular property
prediction, but no prior work has evaluated an agentic framework for pose
ranking against a crystallographic ground-truth benchmark with quantitative
metrics for accuracy, reasoning faithfulness, and decision attribution.

This paper presents AgenticPosesRanker, an agentic AI framework that
couples deterministic, physically grounded analysis tools with
large-language-model reasoning to rank docking poses through
chain-of-thought scientific evaluation. We evaluate the framework on a
curated crystallographic benchmark balanced by construction between
scoring-function successes and failures, quantifying ranking accuracy,
the alignment between stated tool weights and objective metric
separations, and decision attribution, together constituting an
evaluation template absent from prior agentic chemistry studies.
On this ten-system benchmark the agent attains 50.0\% best-pose
accuracy, numerically matching the design-fixed Smina baseline while
significantly exceeding a 7.7\% uniformly random baseline, and
decomposes symmetrically into 80\% retention of Smina's correct
rankings (4/5) and 20\% recovery of its failures (1/5).
Decision-attribution analysis traces the asymmetry to gaps in tool-suite
coverage rather than to deficiencies in the reasoning process itself. This finding establishes a methodological template for
the rigorous evaluation of agentic AI systems against objective ground
truth in the natural sciences. The following
section reviews the relevant background on docking, scoring paradigms,
and agentic AI to contextualise this contribution.

\section{Background and Related Work}
\label{sec:background}

\subsection{Molecular Docking and the Pose-Ranking Problem}
\label{sec:bg-docking}

\paragraph{Docking as a two-stage process.}

Molecular docking predicts the preferred orientation of a small-molecule ligand
within a protein binding site by computationally generating and evaluating candidate
bound conformations, termed \emph{poses}~\cite{warren2006critical}.
The process is conventionally decomposed into two sequential stages.
First, a \emph{sampling} algorithm explores the conformational and orientational
space to produce a diverse ensemble of candidate geometries; modern programs employ
search strategies including stochastic global optimisation~\cite{trott2010vina,
koes2013smina}, genetic algorithms, and incremental-construction methods.
Second, a \emph{scoring function} evaluates each sampled pose and ranks the ensemble,
ideally placing the conformation closest to the experimentally observed binding mode
at the top.

In practice, the scoring function is the weaker of the two stages.
In a comparative evaluation on the Astex Diverse set, sampling algorithms generated
at least one near-native pose (RMSD~$< 2$~\AA{}) in 85--99\% of cases, yet the
corresponding scoring functions ranked that pose first in only 35--73\% of
cases~\cite{gaudreault2015flexaid}.
The gap between sampling and ranking success was present for every program
tested: AutoDock Vina exhibited the largest disconnect (93.4\% sampling
versus 35--40\% ranking, a gap exceeding 50 percentage points), but FlexX
(30~pp), FlexAID (29~pp), and rDock (25--27~pp) all showed ranking success
rates well below their sampling rates~\cite{gaudreault2015flexaid}.
A broader benchmark of ten programs across 2\,002 complexes confirmed that the best
sampling program is not the best ranking program and that no single approach excels
at both tasks~\cite{wang2016comprehensive}.
These observations establish that the principal bottleneck in structure-based pose
prediction lies not in the search algorithm but in the scoring function, a limitation
that motivates the rescoring and multi-criteria approaches discussed in the following
subsections.

\paragraph{Near-native pose identification and RMSD.}

The accuracy of a docking prediction is conventionally quantified by the
root-mean-square deviation (RMSD) between the heavy atoms of the predicted pose
and the crystallographically determined binding mode.
A pose with RMSD~$< 2.0$~\AA{} to the native structure is classified as
\emph{near-native}, reproducing the experimentally observed ligand placement;
poses in the range $2.0$--$4.0$~\AA{} occupy an ambiguous intermediate zone,
while those exceeding 4.0~\AA{} are considered
non-native~\cite{warren2006critical, su2019comparative}.
The 2.0~\AA{} near-native threshold has become the de facto standard across
major benchmarking campaigns, including the Comparative Assessment of Scoring
Functions (CASF)~\cite{su2019comparative} and evaluations on the Astex Diverse
set~\cite{hartshorn2007diverse}.
For molecules that contain chemically equivalent atom permutations, such as
symmetric aromatic rings or carboxylate groups, naive RMSD calculations can yield
artificially inflated values; symmetry-corrected variants based on graph
isomorphism resolve this problem~\cite{meli2020spyrmsd}.

Because RMSD compares atomic coordinates directly against an experimental
reference, it provides a geometry-based ground truth that is independent of any
scoring function.
This independence is its principal strength: it allows sampling success (whether
a near-native pose exists in the ensemble) and ranking success (whether the
scoring function places it first) to be assessed separately, the distinction
exploited in the benchmarks discussed above.
Nevertheless, RMSD has recognised limitations.
Poses within the 2.0~\AA{} threshold can exhibit substantially different
calculated protein--ligand interaction energies~\cite{greenidge2014mmgbsa},
indicating that geometric proximity to the native mode does not guarantee
energetically favourable binding.
These caveats motivate the complementary quality metrics defined alongside
each analysis tool in the Methods
(Sections~\ref{sec:plip}--\ref{sec:derived-tools}).

\paragraph{The pose-ranking problem as multi-criteria decision-making.}

In principle, the Gibbs free energy of binding would be the single correct
criterion for pose ranking: if the enthalpic and entropic contributions could
be computed exactly, the pose with the most favourable $\Delta G$ would, by
definition, represent the best prediction. Current scoring functions,
however, approximate this free energy through heavily simplified functional
forms. They typically collapse van~der~Waals, electrostatic, desolvation,
and torsional contributions into a small number of weighted terms whose
coefficients are fitted to reproduce experimental binding
affinities~\cite{trott2010vina, koes2013smina}. Entropic effects (ligand
conformational entropy, the reorganisation of solvent molecules upon binding)
and desolvation penalties for buried polar groups are either absorbed into
global fitting weights or neglected
entirely~\cite{greenidge2014mmgbsa, bissantz2010guide}. Because the
resulting score is a coarse approximation rather than a rigorous free-energy
estimate, it can assign high ranks to poses that violate basic physical
constraints and penalise poses that are thermodynamically sound.
Recognising these limitations motivates an alternative strategy: rather than
relying on a single approximate $\Delta G$, one can compute several
complementary physicochemical descriptors from the docked structure and
reason over them jointly. These include non-covalent interaction quality
(hydrogen bonds, hydrophobic contacts, salt
bridges~\cite{bissantz2010guide}), conformational
strain~\cite{gu2021strain}, steric clashes, binding-site burial, and
satisfaction of buried polar atoms through compensatory hydrogen
bonds~\cite{buttenschoen2024posebusters}.
These descriptors, however, do not represent independent, additive
contributions to the binding free energy. The thermodynamic value of a
hydrogen bond depends on the desolvation cost of the participating groups,
which is itself a function of burial depth; conformational strain modulates
both interaction geometry and solvent exposure
simultaneously~\cite{bissantz2010guide, greenidge2014mmgbsa}. Scoring
functions that model $\Delta G$ as a fixed-weight linear combination of such
terms assume additivity where the underlying physics is coupled, and fix
the relative importance of each term where the balance shifts with
binding-site context.
This coupling also means that the descriptors can conflict at the observable
level: a deeply buried pose may exhibit elevated conformational strain, while
a conformation rich in hydrogen bonds may simultaneously introduce steric
overlaps with the protein. A fixed-weight composite score cannot arbitrate
these trade-offs, because neither the linear functional form nor the training
objective (affinity prediction) captures the non-linear, context-dependent
relationships among the
descriptors~\cite{warren2006critical, wang2016comprehensive}.
The gap between what a rigorous free-energy calculation would provide and
what current scoring functions actually deliver motivates the detailed
examination of scoring-function paradigms in Section~\ref{sec:bg-scoring}
and of reasoning-based approaches that can integrate these coupled descriptors
through context-adaptive reasoning rather than predetermined weights, as
explored in Section~\ref{sec:bg-agentic}.

\subsection{Scoring Function Limitations and the Interpretability Gap}
\label{sec:bg-scoring}

Scoring functions fall into three broad paradigm families.
\emph{Empirical} scorers, such as AutoDock Vina~\cite{trott2010vina} and
Smina~\cite{koes2013smina}, model the binding free energy as a weighted sum
of steric, hydrophobic, hydrogen-bonding, and torsional terms calibrated
against experimental affinities.
\emph{Knowledge-based} methods instead derive distance-dependent
pseudo-energies from observed protein--ligand contact frequencies in the
Protein Data Bank~\cite{gohlke2000knowledge, muegge1999general}.
\emph{Machine-learning} approaches learn non-linear mappings directly from
structural data: Gnina trains a CNN ensemble on 22.5~million poses and
raises top-1 redocking accuracy from 58\,\% (Vina) to
73\,\%~\cite{mcnutt2021gnina, francoeur2020crossdock}, while diffusion
models such as DiffDock frame docking as sampling over $\mathrm{SE}(3)$
manifolds~\cite{corso2023diffdock}.
\emph{Consensus scoring} combines rankings from multiple functions to
reduce false positives, raising pose-prediction success from 66--76\,\%
individually to approximately 80\,\% or
higher~\cite{charifson1999consensus, wang2003comparative}, but applies a
fixed aggregation rule that cannot adapt to binding-site
context~\cite{yang2005consensus}.
A detailed comparative analysis of scoring-function performance across
major benchmarks is provided in Supplementary
Section~\ref{sec:si-scoring-benchmarks}.

All three paradigm families, and the consensus combinations built on top of
them, share a structural limitation.
Because their functional forms omit or crudely approximate conformational
entropy, solvent reorganisation, and desolvation penalties for buried polar
groups~\cite{greenidge2014mmgbsa, bissantz2010guide}, the aggregate score
is an unreliable surrogate for the true binding free energy.
A medicinal chemist cannot determine from a Vina
score whether the top-ranked pose was preferred because of strong hydrogen
bonds despite moderate strain, or because the simplified functional form
rewarded a large hydrophobic contact
area~\cite{bissantz2010guide}.
Machine-learning methods compound this opacity: CNN confidence scores and
diffusion likelihoods provide no reasoning trace, and the PoseBusters
validation framework showed that many deep-learning methods generate poses
with steric clashes, incorrect stereochemistry, or non-standard bond
geometries~\cite{buttenschoen2024posebusters}.
No existing benchmark assesses whether a method can articulate \emph{which}
observable properties drove its ranking and how competing criteria were
weighed against one another~\cite{su2019comparative, warren2006critical}.
Across all paradigm families, scoring functions therefore operate as
recommendation systems without explanations, a property that limits their
utility for the iterative, hypothesis-driven decisions that characterise
structure-based drug design.

Bridging this interpretability gap requires two components: (i)~a set of
independently computed physicochemical descriptors that characterise each
pose along multiple dimensions, and (ii)~a reasoning mechanism that
integrates these descriptors through context-adaptive logic rather than
predetermined weights.
The descriptors used in this work (non-covalent interaction profiles,
solvent exposure and burial, conformational strain, steric clash counts,
and unsatisfied polar atom penalties) are defined and formalised in the
Methods (Sections~\ref{sec:plip}--\ref{sec:derived-tools}), where each
tool's algorithm, geometric criteria, and scientific basis are presented
together.
The reasoning mechanism is introduced in the following subsection.

\subsection{Large Language Models and Agentic AI for Scientific Reasoning}
\label{sec:bg-agentic}

The interpretability gap identified above leaves a question: how can multiple
physicochemical descriptors be integrated without collapsing them into a
single scalar?

\paragraph{Tool-augmented LLM agents.}
A tool-augmented large-language-model (LLM) agent pairs a foundation model with
a suite of external tools, computational programs, databases, or instrument
interfaces, that the model invokes through a structured
\emph{observe--reason--act} loop~\cite{yao2023react}. At each step the model
inspects tool outputs, formulates a reasoning trace, and selects the next
action, iterating until the task objective is met. This paradigm differs from
two established alternatives: end-to-end machine-learning scoring, where a
single model learns the complete mapping from molecular inputs to predicted
outputs, and retrieval-augmented generation, where the model retrieves and
summarises textual passages but does not invoke computational procedures. In
chemistry, ChemCrow demonstrated the paradigm by coupling GPT-4 with eighteen
chemistry tools in a ReAct-style orchestration layer, enabling synthesis
planning, insect-repellent design, and chromophore discovery; expert evaluators
judged its outputs more accurate than those of the unaugmented
model~\cite{bran2024chemcrow}. Coscientist extended the approach to autonomous
experimental orchestration, directing robotic platforms to optimise
palladium-catalysed cross-coupling reactions through iterative search, code
execution, and documentation lookup~\cite{boiko2023coscientist}. These systems
are distinct from domain-specific fine-tuned models such as MoleculeGPT, where
the language model \emph{is} the predictor; in a tool-augmented agent the model
serves as a reasoning orchestrator while computation is delegated to verified
external tools. Despite growing adoption, existing agentic chemistry systems
evaluate task-completion accuracy on domain-specific benchmarks
(molecular-property prediction, synthesis feasibility, drug--target
interaction scoring); none has been applied to docking pose ranking,
where accuracy is defined by RMSD agreement with a crystallographic
reference structure rather than by a predicted property or categorical
answer.
To date, no tool-augmented chemistry agent has therefore reported
quantitative pose-ranking accuracy, faithfulness, or
decision-attribution metrics against such a benchmark, leaving open
the question of whether agentic reasoning can match or complement
physics-based scoring in structural pose evaluation.

\paragraph{Chain-of-thought reasoning and transparency.}
Chain-of-thought (CoT) prompting elicits explicit step-by-step reasoning from
large language models by providing exemplars that demonstrate intermediate
reasoning steps before arriving at a final
answer~\cite{wei2022chainofthought}. On arithmetic, commonsense, and
symbolic-reasoning benchmarks, CoT prompting improved accuracy by substantial
margins, up to 17 percentage points on GSM-8K for PaLM~540B, by decomposing
complex queries into sequential sub-goals that the model addresses
incrementally~\cite{wei2022chainofthought}. Subsequent work revealed that
models retain 80--90\,\% of their CoT-elicited performance even when the
exemplar reasoning steps are logically invalid, indicating that much of the
benefit derives from adopting a structured output format rather than from
faithfully executing each intermediate step~\cite{wang2023towards}. Coupling
the natural-language trace with deterministic symbolic solvers, an approach
termed \emph{Faithful CoT}, recovered both accuracy (state-of-the-art on 7
of 10 evaluation tasks) and verifiability by ensuring that the computational
portion of the reasoning chain is symbolically
correct~\cite{lyu2023faithful}.
Beyond accuracy, CoT reasoning offers a transparency mechanism absent from
the single-scalar scoring paradigms surveyed in
Section~\ref{sec:bg-scoring}: the reasoning trace makes the model's stated
decision process inspectable and auditable. In an analytical context, such a
trace can reference specific measurements, thresholds, and
trade-offs, producing a structured argument analogous to a domain expert's
written assessment rather than an opaque aggregate score. The scientific
utility of this transparency, however, depends critically on whether the
verbalised reasoning faithfully represents the factors that actually
determined the model's output, a concern that recent work on CoT
faithfulness has shown to be well-founded~\cite{turpin2023language}.

\paragraph{Reasoning faithfulness and hallucination risks.}
The transparency benefits of chain-of-thought reasoning are contingent on a
non-trivial assumption: that the verbalised reasoning trace faithfully
reflects the factors that actually determined the model's
output~\cite{turpin2023language}. Turpin et~al.\ demonstrated that
introducing biasing features, such as a suggested answer from a purportedly
authoritative source, caused accuracy drops of up to 36~percentage points,
yet the resulting chain-of-thought explanations almost never mentioned these
biases; in 73\,\% of cases the model produced a plausible but fabricated
justification that supported the bias-influenced
answer~\cite{turpin2023language}. Lanham et~al.\ formalised this concern
through a suite of intervention-based faithfulness tests, truncating the
chain midway, injecting deliberate mistakes, and inserting filler tokens, and
found that, counterintuitively, larger and more capable models were often
\emph{less} faithful: on seven of eight evaluation tasks, a
13-billion-parameter model exhibited higher faithfulness scores than its
175-billion-parameter counterpart~\cite{lanham2023measuring}. These findings
indicate that the mere presence of a reasoning trace does not guarantee its
explanatory validity; models can produce coherent post-hoc rationalisations
that bear little relation to their internal decision
process~\cite{wang2023towards}.
A related concern is hallucination: large language models can generate
plausible-sounding but factually incorrect claims, a risk that extends to
scientific contexts where the model may report interactions, properties, or
trends that no underlying computation actually produced. Taken together,
unfaithful reasoning and hallucination represent fundamental reliability
challenges for any framework that presents LLM-generated reasoning as an
auditable scientific justification. Mitigating these risks requires
architectural safeguards, grounding the model's reasoning in verified tool
outputs rather than parametric memory, and providing quantitative means to
assess whether stated decision factors align with objective evidence, design
considerations explored in Section~\ref{sec:agent-architecture}.

\paragraph{Positional bias and evaluation fairness.}
When large language models process ordered lists, their attention to individual
items depends systematically on position. Liu et al.\ demonstrated this
``lost in the middle'' phenomenon: models retrieve information far more
reliably from the beginning and end of their input context than from
intermediate positions, producing a characteristic U-shaped performance curve
across question-answering and retrieval tasks~\cite{liu2024lost}. The effect
extends directly to evaluation settings. In multiple-choice benchmarks
spanning 20~models, relocating the correct answer from position~A to
position~D reduced accuracy by up to 15~percentage points, and
option-selection proportions dropped monotonically from 34.6\% for the first
option to 15.8\% for the last, revealing a strong primacy
bias~\cite{zheng2024large}. For listwise ranking, permutation
self-consistency, shuffling the input list, collecting independent rankings,
and aggregating by majority vote, improved accuracy by 7--18\% for GPT-3.5
and 8--16\% for LLaMA-2~70B, confirming that positional bias materially
distorts rank-order judgements~\cite{tang2024found}.
These findings have direct implications for any framework that asks an LLM to
rank molecular poses. If the model systematically favours poses presented
early in its context window, rankings will reflect input ordering rather than
physicochemical merit. To mitigate this risk, our system applies a
deterministic anonymisation protocol that replaces all pose identifiers with
SHA-256-derived codes and sanitises structural metadata before the model
encounters the ranking task (Section~\ref{sec:anonymisation}). Anonymisation
eliminates label-based bias, the model cannot infer original rank from an
opaque identifier, but neither the anonymisation protocol nor the current
pipeline implementation randomises presentation order: poses are presented
in a fixed sequence across runs, and positional effects therefore remain an
uncontrolled confound (Section~\ref{sec:disc-anonymisation}). It is best
understood as a necessary but not sufficient safeguard, complemented by
explicit anti-bias instructions in the agent's system prompt.

\subsection{Evaluation Frameworks and Benchmarks for Docking}
\label{sec:bg-benchmarks}

Sections~\ref{sec:bg-scoring} and~\ref{sec:bg-agentic} together define the
components of a reasoning-based pose-ranking system: the interpretability gap
that motivates multi-descriptor evaluation, and an LLM agent that integrates
these descriptors through transparent reasoning. Evaluating such a system
requires benchmarks with crystallographic ground truth and sufficient
structural diversity.

\paragraph{PDBbind and CASF benchmarks.}
The PDBbind database, introduced by Wang et~al.~\cite{wang2004pdbbind, wang2005pdbbind}, provides the largest curated
collection of protein--ligand complexes for which both three-dimensional
crystal structures and experimentally determined binding
affinities, dissociation constants ($K_d$) or inhibition constants
($K_i$), are available. Updated
annually since its first public release in 2004, PDBbind organises its entries
into three nested tiers of increasing quality: a \emph{general set}
encompassing all qualifying complexes, a \emph{refined set} filtered for
crystallographic resolution better than 2.5~\AA{}, unambiguous binding data,
and binary complex topology, and a \emph{core set} in which entries are
further clustered by protein-sequence similarity to ensure target
diversity~\cite{su2019comparative}. This hierarchical design serves dual
roles: the general and refined sets supply training data for scoring-function
development, while the core set provides a compact, non-redundant evaluation
benchmark.
The Comparative Assessment of Scoring Functions (CASF) is the companion
benchmarking protocol built on successive PDBbind core sets. CASF evaluates
scoring functions along four complementary tasks, scoring power (correlation
with experimental affinities), ranking power (correct ordering of ligands
within a target cluster), docking power (identification of a near-native pose
among decoys), and screening power (enrichment of true binders from a mixed
library)~\cite{su2019comparative}. The most recent iteration, CASF-2016,
tested 25~scoring functions on 285~core-set complexes partitioned into
57~target clusters and confirmed that no single function excels across all
four tasks, underscoring the multifaceted nature of scoring-function
quality~\cite{su2019comparative}. Francoeur et~al.~\cite{francoeur2020crossdock}
extended the PDBbind ecosystem by generating large-scale docking ensembles with
Smina for thousands of PDBbind systems, including both self-docking
(re-docking) and cross-docking configurations; these pre-computed pose sets,
with crystal-structure ground truth, have since been widely reused for
scoring-function evaluation and machine-learning
training~\cite{mcnutt2021gnina}. The present benchmark draws its protein
structures from PDBbind~2016 and its Smina-generated docked poses from the
Francoeur et~al.\ dataset (Section~\ref{sec:benchmark-construction}).

\paragraph{Evaluation metrics and the reasoning gap.}
Among the CASF evaluation dimensions, docking power, the percentage of
systems for which the top-ranked pose falls within 2.0~\AA{} RMSD of the
crystal structure, is the task most directly comparable to pose-ranking
accuracy, with values ranging from approximately 60\,\% to 90\,\% across the
25~scoring functions tested in CASF-2016 under self-docking
conditions~\cite{su2019comparative}. Self-docking success rates for the
best-performing programs typically range from 60\,\% to 80\,\% across large
PDBbind-derived test
sets~\cite{wang2016comprehensive, su2019comparative}; under cross-docking
conditions, where a ligand is docked into a receptor conformation crystallised
with a different ligand, success rates decline substantially due to
receptor-conformation
mismatch~\cite{francoeur2020crossdock, mcnutt2021gnina}. However, all four
CASF tasks quantify \emph{what} a scoring function selects, a top pose, an
affinity estimate, a rank, or an enriched compound set, without examining
\emph{why} that selection was made. Inspecting the individual terms of an
empirical scoring function (van~der~Waals, hydrogen-bonding, torsional
contributions) does reveal which fitted energy components favoured a given
pose~\cite{trott2010vina, greenidge2014mmgbsa}, but these terms report the
output of simplified functional forms rather than independently verified
structural properties: a favourable hydrogen-bonding term does not indicate
whether the underlying donor--acceptor geometries are well-formed, nor whether
the associated desolvation cost has been accounted
for~\cite{bissantz2010guide}. No existing benchmarking protocol evaluates
whether the factors cited for a ranking, whether expressed as a scoring-term
decomposition or as an explicit decision trace, align with physicochemical
evidence computed independently of the scoring function, leaving the
faithfulness of ranking decisions outside the scope of current evaluation
frameworks.

\paragraph{Statistical considerations for small-sample evaluation.}
Docking-benchmark design forces a trade-off between statistical power
and evaluation depth. Large-scale benchmarks such as
CASF-2016~\cite{su2019comparative} evaluate hundreds of complexes, yielding
narrow confidence intervals around aggregate metrics, top-1 success rates,
Pearson correlations, Spearman rank coefficients, but the sheer volume of
systems restricts assessment to a single numerical score per complex.
Conversely, small, carefully curated benchmarks can support detailed
per-system analyses, complete reasoning traces, metric decomposition across
physical observables, and faithfulness measurement between stated rationales
and the underlying evidence, yet the resulting sample sizes introduce
substantial uncertainty into any point estimate of accuracy.
For success rates estimated from small benchmarks, the standard normal
(Wald) approximation to the binomial confidence interval is known to
produce coverage well below the nominal level, particularly when the
true proportion lies near zero or one~\cite{brown2001interval}. The
Wilson score interval~\cite{wilson1927probable} corrects this
deficiency by inverting the score test rather than the Wald test,
guaranteeing coverage that remains close to the stated confidence level
even for sample sizes as low as ten. Complementarily, an exact
one-sided binomial test can evaluate whether an observed success count
is significantly greater than a random-selection baseline, providing a
formal hypothesis test that does not rely on asymptotic assumptions.
Reporting such uncertainty ranges alongside point estimates is
essential whenever the number of evaluated systems is small, because a
single additional success or failure can shift the estimated accuracy
by a large margin.
The present study deliberately restricts the benchmark to a small number of
systems (Section~\ref{sec:benchmark-construction}) in order to permit the
depth of evaluation, complete reasoning traces, decision attribution, and
faithfulness quantification, that would be infeasible at the scale of
CASF-2016's 285~complexes. This design choice is accompanied throughout
by Wilson confidence intervals and exact binomial tests so that the
statistical limitations of the sample size remain transparent.

The following section describes the concrete realisation of this
approach: the analysis tools, agent architecture, benchmark construction,
and evaluation protocol.

\section{Methods}

\subsection{System Overview}
\label{sec:system-overview}

\begin{figure}[htbp]
    \centering
    \includegraphics[width=\textwidth]{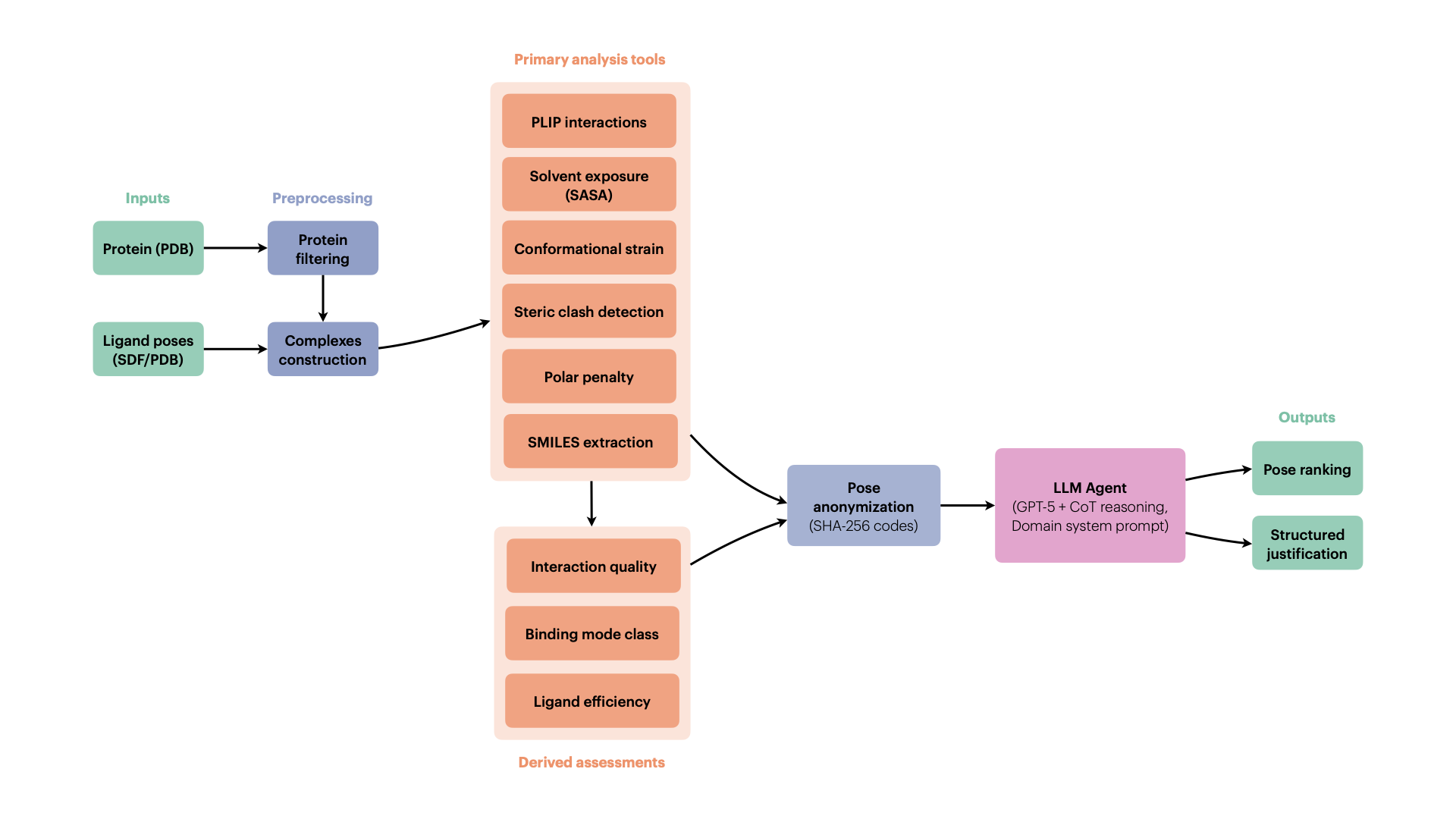}
    \caption{
        \textbf{Architecture of the AgenticPosesRanker framework.}
        The pipeline accepts a protein structure (PDB) and one or more docked ligand poses (SDF) as inputs. After protein filtering and complex construction, each pose is evaluated by six primary analysis tools covering non-covalent interactions, solvent exposure, conformational strain, steric clashes, polar desolvation penalties, and chemical identity verification (via SMILES extraction, not used for ranking). Three derived tools then assess interaction quality, binding mode, and ligand efficiency from the primary outputs. All tool results are written to anonymised analysis files identified only by eight-character SHA-256-derived codes, preventing the LLM from inferring pose identity or ordering. The anonymised files are submitted together with a domain-specific system prompt to a GPT-5 agent with chain-of-thought reasoning enabled. The agent returns a ranked list of poses with a structured justification referencing specific metric values. Anonymous codes are automatically replaced with original pose names in the final output.
    }
    \label{fig:system_overview}
\end{figure}

Rather than learning a single composite scoring function, the system delegates each physical observable to a dedicated, independently validated computational chemistry tool and then asks an LLM to \emph{reason} over the collected evidence. The overall architecture is summarised in Figure~\ref{fig:system_overview} and the end-to-end pipeline is described below.

\paragraph{Inputs.}
The pipeline accepts two categories of input files: (i)~a protein structure in PDB format and (ii)~one or more candidate ligand poses in SDF format (PDB format is also supported; SDF files are converted to PDB internally via RDKit~\cite{rdkit2025}). Each pose file contains a single docked conformation of the ligand.

\paragraph{Preprocessing.}
Before any analysis is performed, the protein structure undergoes a filtering step implemented in a dedicated module. This step removes all non-essential HETATM records, including co-crystallised ligands, crystallisation additives, detergents, glycans, and buffer components, while preserving water molecules and catalytically relevant metal ions (Zn, Mg, Ca, Fe, Mn, Co, Ni, Cu). The rationale is to prevent co-crystallised ligands or other heteroatoms from interfering with the downstream PLIP interaction analysis, which should detect contacts only between the protein and the docked pose under evaluation.

For each pose, a protein-ligand complex file is then constructed by concatenating the filtered protein coordinates with the ligand atoms. During this step, ligand atoms are assigned a chain identifier (required by BioPython's SASA calculator) and atom serial numbers are renumbered to continue from the protein's last serial, ensuring valid PDB formatting. All CONECT records are placed at the end of the file to satisfy BioPython's parser requirements.

\paragraph{Analysis tools.}
Each complex is passed through six primary analysis tools that compute physical and structural observables. These tools are:

\begin{enumerate}
    \item \textbf{PLIP interaction profiling}: detection of hydrogen
          bonds, hydrophobic contacts, $\pi$-stacking, salt bridges,
          and water bridges with full geometric parameters;
    \item \textbf{Solvent exposure (SASA)}: Solvent-accessible surface area and burial ratio;
    \item \textbf{Conformational strain}: MMFF94 energy difference
          between the bound conformation and a force-field-optimised
          reference;
    \item \textbf{Steric clash detection}: van der Waals overlap
          count, energy, and severity;
    \item \textbf{Unsatisfied polar penalty}: identification of
          buried polar atoms lacking compensatory hydrogen bonds;
    \item \textbf{SMILES extraction}: canonical isomeric SMILES for
          chemical identity verification (not used for ranking).
\end{enumerate}

Three additional \emph{derived assessment tools} then operate on the
outputs of the primary tools:

\begin{enumerate}
    \setcounter{enumi}{6}
    \item \textbf{Interaction quality assessment}: geometry-based
          quality scoring of each interaction detected by PLIP;
    \item \textbf{Binding mode classification}: classification of
          each pose as \texttt{deep\_pocket}, \texttt{partial\_pocket},
          or \texttt{surface} based on ligand burial ratio;
    \item \textbf{Ligand efficiency assessment}: normalisation of
          interaction counts and quality by the number of heavy atoms.
\end{enumerate}

All tools are executed sequentially for each pose. The primary tools each write their results to both a human-readable log file (retaining the original pose name) and an anonymised analysis file (see below). After all primary tools have completed, the three derived tools read the existing analysis files, compute their metrics, and append the results before the analysis files are finalised.

\paragraph{Pose anonymisation.}
To prevent the LLM from inferring pose identity or ordering from filenames (e.g.\ \texttt{pose\_01.sdf} versus \texttt{pose\_09.sdf}), every pose is assigned a deterministic eight-character base-36 code generated by hashing the concatenation of a system identifier and the pose name with SHA-256. The analysis files are named after these anonymous codes and are purged of any residual references to the original filenames before being passed to the LLM.

\paragraph{LLM reasoning.}
The anonymised analysis files, one per pose, are concatenated into a single textual context and submitted to the LLM together with a domain-specific system prompt. We use OpenAI's GPT-5 model accessed through the OpenAI Agents SDK (version~0.5.0), with chain-of-thought (CoT) reasoning enabled at \texttt{effort="high"} and reasoning summary set to \texttt{"detailed"}, together with model verbosity set to \texttt{"high"}. All other sampling parameters were left at the SDK's defaults for reasoning models: temperature, top-$p$, and frequency and presence penalties were not explicitly set (GPT-5's reasoning endpoint does not accept user-supplied values for these parameters at the time of evaluation), no random seed was supplied, and no explicit \texttt{max\_tokens} limit was imposed, relying on the model's internal length budget under the high-effort and high-verbosity settings.

The system prompt (${\sim}7\,800$ words) encodes: (i)~complete documentation of each tool's algorithm, output format, and interpretation guidelines; (ii)~a hierarchical decision framework that prioritises binding mode and interaction quality over raw interaction counts, and uses conformational strain, steric clashes, and polar penalties as tiebreakers; (iii)~explicit anti-bias instructions (e.g.\ "surface binders should never rank first, even with high interaction counts"); (iv)~structured output requirements specifying the expected sections of the response (best pose selection, comparative analysis, confidence assessment, and recommendations). The prompt was refined iteratively against exploratory PDBbind~2016 systems during development. These exploratory systems were not formally held out from the ten benchmark systems reported in Table~\ref{tab:selected-systems}, and we cannot guarantee full disjointness between the prompt-tuning set and the evaluation set; this transparency disclosure is revisited in Section~\ref{sec:disc-limitations}.

The agent processes all analysis files in a single inference call and returns (a)~a chain-of-thought reasoning trace and (b)~a structured final answer containing the selected best pose, a metric-by-metric justification, a comparative analysis of runner-up poses, and a self-assessed confidence level. After inference, the anonymous codes in the agent's output are automatically mapped back to the original pose names for human-readable reporting.

\paragraph{Outputs.}
The pipeline produces three categories of output: (i)~per-pose analysis log files with the full tool outputs; (ii)~scientific visualisation plots summarising the metric distributions across all poses (generated using Matplotlib and Seaborn); and (iii)~a comprehensive ranking log that records the system prompt, the anonymised analysis context sent to the model, the chain-of-thought reasoning, and the final ranking with justification.

\paragraph{User interfaces.}
AgenticPosesRanker is accessible through two complementary front-ends, a Streamlit~\cite{streamlit2024} web application and a Typer-based command-line interface, both of which share the same analysis and ranking back-end. The web application provides an interactive graphical environment in which users can upload structures, visualise protein-ligand complexes in 3D via an embedded Mol*~\cite{sehnal2021molstar} viewer, monitor pipeline progress in real time, and conduct follow-up conversations with the agent after the initial ranking. The interface is described in Section~\ref{sec:user-interface}, with full layout and configuration details in Supplementary Section~\ref{sec:si-user-interface}.

\paragraph{Design principles.}
Two principles guided the design of the framework:

\begin{itemize}
    \item \textbf{Tool-augmented reasoning:} The LLM does not predict binding scores or learn a scoring function. Instead, it reasons over pre-computed observables produced by established computational chemistry tools (RDKit, PLIP, BioPython). This decouples the physical evaluation, which is deterministic, from the multi-criteria decision-making, which benefits from the LLM's ability to handle conflicting signals (e.g.\ high strain compensated by a strong interaction network).
    \item \textbf{Explainability by design:} Every ranking decision is accompanied by chain-of-thought reasoning and a structured justification that explicitly references specific metric values. This transparency enables domain experts to audit, critique, and override the agent's conclusions, a property absent from conventional black-box scoring functions.
\end{itemize}

\subsection{Analysis Tools}

\subsubsection{Protein-Ligand Interaction Profiling (PLIP)}
\label{sec:plip}

We characterise the non-covalent interaction network between each docked pose and the protein using the Protein-Ligand Interaction Profiler (PLIP) version~3.0~\cite{salentin2015plip, adasme2021plip, schake2025plip}. PLIP applies a deterministic, rule-based algorithm that identifies interactions through geometric criteria (distance and angle cutoffs) derived from the structural biology literature. Because no trainable parameters are involved, the analysis is fully deterministic.

We invoke PLIP through its Python API and load the pre-assembled protein-ligand complex PDB file produced during the preprocessing step (Section~\ref{sec:system-overview}). After loading, the module automatically detects all heteroatom groups present in the structure. To avoid spurious interaction reports, we skip common non-ligand heteroatom groups, specifically HOH, WAT, CA, MG, ZN, NA, CL, SO\textsubscript{4}, and PO\textsubscript{4}, before calling \texttt{characterize\_complex()} on the first valid ligand detected. Only one ligand per complex is analysed, corresponding to the single docked pose under evaluation.

Five categories of non-covalent interactions are extracted, each accompanied by per-contact geometric parameters:

\begin{enumerate}
    \item Hydrogen bonds: Both ligand-donor and protein-donor orientations are tracked separately. For each hydrogen bond, the donor-acceptor distance $d_{\mathrm{AD}}$ (\AA) and the donor-H-acceptor angle $\theta$ (\textdegree) are recorded.
    \item Hydrophobic contacts: Non-polar atom pairs within a distance cutoff of 4.0~\AA{} are reported with their inter-atomic distance (\AA).
    \item $\pi$-Stacking interactions: Ring-ring contacts between aromatic systems on the ligand and the protein are detected and classified as \emph{parallel} (face-to-face; inter-ring angle ${<}\,30$\textdegree) or \emph{T-shaped} (edge-to-face; angle ${\geq}\,30$\textdegree). The centroid-centroid distance (\AA), inter-ring angle (\textdegree), and lateral offset (\AA) are recorded for each contact.
    \item Salt bridges: Ionic interactions between oppositely charged groups are recorded with their distance (\AA), distinguishing ligand-negative/protein-positive from ligand-positive/protein-negative pairings.
    \item Water bridges: Water-mediated hydrogen bonds that bridge ligand and protein atoms are reported with two distances: the ligand-water distance $d_{\mathrm{AW}}$ (\AA) and the water-protein distance $d_{\mathrm{DW}}$ (\AA).
\end{enumerate}

For each pose, the tool produces two complementary outputs: (i)~a formatted text report listing every detected contact with its participating residues and full geometric parameters, and (ii)~a structured data dictionary that organises interaction details per ligand and provides aggregate counts (total interactions and per-type subtotals for hydrogen bonds, hydrophobic contacts, $\pi$-stacking, salt bridges, and water bridges). The structured output is used by the downstream interaction quality and ligand efficiency assessment tools (Section~\ref{sec:derived-tools}), while the formatted report is written to the anonymised analysis file consumed by the LLM.

The geometric parameters recorded by PLIP serve a dual purpose in the agentic pipeline. First, they enable the agent to distinguish high-quality interactions (e.g.\ hydrogen bonds with $d_{\mathrm{AD}} \approx 2.8$-$3.2$~\AA{} and $\theta \approx 160$-$180$\textdegree) from marginal ones that satisfy PLIP's detection cutoffs but contribute little to binding stability. Second, directionality and redundancy information, such as whether a single atom participates in multiple hydrogen bonds simultaneously, is available for the agent to assess network plausibility rather than relying on raw interaction counts.

\subsubsection{Solvent Exposure Assessment (SASA)}
\label{sec:sasa}

We quantify the degree to which each docked ligand is buried within the protein binding pocket by computing atomic solvent-accessible surface areas (SASA) using the Shrake-Rupley algorithm~\cite{shrake1973environment} as implemented in BioPython~\cite{cock2009biopython} version~1.86. Ligand burial is a necessary prerequisite for specific binding: poses that remain largely solvent-exposed are almost invariably docking artefacts rather than productive binding modes.

The calculation is performed on the full protein-ligand complex PDB file produced during preprocessing (Section~\ref{sec:system-overview}). After parsing the structure with BioPython, the \texttt{ShrakeRupley} calculator is invoked at atomic resolution (\texttt{level="A"}). SASA is computed on the complete protein-ligand complex, so that per-atom values reflect the combined shielding from the protein surface and from neighboring ligand atoms. This yields the effective solvent accessibility in the bound state, which is the quantity consumed by the downstream polar penalty tool (Section~\ref{sec:polar-penalty}). The approach does not, however, distinguish protein-contributed burial from intramolecular self-occlusion: atoms in concave regions of the ligand may register low SASA even without protein shielding, inflating the apparent burial ratio. A~$\Delta$SASA decomposition (subtracting complex SASA from isolated-ligand SASA per atom) would isolate the protein-specific contribution and produce a more precise burial metric for both the polar penalty assessment and the binding-mode classification, but is not implemented in the current version.

Ligand atoms are then identified by excluding standard amino-acid residues (detected via BioPython's \texttt{is\_aa()}) and the same set of non-ligand heteroatom groups excluded by the PLIP analysis (HOH, WAT, CA, MG, ZN, NA, CL, SO\textsubscript{4}, PO\textsubscript{4}). Only the first valid ligand residue is processed, corresponding to the single docked pose under evaluation. From the per-atom SASA values we derive six summary statistics:

\begin{itemize}
    \item Total SASA: the sum of SASA values across all ligand
          atoms (\AA$^{2}$).
    \item Average SASA: the mean SASA per ligand atom
          (\AA$^{2}$), serving as the principal burial indicator.
    \item Maximum and minimum SASA: the highest and lowest
          per-atom SASA values (\AA$^{2}$), indicating heterogeneity in
          ligand exposure.
    \item Exposed and buried atom counts: atoms are classified
          as buried if their SASA is below a threshold of
          1.0~\AA$^{2}$ and as exposed otherwise.
    \item Burial ratio: the fraction of ligand atoms classified
          as buried, expressed as a percentage.
\end{itemize}

The average SASA is further mapped to one of three qualitative categories that are reported alongside the numerical metrics: \emph{highly buried} (average SASA ${<}\,5.0$~\AA$^{2}$), \emph{moderately buried} ($5.0$-$15.0$~\AA$^{2}$), or \emph{significantly exposed} (${>}\,15.0$~\AA$^{2}$). These thresholds were chosen to provide the LLM agent with an interpretable summary that correlates with binding-site occupancy: highly buried ligands typically fill a well-defined pocket, moderately buried ones occupy a partial pocket with solvent-exposed substituents, and significantly exposed ligands are likely surface-adsorbed artefacts.

The burial ratio additionally serves as the input for the downstream binding mode classification tool (Section~\ref{sec:derived-tools}), which assigns each pose to one of three binding modes (\texttt{deep\_pocket}, \texttt{partial\_pocket}, or \texttt{surface}) based on defined burial-ratio thresholds. This two-stage design separates the quantitative SASA calculation from the categorical interpretation, allowing the agent to reason over both raw numbers and high-level classifications when comparing poses.

\subsubsection{Conformational Strain Analysis}
\label{sec:strain}

We quantify the energetic penalty associated with the bound-state ligand conformation by comparing its molecular mechanics energy to that of a force-field-optimised reference conformation. Conformational strain is a well-established indicator of pose quality: productive binding modes typically exhibit moderate conformational strain offset by favorable protein-ligand interactions, whereas docking artefacts often show excessive strain without compensatory binding contacts~\cite{gu2021strain}.

The calculation is implemented using the MMFF94 force field~\cite{halgren1996mmff} as provided in RDKit~\cite{rdkit2025,tosco2014mmff_rdkit} and operates on the raw pose file rather than on the protein-ligand complex, as detailed in Algorithm~\ref{alg:strain}.

\begin{algorithm}[ht]
\setlength{\baselineskip}{1.3\baselineskip}
\caption{Conformational Strain Energy Calculation}
\label{alg:strain}
\KwIn{Bound-state ligand conformation $M_{\mathrm{pose}}$}
\KwOut{Strain energy $\Delta E_{\mathrm{strain}}$ (kcal/mol), structural deviation RMSD (\AA)}

$M_{\mathrm{ref}} \gets \text{Copy}(M_{\mathrm{pose}})$\;
Add explicit hydrogens with 3D coordinates to $M_{\mathrm{pose}}$ and $M_{\mathrm{ref}}$ if absent\;

\eIf{MMFF94 parameterisation succeeds for $M_{\mathrm{ref}}$}{
    Minimise $M_{\mathrm{ref}}$ with MMFF94 (max 500 iterations)\;
}{
    Minimise $M_{\mathrm{ref}}$ with UFF (max 500 iterations)\;
}

$E_{\mathrm{pose}} \gets \text{ForceFieldEnergy}(M_{\mathrm{pose}})$\;
$E_{\mathrm{ref}} \gets \text{ForceFieldEnergy}(M_{\mathrm{ref}})$\;
$\Delta E_{\mathrm{strain}} \gets E_{\mathrm{pose}} - E_{\mathrm{ref}}$\;

$\text{RMSD} \gets \text{AlignAndMeasure}(M_{\mathrm{ref}},\; M_{\mathrm{pose}})$\;

\Return{$\Delta E_{\mathrm{strain}}$, RMSD}\;
\end{algorithm}

The strain energy $\Delta E_{\mathrm{strain}} = E_{\mathrm{pose}} - E_{\mathrm{ref}}$ is computed in kcal/mol, where a positive value indicates that the bound conformation is higher in energy than the relaxed reference, with the magnitude reflecting the thermodynamic cost of adopting the bound-state geometry. The RMSD complements the energetic metric by revealing whether the strain arises from localised distortions (low RMSD despite non-negligible strain) or from large-scale conformational rearrangements (high RMSD).

The strain energy is mapped to one of five qualitative categories to provide the LLM agent with an interpretable summary: \emph{excellent} ($\Delta E_{\mathrm{strain}} < 1$~kcal/mol), \emph{good} ($1$-$3$~kcal/mol), \emph{acceptable} ($3$-$6$~kcal/mol), \emph{concerning} ($6$-$10$~kcal/mol), or \emph{poor} (${>}\,10$~kcal/mol, likely indicative of a docking artefact). Analogously, the RMSD is classified as \emph{minimal structural deviation} (${<}\,0.5$~\AA), \emph{small structural changes} ($0.5$-$1.0$~\AA), \emph{moderate conformational changes} ($1.0$-$2.0$~\AA), or \emph{significant conformational rearrangement} (${>}\,2.0$~\AA).

If the MMFF94 parameterisation fails for either the bound or reference molecule, typically because the ligand contains atom types outside the force field's coverage, the tool raises a structured error rather than returning unreliable energies. This fail-safe prevents silent propagation of incorrect strain values into the agent's decision context.

The agent's system prompt instructs the LLM to interpret strain in combination with other metrics rather than in isolation. In particular, moderate strain ($3$-$6$~kcal/mol) can be acceptable if offset by a strong interaction network, a principle consistent with the observation that protein-bound ligands frequently adopt conformations above their global minimum in exchange for complementary interactions with the binding site~\cite{gu2021strain}. Conversely, very low strain alone does not guarantee a productive binding mode if the pose lacks meaningful protein-ligand contacts.

\paragraph{Caveat on absolute strain magnitudes.}
The reader will encounter raw $\Delta E_{\mathrm{strain}}$ values in the range $44$-$97$~kcal/mol in the representative-system analyses (Section~\ref{sec:results-representative}) and the case-study tables that follow. These absolute magnitudes exceed by roughly an order of magnitude what crystallographic surveys of protein-bound ligands would predict (${\lesssim}\,5$~kcal/mol for the majority of complexes)~\cite{perola2004conformational}, consistent with known tendencies of MMFF94 to overestimate small-molecule strain~\cite{halgren1996mmff} and with the use of a static, unminimised bound-state geometry rather than a restrained complex minimisation. We therefore flag here, before the absolute values are used in downstream analyses, that the raw numbers are not interpretable as physical strain penalties. They enter the pipeline as \emph{relative} discriminators only: across poses of the same ligand in the same system, ranking differences in $\Delta E_{\mathrm{strain}}$ remain informative even when the baseline is inflated, and the agent's system prompt makes this relative-only interpretation explicit (Section~\ref{sec:agent-architecture}). Validating the absolute scale, for example by restrained minimisation of the bound-state complex or by substituting a higher-fidelity force field or a quantum-mechanical torsion scan, is deferred to future work (Section~\ref{sec:disc-future-directions}).

\subsubsection{Steric Clash Detection}
\label{sec:clashes}

We detect van~der~Waals (vdW) overlaps between ligand and protein atoms as a direct indicator of geometrically implausible docking poses. Poses with numerous or deep steric clashes violate fundamental physical constraints and are almost invariably docking artefacts, making clash count and severity valuable discriminators during multi-criteria ranking.

The clash detection algorithm operates on two inputs: (i)~the protein-ligand complex PDB file produced during preprocessing (Section~\ref{sec:system-overview}), from which protein atom coordinates are extracted, and (ii)~the raw ligand pose file, from which ligand atom coordinates and element types are obtained via RDKit~\cite{rdkit2025}. Only standard protein residues are included in the analysis; water molecules and metal ions (the only HETATM records retained after preprocessing) are excluded because displaced waters are expected at the binding site and because metal-ligand coordination distances (${\sim}2.0$-$2.5$~\AA) fall below the vdW overlap threshold and would generate false-positive clashes.

The procedure is formalised in Algorithm~\ref{alg:clash}.

\begin{algorithm}[!ht]
\setlength{\baselineskip}{1.3\baselineskip}
\DontPrintSemicolon
\SetAlgoLined
\SetKwInOut{Input}{Input}
\SetKwInOut{Output}{Output}
\Input{Ligand atom set $\mathcal{L} = \{l_1, \dots, l_N\}$ with coordinates and element types;\newline
       Protein atom set $\mathcal{P} = \{p_1, \dots, p_M\}$ (standard residues only);\newline
       Tolerance factor $\alpha = 0.75$}
\Output{Clash count $C$, clash energy $E$, clash severity $S$, clashing pairs $\mathcal{K}$}
\BlankLine
\ForEach{atom $a \in \mathcal{L} \cup \mathcal{P}$}{
    $r_{\mathrm{vdW}}^{a} \leftarrow$ tabulated vdW radius for element of $a$\;
    \lIf{element not parameterised}{$r_{\mathrm{vdW}}^{a} \leftarrow 1.7$~\AA}
}
\BlankLine
$\mathbf{D} \leftarrow$ pairwise distance matrix between $\mathcal{L}$ and $\mathcal{P}$\;
$\mathcal{K} \leftarrow \emptyset$\;
\BlankLine
\For{$i \leftarrow 1$ \KwTo $N$}{
    \For{$j \leftarrow 1$ \KwTo $M$}{
        $d_{\mathrm{thresh}} \leftarrow \alpha \cdot \bigl(r_{\mathrm{vdW}}^{l_i} + r_{\mathrm{vdW}}^{p_j}\bigr)$\;
        \If{$\mathbf{D}_{ij} < d_{\mathrm{thresh}}$}{
            $\delta \leftarrow d_{\mathrm{thresh}} - \mathbf{D}_{ij}$\;
            $\mathcal{K} \leftarrow \mathcal{K} \cup \{(i, j, \delta)\}$\;
        }
    }
}
\BlankLine
$C \leftarrow |\mathcal{K}|$\;
$E \leftarrow \displaystyle\sum_{(i,j,\delta)\,\in\,\mathcal{K}} \delta^{2}$\;
$S \leftarrow \min\!\bigl(1,\; E \,/\, (N \times 0.25)\bigr)$\;
\BlankLine
\Return{$C,\; E,\; S,\; \mathcal{K}$}\;
\caption{Steric clash detection between ligand and protein atoms.}
\label{alg:clash}
\end{algorithm}

The tolerance factor $\alpha = 0.75$ follows PoseBusters~\cite{buttenschoen2024posebusters}, which uses the same threshold for its intermolecular clash test, a value at which nearly all experimentally resolved crystal structures pass. This choice accounts for the "softness" of vdW surfaces, so that only contacts representing genuine steric violations, rather than routine close contacts, are flagged. The clash energy $E$ penalises deep penetrations quadratically, while the severity $S$ normalises $E$ by the maximum reasonable overlap for the ligand (corresponding to a 0.5~\AA{} overlap per ligand atom) and clamps the result between 0 and 1.

The clash count is mapped to one of five qualitative categories: \emph{excellent} (0~clashes), \emph{good} (${\leq}\,2$), \emph{acceptable} ($3$-$5$), \emph{concerning} ($6$-$10$), or \emph{poor} (${>}\,10$~clashes, strongly indicative of a false-positive pose). Additionally, the ten most severe clashes, ranked by overlap magnitude, are reported with their ligand atom index, protein residue identity, and overlap distance, enabling the LLM agent to assess whether clashes occur in the core binding region or in peripheral, flexible segments of the ligand.

The agent's system prompt provides guidelines for interpreting clash information in context. In particular, the agent is instructed to distinguish peripheral clashes (solvent-exposed tails) from clashes near key interaction sites, and to consider the correlation between clashes and conformational strain: a high clash count coupled with high strain is especially problematic, whereas modest clashes with low strain may be resolvable through minor rigid-body adjustment.

\subsubsection{Unsatisfied Polar Atom Penalty}
\label{sec:polar-penalty}

When a polar atom on the ligand is buried inside the protein binding site without forming a compensating hydrogen bond or electrostatic contact, the desolvation cost of stripping its hydration shell is not offset, incurring a free-energy penalty estimated at several kcal/mol per unsatisfied group~\cite{bissantz2010guide}. The same penalty applies in reverse when the ligand occludes a protein polar atom without forming a compensating contact, but the current implementation evaluates only the ligand side; extending the analysis to protein-side desolvation is left as future work. Even a single such penalty can negate a substantial portion of the binding free energy, making the count of unsatisfied buried polar atoms a sensitive discriminator between productive binding modes and docking artefacts.

We quantify this effect by combining SMARTS-based identification of ligand polar atoms with atomic-resolution SASA burial assessment and distance-based proximity checking against protein polar atoms. The algorithm operates on two inputs: (i)~the protein-ligand complex PDB file, from which protein polar-atom coordinates and per-atom SASA values are obtained, and (ii)~the raw ligand pose file, from which polar atoms are identified via RDKit~\cite{rdkit2025}. It proceeds in four stages, formalised in Algorithm~\ref{alg:polar}.

\paragraph{Stage 1: ligand polar-atom identification.}
Hydrogen-bond acceptor (HBA) atoms are identified by three SMARTS patterns applied to the ligand molecule: (i)~nitrogen acceptors \texttt{[N;!\$(N-[SX4](=O)(=O));!\$(N-C(=O)-O);!\$(N=*)]} (excluding nitro groups, carbamates, and double-bonded nitrogens), (ii)~terminal oxygen atoms \texttt{[O;D1]} (e.g.\ carbonyls), and (iii)~divalent oxygen atoms \texttt{[O;D2]} (e.g.\ ethers and hydroxyls). Hydrogen-bond donor (HBD) atoms are identified by a single SMARTS pattern \texttt{[NX3,NX4,OX2,SX2;!H0]}, which matches any nitrogen, oxygen, or sulfur atom bearing at least one explicit or implicit hydrogen. The union of HBA and HBD atom indices constitutes the set of ligand polar atoms $\mathcal{A}$.

\paragraph{Stage 2: burial assessment.}
Per-atom SASA values are computed on the full protein-ligand complex using the Shrake-Rupley algorithm~\cite{shrake1973environment} via BioPython~\cite{cock2009biopython} at atomic resolution (\texttt{level="A"}), following the same procedure and residue-exclusion criteria described in Section~\ref{sec:sasa}. A ligand polar atom $a \in \mathcal{A}$ is classified as \emph{buried} if its SASA falls below a threshold of $\tau_{\mathrm{burial}} = 1.0$~\AA$^{2}$. Atoms whose SASA could not be determined (e.g.\ due to naming mismatches) are assigned a default SASA of 100.0~\AA$^{2}$, conservatively treating them as exposed and thus excluding them from the penalty.

\paragraph{Stage 3: interaction checking.}
For each buried ligand polar atom, we test whether a compensating interaction with the protein exists by measuring the Euclidean distance to every protein polar atom. The protein polar-atom set $\mathcal{P}$ comprises all backbone nitrogen and oxygen atoms of standard residues plus the following side-chain atoms: Ser~O$\gamma$, Thr~O$\gamma$1, Tyr~O$\eta$, Cys~S$\gamma$, His~N$\delta$1 and N$\varepsilon$2, Lys~N$\zeta$, Arg~N$\varepsilon$, N$\eta$1, and N$\eta$2, Asn~O$\delta$1 and N$\delta$2, Gln~O$\varepsilon$1 and N$\varepsilon$2, Asp~O$\delta$1 and O$\delta$2, and Glu~O$\varepsilon$1 and O$\varepsilon$2. A buried polar atom is considered \emph{satisfied} if any protein polar atom lies within $d_{\mathrm{int}} = 3.5$~\AA; otherwise it is marked as \emph{unsatisfied}. This criterion evaluates direct protein contacts only; satisfaction through a water-mediated hydrogen bond, which PLIP detects as a water bridge (Section~\ref{sec:plip}), is not accounted for in the current penalty calculation.

\paragraph{Stage 4: penalty scoring.}
Each unsatisfied buried polar atom incurs a penalty of 0.5 units. The raw penalty is capped at a maximum of 10.0 to prevent a single severely penalised pose from dominating the multi-criteria comparison. A burial ratio $\rho = n_{\mathrm{unsat}} / |\mathcal{A}|$ is additionally reported, expressing the fraction of the ligand's total polar atoms that are both buried and unsatisfied.

\begin{algorithm}[!ht]
\setlength{\baselineskip}{1.3\baselineskip}
\DontPrintSemicolon
\SetAlgoLined
\SetKwInOut{Input}{Input}
\SetKwInOut{Output}{Output}
\Input{Ligand polar atom set $\mathcal{A}$ (HBA $\cup$ HBD, identified by SMARTS);\newline
       Protein polar atom set $\mathcal{P}$ (backbone N, O $+$ side-chain polar atoms);\newline
       Per-atom SASA values from complex;\newline
       Burial threshold $\tau_{\mathrm{burial}} = 1.0$~\AA$^{2}$;\newline
       Interaction distance $d_{\mathrm{int}} = 3.5$~\AA;\newline
       Penalty weight $w = 0.5$; maximum penalty $P_{\max} = 10.0$}
\Output{Unsatisfied count $n_{\mathrm{unsat}}$, penalty $P$, burial ratio $\rho$}
\BlankLine
$n_{\mathrm{unsat}} \leftarrow 0$\;
\BlankLine
\ForEach{atom $a \in \mathcal{A}$}{
    $\mathrm{SASA}_{a} \leftarrow$ atomic SASA of $a$ in the complex\;
    \lIf{$\mathrm{SASA}_{a}$ unavailable}{$\mathrm{SASA}_{a} \leftarrow 100.0$}
    \If{$\mathrm{SASA}_{a} < \tau_{\mathrm{burial}}$}{
        $d_{\min} \leftarrow \displaystyle\min_{p \,\in\, \mathcal{P}} \|\mathbf{x}_{a} - \mathbf{x}_{p}\|$\;
        \If{$d_{\min} \geq d_{\mathrm{int}}$}{
            $n_{\mathrm{unsat}} \leftarrow n_{\mathrm{unsat}} + 1$\;
        }
    }
}
\BlankLine
$P \leftarrow \min(P_{\max},\; n_{\mathrm{unsat}} \times w)$\;
$\rho \leftarrow n_{\mathrm{unsat}} \,/\, |\mathcal{A}|$\;
\BlankLine
\Return{$n_{\mathrm{unsat}},\; P,\; \rho$}\;
\caption{Unsatisfied polar atom penalty calculation.}
\label{alg:polar}
\end{algorithm}

The penalty score is mapped to one of five qualitative categories: \emph{excellent} ($P = 0$), \emph{good} ($P \leq 1.0$), \emph{acceptable} ($1.0 < P \leq 2.5$), \emph{concerning} ($2.5 < P \leq 5.0$), or \emph{poor} ($P > 5.0$). In addition to the per-atom classification, the tool reports each unsatisfied atom's index, element type, and SASA value, allowing the LLM agent to assess whether a penalty arises from deeply buried pharmacophoric groups (more severe) or from marginally buried, solvent-accessible atoms (potentially less consequential).

The agent's system prompt instructs the LLM to weigh polar penalties in concert with other metrics. Buried unsatisfied \emph{charged} groups (e.g.\ amines, carboxylates) carry a higher implicit desolvation cost than neutral polar atoms, and the agent is directed to flag such cases as especially problematic. Conversely, a moderate penalty score may be acceptable if the corresponding atoms lie at the pocket periphery and the pose otherwise exhibits strong interaction quality and low conformational strain.

\subsubsection{SMILES Extraction}
\label{sec:smiles}

The sixth primary tool extracts a canonical isomeric SMILES string for each ligand pose using RDKit~\cite{rdkit2025}. Unlike the five preceding tools, SMILES extraction does not produce a ranking metric; it serves as a chemical-identity check that determines whether the poses under comparison are conformers of the same molecule or distinct ligands. All ten benchmark systems in this study involve a single ligand docked in multiple conformations, so the SMILES check reduces to a no-op diversity verification and the SMILES block is omitted from every anonymised analysis file passed to the LLM; the multi-ligand code path is an architectural provision that becomes active only in use cases such as virtual-screening hit-list triage. Algorithmic detail (canonicalisation settings, hydrogen handling, and the diversity-check logic) is provided in Supplementary Section~\ref{sec:si-smiles}.

\subsubsection{Derived Assessment Tools}
\label{sec:derived-tools}

The six primary tools described in Sections~\ref{sec:plip} to~\ref{sec:smiles} each probe a single physicochemical property of the protein-ligand complex. Three additional \emph{derived} tools (Interaction Quality Assessment, Binding Mode Classification, and Ligand Efficiency Assessment), described in the remainder of this section, are executed sequentially after all primary analyses have completed. Rather than re-running molecular simulations, each derived tool reads the structured analysis files already written by the primary tools and computes higher-level metrics that integrate information across multiple properties. This design avoids redundant computation while enabling the agent to reason about composite quantities such as the geometric quality of a hydrogen bond or the number of interactions per heavy atom.

The three derived tools are executed in a fixed order: interaction quality first, binding mode classification second, and ligand efficiency last. This ordering is deliberate: the ligand efficiency tool requires the total quality score produced by the interaction quality tool, so the latter must complete first. All derived metrics are appended to the corresponding analysis files before the final anonymisation and presentation to the LLM agent.

\paragraph{Interaction Quality Assessment.}
\label{par:interaction-quality}

The PLIP tool (Section~\ref{sec:plip}) reports the \emph{count} and geometric parameters of detected protein-ligand interactions, but it does not distinguish a strong, near-ideal hydrogen bond from a marginal one that barely satisfies the detection threshold. The Interaction Quality Assessment tool closes this gap by assigning a continuous quality score from 0 to 1 to every individual interaction based on its geometric optimality, and then aggregating these scores into a total quality metric for the pose.

Three interaction types are scored: hydrogen bonds, hydrophobic contacts, and salt bridges.

\subparagraph{Hydrogen-bond quality.}
Hydrogen-bond geometry is evaluated along two dimensions: heavy-atom donor-acceptor distance~$d$ and donor-hydrogen-acceptor angle~$\theta$. Each dimension is mapped to a score $s_{d}$ and $s_{\theta}$, respectively, via piecewise-constant functions informed by crystallographic statistics of hydrogen bonds in organic and protein crystals~\cite{jeffrey1997hydrogen,steiner2002hydrogen}:

\begin{equation}
\label{eq:hbond-dist-score}
s_{d} =
\begin{cases}
1.0 & \text{if } d \leq 3.0\;\text{\AA}, \\
0.9 & \text{if } 3.0 < d \leq 3.2\;\text{\AA}, \\
0.6 & \text{if } 3.2 < d \leq 3.5\;\text{\AA}, \\
0.3 & \text{if } 3.5 < d \leq 3.8\;\text{\AA}, \\
0.1 & \text{otherwise},
\end{cases}
\end{equation}

\begin{equation}
\label{eq:hbond-angle-score}
s_{\theta} =
\begin{cases}
1.0 & \text{if } \theta \geq 150^{\circ}, \\
0.9 & \text{if } 140^{\circ} \leq \theta < 150^{\circ}, \\
0.7 & \text{if } 130^{\circ} \leq \theta < 140^{\circ}, \\
0.5 & \text{if } 120^{\circ} \leq \theta < 130^{\circ}, \\
0.3 & \text{if } 110^{\circ} \leq \theta < 120^{\circ}, \\
0.1 & \text{otherwise}.
\end{cases}
\end{equation}

\noindent Both functions assign a floor of 0.1 rather than zero because they are applied exclusively to interactions that PLIP has already detected as hydrogen bonds using its own, broader geometric thresholds. A contact that passes PLIP's detection criteria but falls outside the quality scorer's optimal ranges still receives a small positive contribution to $Q$, preserving consistency with the interaction count $n_{\mathrm{total}}$ (which includes every PLIP-detected contact regardless of geometry).

\noindent The combined score for a single hydrogen bond is the geometric mean of the two components:

\begin{equation}
\label{eq:hbond-quality}
q_{\mathrm{hb}} = \sqrt{s_{d} \cdot s_{\theta}}\,.
\end{equation}

\noindent The geometric mean is preferred over the arithmetic mean because it is more sensitive to imbalances: a hydrogen bond with an excellent distance but a poor angle (or vice versa) receives a lower score than one with moderate values in both dimensions. To facilitate the agent's qualitative reasoning, each hydrogen bond is additionally labelled as \emph{excellent} ($q_{\mathrm{hb}} \geq 0.9$), \emph{good} ($0.7 \leq q_{\mathrm{hb}} < 0.9$), \emph{acceptable} ($0.5 \leq q_{\mathrm{hb}} < 0.7$), \emph{weak} ($0.3 \leq q_{\mathrm{hb}} < 0.5$), or \emph{poor} ($q_{\mathrm{hb}} < 0.3$).

\subparagraph{Hydrophobic contact quality.}
The quality of a hydrophobic contact depends primarily on the distance between the interacting non-polar atoms, with shorter contacts indicating tighter non-polar packing and correspondingly stronger van~der~Waals attraction~\cite{bissantz2010guide}. The distance-based score is:

\begin{equation}
\label{eq:hydro-score}
q_{\mathrm{hp}} =
\begin{cases}
1.0  & \text{if } d \leq 3.5\;\text{\AA}, \\
0.85 & \text{if } 3.5 < d \leq 3.8\;\text{\AA}, \\
0.7  & \text{if } 3.8 < d \leq 4.0\;\text{\AA}, \\
0.5  & \text{if } 4.0 < d \leq 4.3\;\text{\AA}, \\
0.3  & \text{if } 4.3 < d \leq 4.5\;\text{\AA}, \\
0.15 & \text{otherwise}.
\end{cases}
\end{equation}

\noindent Contacts scoring $q_{\mathrm{hp}} \geq 0.85$ are labelled \emph{strong}, those with $0.3 \leq q_{\mathrm{hp}} < 0.85$ as \emph{moderate}, and the remainder as \emph{weak}.

\subparagraph{Salt-bridge quality.}
Salt bridges are scored by the distance between the centres of the interacting charged groups, informed by established distance criteria for ion pairs in proteins~\cite{kumar1999salt}:

\begin{equation}
\label{eq:salt-score}
q_{\mathrm{sb}} =
\begin{cases}
1.0 & \text{if } d \leq 3.5\;\text{\AA}, \\
0.8 & \text{if } 3.5 < d \leq 4.0\;\text{\AA}, \\
0.5 & \text{if } 4.0 < d \leq 4.5\;\text{\AA}, \\
0.3 & \text{if } 4.5 < d \leq 5.0\;\text{\AA}, \\
0.1 & \text{otherwise}.
\end{cases}
\end{equation}

\noindent Salt bridges scoring $q_{\mathrm{sb}} \geq 0.8$ are classified as \emph{strong}, those with $0.3 \leq q_{\mathrm{sb}} < 0.8$ as \emph{moderate}, and the rest as \emph{weak}.

\subparagraph{Aggregation.}
The total quality score for a pose is the sum of all individual interaction quality scores:

\begin{equation}
\label{eq:total-quality}
Q = \sum_{i=1}^{n_{\mathrm{hb}}} q_{\mathrm{hb},i}
  + \sum_{j=1}^{n_{\mathrm{hp}}} q_{\mathrm{hp},j}
  + \sum_{k=1}^{n_{\mathrm{sb}}} q_{\mathrm{sb},k}\,.
\end{equation}

\noindent Unlike the individual scores, $Q$ is not normalised to 0 to 1 and grows with the number of interactions. This is intentional: a pose forming many geometrically favourable interactions should receive a higher composite score than one forming only a few. Because all poses within a single ranking call represent conformations of the same compound, they share the same heavy-atom count; $Q$ differences between poses therefore reflect genuine differences in interaction networks, not molecular size. When molecules of different sizes must be compared, the quality efficiency $\mathrm{QE} = Q / N_{\mathrm{HA}}$ (Section~\ref{par:ligand-efficiency}) provides the size-normalised counterpart. The agent's system prompt provides interpretive guidance: $Q > 10$ indicates high overall quality, $5 \leq Q \leq 10$ moderate quality, and $Q < 5$ low quality. Per-type averages and counts of excellent, good, and poor interactions are also reported, enabling the agent to identify, for example, a pose with many interactions of which the majority are geometrically poor.

\paragraph{Binding Mode Classification.}
\label{par:binding-mode}

The SASA-based burial ratio computed in Section~\ref{sec:sasa} quantifies how much of the ligand's surface area is shielded from solvent in the bound state (by the protein and, to a lesser extent, by intramolecular self-occlusion), but it does not directly answer the pharmacologically relevant question of \emph{where} the ligand sits in relation to the protein surface. The Binding Mode Classification tool maps the burial ratio and the average per-atom SASA to one of three discrete binding modes, each associated with a confidence score and a quality estimate:

\begin{itemize}
  \item \textbf{DEEP\_POCKET.} The ligand is substantially enclosed within the protein interior. Two threshold pairs can trigger this classification, evaluated in order of decreasing stringency: the strict pair ($\rho_{\mathrm{b}} \geq 70\%$ and $\bar{s} < 2.0$~\AA$^{2}$) is tested first and yields a confidence score between 0.70 and 0.95; if it fails, the relaxed pair ($\rho_{\mathrm{b}} \geq 60\%$ and $\bar{s} < 3.0$~\AA$^{2}$) is tested and yields a confidence between 0.60 and 0.85. Deep-pocket binding is characteristic of well-docked poses and carries the highest quality scores.

  \item \textbf{PARTIAL\_POCKET.} The ligand is partially enclosed by the protein. Again, two threshold pairs are tested in order: the dual-criteria condition ($\rho_{\mathrm{b}} \geq 50\%$ and $\bar{s} < 4.0$~\AA$^{2}$, confidence 0.60) is evaluated first; if it fails, a burial-ratio-only fallback ($\rho_{\mathrm{b}} \geq 40\%$, confidence 0.50) applies. Partial binding may reflect an extended binding site or a ligand that protrudes from the pocket; the quality score decreases as either metric deteriorates.

  \item \textbf{SURFACE.} The ligand lies predominantly on the protein surface with minimal burial. This mode is assigned when $\rho_{\mathrm{b}} < 40\%$. Surface binding almost always indicates a docking artefact, and the classification carries the lowest quality score.
\end{itemize}

\noindent The dual-criteria design, requiring both a burial ratio threshold \emph{and} a per-atom SASA constraint for the higher-quality modes, prevents misclassification of large ligands that achieve a high nominal burial ratio while still presenting substantial solvent-exposed surface area per atom. A fourth label, \textsc{UNCERTAIN}, is assigned when the required SASA data are unavailable.

The agent's system prompt instructs the LLM to treat binding mode as a strong discriminator: surface-bound poses should never rank first regardless of their interaction counts, because a high number of interactions on the protein surface almost certainly reflects non-specific contacts rather than genuine binding. This rule assumes the target has a well-defined binding pocket; for targets with shallow grooves, solvent-exposed allosteric sites, or protein-protein interaction hotspots, surface binding can be the legitimate mode, and the directive would penalise the correct pose (Section~\ref{sec:disc-limitations}). Conversely, a deep-pocket classification reinforces the plausibility of a pose and lends additional weight to its interaction profile and complementarity metrics.

\paragraph{Ligand Efficiency Assessment.}
\label{par:ligand-efficiency}

Larger ligands tend to form more interactions simply by virtue of having more atoms, which can bias naive ranking schemes toward bulkier molecules or toward poses that happen to extend into secondary sites. The Ligand Efficiency Assessment tool corrects for this size effect by normalising interaction counts and quality scores by the number of heavy (non-hydrogen) atoms~$N_{\mathrm{HA}}$, inspired by the ligand-efficiency concept introduced by Hopkins~et~al.~\cite{hopkins2004ligand}.

Two efficiency metrics are computed:

\begin{equation}
\label{eq:le}
\mathrm{LE} = \frac{n_{\mathrm{total}}}{N_{\mathrm{HA}}}\,,
\end{equation}

\noindent where $n_{\mathrm{total}}$ is the total number of detected interactions (hydrogen bonds, hydrophobic contacts, and salt bridges); and, when the interaction-quality total $Q$ is available from the preceding tool:

\begin{equation}
\label{eq:quality-eff}
\mathrm{QE} = \frac{Q}{N_{\mathrm{HA}}}\,.
\end{equation}

\noindent The burial ratio~$\rho_{\mathrm{b}}$ (Section~\ref{sec:sasa}) is not normalised by~$N_{\mathrm{HA}}$ because it is already defined as the fraction of ligand atoms classified as buried and is therefore inherently size-independent.

The ligand efficiency~$\mathrm{LE}$ is mapped to a qualitative rating: \emph{excellent} ($\mathrm{LE} \geq 0.4$), \emph{good} ($0.3 \leq \mathrm{LE} < 0.4$), \emph{acceptable} ($0.2 \leq \mathrm{LE} < 0.3$), or \emph{poor} ($\mathrm{LE} < 0.2$). A per-type breakdown is additionally reported, expressing the number of hydrogen bonds, hydrophobic contacts, and salt bridges per heavy atom. This breakdown helps the agent distinguish, for example, two poses with identical~$\mathrm{LE}$ values where one achieves its efficiency through many weak hydrophobic contacts and the other through a smaller number of geometrically strong hydrogen bonds.

The agent is instructed to use ligand efficiency as a tie-breaking criterion: when two poses exhibit similar raw interaction counts and quality scores, the pose with the higher~$\mathrm{LE}$ is preferred, as it achieves comparable recognition with fewer atomic degrees of freedom. Because all ten benchmark systems involve a single ligand, every pose within a given system shares the same~$N_{\mathrm{HA}}$; both~$\mathrm{LE}$ and~$\mathrm{QE}$ are therefore rank-equivalent to their unnormalised counterparts ($n_{\mathrm{total}}$ and~$Q$, respectively) and cannot change the ordering that raw metrics already determine. Size normalisation would become informative only in multi-ligand comparisons where molecules with different heavy-atom counts compete directly.

\subsection{Pose Anonymisation Protocol}
\label{sec:anonymisation}

Large language models can exhibit positional bias, a tendency to favour items presented earlier or later in the input context, irrespective of their content~\cite{liu2024lost}. In a pose-ranking scenario, this bias could manifest as a systematic preference for poses that appear first in the analysis context or whose filenames suggest a lower ordinal number (e.g.\ \texttt{pose\_01} over \texttt{pose\_09}). Because the docking engine typically assigns pose names in order of decreasing docking score, exposing these names to the LLM would allow it to infer, consciously or implicitly, the scoring-function ranking, defeating the purpose of an independent re-evaluation.

We mitigate both positional and naming bias through a deterministic anonymisation protocol with three stages. First, each pose is assigned an eight-character alphanumeric code by hashing the concatenation of the system identifier and pose name:

\begin{equation}
\label{eq:anon-code}
\mathrm{code}(s, p) = \mathrm{Base36}\!\bigl(\mathrm{SHA\text{-}256}(s \,\|\, p)\;[0{:}8]\bigr)\,,
\end{equation}

\noindent where $s$ is the system identifier, $p$ the pose name, and $\|$ denotes string concatenation. SHA-256's avalanche property ensures that consecutively numbered poses produce codes with no discernible pattern. Second, all tool outputs are passed through a sanitisation step that replaces every occurrence of the original pose name with the anonymous code, preventing identity leakage into the LLM's context. Third, after inference, a de-anonymisation step maps the codes back to the original pose names for human-readable reporting. Design rationale, implementation details, and context-presentation rules are provided in Supplementary Section~\ref{sec:si-anonymisation}.

\subsection{Agent Architecture}
\label{sec:agent-architecture}

The analysis tools described in Sections~\ref{sec:plip}-\ref{sec:derived-tools} produce a rich but heterogeneous body of evidence for every pose: geometric interaction parameters, burial statistics, strain energies, clash counts, polar penalties, quality scores, binding-mode classifications, and ligand-efficiency metrics. Translating this multi-dimensional evidence into a single, justified ranking is the responsibility of an LLM-based reasoning agent whose architecture is described in this section.

\paragraph{Model and orchestration.}
We employ OpenAI's GPT-5 as the reasoning backbone, accessed through the OpenAI Agents SDK (version~0.5.0). The SDK provides a lightweight abstraction over the model API that supports structured agent definitions, comprising a name, a model identifier, model-level settings, and a system prompt loaded from an external file, together with a streaming execution runner that returns both the final answer and the chain-of-thought (CoT) reasoning trace as separate outputs. Three model-level parameters govern the reasoning behaviour: (i)~reasoning effort is set to \texttt{"high"}, instructing the model to allocate extended internal computation before committing to an answer; (ii)~the reasoning summary mode is set to \texttt{"detailed"}, requesting a verbose CoT trace that exposes the model's intermediate comparisons and trade-offs; and (iii)~model verbosity is set to \texttt{"high"}, encouraging exhaustive metric-by-metric discussion rather than terse conclusions. These settings are chosen to maximise the transparency and auditability of the ranking decision at the cost of higher token usage and latency.

Inference is executed via the SDK's streaming runner, which yields response events incrementally. Two event types are processed: \emph{text-delta} events, which deliver fragments of the structured final answer, and \emph{reasoning-item} events, which expose the model's internal reasoning summary once inference completes. Both streams are accumulated, de-anonymised (Section~\ref{sec:anonymisation}), and persisted to a comprehensive analysis log that records the full system prompt, the anonymised input context, the raw CoT reasoning, and the final ranking.

\paragraph{System prompt architecture.}
The agent's behaviour is governed by a domain-specific system prompt of approximately 7\,800 words, maintained as a standalone Markdown file and loaded at runtime. The prompt is organised into seven major sections, each encoding a distinct aspect of the agent's role:

\begin{enumerate}
    \item \textbf{Anonymous code integrity} (${\sim}500$~words). Rules for handling the eight-character pose identifiers introduced by the anonymisation protocol (Section~\ref{sec:anonymisation}): character-for-character copying, case sensitivity, mandatory cross-checking against analysis-file headers, and worked examples of correct versus incorrect usage. These instructions prevent transcription errors that would break the automatic de-anonymisation step.

    \item \textbf{Analysis data documentation} (${\sim}4\,200$~words). A tool-by-tool reference describing the algorithm, output format, key metrics, interpretation thresholds, and common pitfalls for each of the nine computational tools. For every tool, the section mirrors the information that a domain expert would need to interpret the raw output: threshold values that delineate quality categories (e.g.\ "excellent" versus "poor" strain), caveats about metric reliability (e.g.\ why a high interaction count on a surface-bound pose is misleading), and guidance on when a given metric should override or yield to another.

    \item \textbf{Evaluation criteria} (${\sim}200$~words). Four general principles, consistency over absolute values, multi-metric corroboration, penalty saturation with diminishing returns, and physicochemical plausibility as a hard constraint, that frame the agent's overall evaluation philosophy.

    \item \textbf{Hierarchical decision framework} (${\sim}1\,800$~words). The core ranking logic, described below.

    \item \textbf{Reasoning process} (${\sim}400$~words). Step-by-step instructions for systematic analysis: initial outlier scan, identification of key differentiators, iterative comparison with backtracking to source data, and mandatory identity verification at every step to prevent pose-label confusion.

    \item \textbf{Output format} (${\sim}500$~words). A template specifying five required output sections (see below) with worked examples illustrating the expected level of metric-by-metric detail, explicit weighting rationale, and uncertainty disclosure.

    \item \textbf{Reminders} (${\sim}200$~words). Closing instructions on scientific judgment, honesty about limitations, pose-identity consistency, and optional suggestions for future extensions (e.g.\ explicit water-network modelling, per-residue energy decomposition).
\end{enumerate}

\paragraph{Hierarchical decision framework.}
Rather than combining all metrics into a single composite score, the system prompt encodes a priority-ordered decision hierarchy that the agent is instructed to apply when evaluating and comparing poses. The hierarchy has four levels:

\begin{enumerate}
    \item \textbf{Binding mode (highest priority).}  Surface-bound poses (burial ratio ${<}\,40\%$) are eliminated from top-tier consideration regardless of other metrics. A high interaction count on a surface-bound pose is explicitly flagged as an artefact of non-specific contacts rather than evidence of productive binding.

    \item \textbf{Interaction quality.}  Among poses that pass the binding-mode filter, the agent evaluates the geometry-based quality of each interaction (Section~\ref{par:interaction-quality}) rather than the raw count. A worked example in the system prompt illustrates the arithmetic consequence of the $Q$ aggregation: three hydrogen bonds with an average quality of 0.85 yield $Q = 2.55$, whereas five hydrogen bonds averaging 0.45 yield only $Q = 2.25$. The example is not an externally imposed rule but a direct consequence of the scoring function defined in Equations~\ref{eq:hbond-quality}-\ref{eq:total-quality}, which weights geometric optimality so that a smaller number of near-ideal contacts can outweigh a larger number of marginal ones.

    \item \textbf{Interaction quantity.}  Additional interactions improve the ranking only if quality is maintained. The agent is instructed to compute quality-per-interaction ratios and to prefer poses with fewer but geometrically superior contacts over those with many marginal ones.

    \item \textbf{Tiebreaker metrics.}  Conformational strain, steric clashes, unsatisfied polar penalties, and ligand efficiency serve as tiebreakers when the preceding levels do not clearly separate candidates. Strain ($> 15$~kcal/mol) is flagged as a tiebreaker penalty: among poses with comparable interaction networks, the one with lower strain is preferred, consistent with the thermodynamic rationale that conformational distortion reduces net binding affinity.
\end{enumerate}

\noindent A dedicated anti-bias section (${\sim}600$~words) within the decision framework addresses the "more interactions $=$ better" failure mode that was identified during early development. The section presents a concrete failure example (a surface-bound pose with 10 interactions ranked above a deep-pocket pose with 6 higher-quality interactions), states four principles for avoiding the bias, and provides a decision-making checklist that the agent is expected to internalise: (1)~check binding mode before evaluating counts; (2)~assess interaction quality, not just quantity; (3)~never accept a pose solely because it has the most interactions; (4)~surface binders should never rank first. A worked example demonstrates the expected reasoning pattern when quality and quantity disagree.

\paragraph{Handling conflicting signals.}
The system prompt provides guidance for four categories of metric conflict: (i)~interaction-rich but strained poses, which are flagged as potential over-optimised artefacts; (ii)~flat metric landscapes, where many poses have similar scores and the agent should report low confidence rather than over-interpret small numerical differences; (iii)~single-metric outliers, which warrant balanced interpretation rather than decisive promotion or demotion; and (iv)~cumulative weak signals, where the combination of multiple moderate deficiencies (e.g.\ weak interactions plus moderate strain plus poor burial) is treated as more problematic than any single issue in isolation. Additionally, the prompt encodes a two-step burial assessment protocol that combines absolute thresholds with relative ranking. Absolute thresholds serve as hard filters: poses with burial below~20\% are classified as surface-bound artefacts and penalised regardless of other metrics, while poses above~80\% receive strong priority as deeply pocket-engaged. For poses in the intermediate range (20-80\% burial), a percentile-based ranking is applied: poses are sorted by burial ratio and divided into top, middle, and bottom tiers relative to the current pose set. This hybrid strategy is necessary because acceptable burial levels vary between protein systems, a shallow binding groove may never produce burial ratios above~70\%, whereas a deep enclosed pocket routinely exceeds~90\%, so purely absolute thresholds would systematically misclassify poses in atypical systems, while purely relative ranking would fail to detect runs in which all poses are surface-bound.

\paragraph{Reasoning process.}
The agent is instructed to work iteratively rather than streaming a single-pass analysis. The prescribed reasoning workflow begins with an initial scan for obvious outliers (severe clashes, minimal interactions, extreme penalties), proceeds to identification of key differentiators, then enters a deep-dive comparison phase for closely ranked poses, and concludes with a verification step that cross-checks all conclusions against the original analysis-file data. A mandatory identity-verification rule requires the agent to confirm the exact pose identifier (anonymous code) at every step of the reasoning, re-opening the relevant analysis-file section if any ambiguity is detected. For large input contexts, the agent is permitted to construct intermediate summaries, but it must perform a final verification pass against the source data to guard against summarisation-induced errors.

\paragraph{Structured output format.}
\label{sec:agent-output}
The system prompt mandates five output sections, each with a defined purpose:

\begin{enumerate}
    \item \textbf{Best Pose Selection.}  The selected best pose, a comprehensive rationale explaining why it is superior, and a metric-by-metric breakdown that states the value and the weight assigned to each metric (interactions, burial/SASA, conformational strain, steric clashes, polar penalties). A worked example in the system prompt demonstrates the expected granularity, including statements such as "interaction quality (30\%), binding mode (25\%), burial (20\%), strain (15\%), clashes (5\%), polar penalties (5\%)".

    \item \textbf{Comparative Analysis.}  A brief explanation of why the two to three runner-up poses were not selected, with specific reference to the metrics that were decisive.

    \item \textbf{Confidence Assessment.}  A self-reported confidence level (high, medium, or low) accompanied by factors supporting or undermining confidence, and a list of uncertainties or additional data that could strengthen the decision.

    \item \textbf{Key Findings.}  Observations about the overall quality of the pose set, common patterns (shared strengths or weaknesses), and any systematic docking issues identified across multiple poses.

    \item \textbf{Recommendations.}  Suggestions for refining the selected pose (e.g.\ local minimisation, side-chain repacking) and identification of alternative poses worth considering as backup candidates.
\end{enumerate}

\noindent This structured format was designed to balance two needs: (a)~giving domain experts enough detail to audit, agree with, or override the agent's decision, and (b)~constraining the model's output to prevent rambling or omission of critical justification. The confidence-assessment section, in particular, addresses a limitation common to LLM-based systems: the tendency to present conclusions with false certainty~\cite{xiong2024llm_uncertainty}. By requiring the agent to enumerate its uncertainties, the prompt encourages calibrated rather than overconfident outputs.

\paragraph{Context assembly and input construction.}
Before inference, the anonymised analysis files, one per pose, are loaded, and each file's content is introduced to the model under its anonymous code (e.g.\ "Analysis file for pose: K7M9N2P4"). A preamble precedes the concatenated analysis data, explicitly stating that pose identifiers are randomised and that ranking must be based solely on the computational metrics. The assembled context, together with the system prompt, constitutes the full input to a single inference call; no multi-turn retrieval or tool-use loop is required during reasoning, because all relevant evidence has been pre-computed and embedded in the prompt.

The framework additionally supports follow-up queries through a conversation-history mechanism. After the initial ranking, users may pose clarifying questions (e.g.\ "Why was pose X ranked below pose Y despite having more interactions?"). In this mode, the full analysis context is re-included alongside the conversation history to ensure that the agent retains access to the source data when formulating its response, rather than relying on potentially lossy internal summaries of earlier turns.

\subsection{User Interface}
\label{sec:user-interface}

AgenticPosesRanker is deployed as a web application built with the Streamlit framework~\cite{streamlit2024}, offering an interactive graphical interface that exposes the full analysis pipeline without requiring command-line expertise (Supplementary Figure~\ref{fig:streamlit_ui}). The interface is designed around three principles: (i)~minimal configuration, users supply only a protein structure, a set of docked poses, and an API key; (ii)~transparency, every stage of the analysis is visible as it executes; and (iii)~interactivity, after the initial ranking, users can ask follow-up questions through a conversational chat, and the agent retains full access to the analysis data when formulating its responses. The application is also available as a command-line interface (CLI) for batch processing and scriptable workflows.

Implementation details, including the configuration workflow, two-column layout, progress feedback mechanism, follow-up conversation design, and error-handling strategy, are described in Supplementary Section~\ref{sec:si-user-interface}.

\subsection{Benchmark Construction}
\label{sec:benchmark-construction}

\subsubsection{Source Dataset}
\label{sec:source-dataset}

We derive our benchmark from
PDBbind~2016~\cite{wang2004pdbbind,wang2005pdbbind}, a curated
database of protein-ligand complexes for which both crystal
structures and experimentally determined binding affinities are
available. PDBbind applies quality filters including unambiguous
binding data and binary (single-protein, single-ligand) complex
topology, making it a widely adopted reference for scoring-function
development and evaluation. The 2016 release comprises 11{,}259
system directories, each containing the protein structure (PDB
format), the co-crystallised ligand (SDF format), and associated
binding-affinity annotations.

\paragraph{Docked poses.}
The docked poses used in this work were not generated by us; they were
obtained from the cross-docked dataset assembled by Francoeur
et~al.~\cite{francoeur2020crossdock} to train the Gnina~1.0 molecular
docking program~\cite{mcnutt2021gnina}. In that dataset, each system
was docked with Smina~\cite{koes2013smina}, a fork of AutoDock
Vina~\cite{trott2010vina}. Each docking run produced multiple candidate
poses with per-pose energy estimates (\texttt{minimizedAffinity} scores);
file format and directory structure details are given in Supplementary
Section~\ref{sec:si-data-files}.

\paragraph{Native pose generation.}
To enable atom-by-atom RMSD comparison between docked poses and the
crystallographic binding mode, we regenerated a \emph{native pose} for each
system by passing the co-crystallised ligand through Smina in scoring-only
mode, producing an output file with the same atom ordering and hydrogen
convention as the docked conformations. The full procedure is described in
Supplementary Section~\ref{sec:si-native-pose}.

\subsubsection{Pose Classification}
\label{sec:pose-classification}

Each docked pose was classified according to its structural deviation from the
crystallographic binding mode. We quantified deviation as the root-mean-square
deviation (RMSD) between the heavy atoms of the docked pose and the
corresponding native pose generated by Smina
(Section~\ref{sec:source-dataset}).

RMSD values were computed with \texttt{spyrmsd}~\cite{meli2020spyrmsd}, a
Python library that accounts for molecular symmetry through graph
isomorphism-based atom matching; computational details are given in
Supplementary Section~\ref{sec:si-rmsd-details}.

Based on the computed RMSD, each pose was assigned to one of three categories:

\begin{itemize}
    \item \textbf{Binder} (class~0): RMSD $< 2.0$~\AA{}, the pose closely
    reproduces the crystallographic binding mode.
    \item \textbf{Ambiguous} (class~1): $2.0 \leq \text{RMSD} \leq 4.0$~\AA{},
    intermediate deviation that precludes confident assignment.
    \item \textbf{Non-binder} (class~2): RMSD $> 4.0$~\AA{}, the pose
    departs substantially from the native geometry.
\end{itemize}

The 2.0~\AA{} threshold for near-native poses follows the convention widely
adopted in molecular docking benchmarks~\cite{warren2006critical}. The
intermediate category (2.0-4.0~\AA{}) serves as a buffer zone between binders
and non-binders. Only systems containing at least one binder
(RMSD~${<}\,2.0$~\AA{}) qualified for the scoring-function evaluation
(Section~\ref{sec:scoring-failures}), and only binder poses were eligible to
serve as the ground-truth reference (Section~\ref{sec:ground-truth}).
Ambiguous poses remain in the candidate pools presented to the agent and may
compete for selection, but they do not define the evaluation target.

Before RMSD calculation, the number of heavy atoms in each docked pose was
verified against the native structure to confirm molecular identity. An
all-or-nothing policy was enforced: if any individual pose within a system could
not be processed, due to file corruption, atom-count mismatch, or RMSD
calculation failure, the entire system was discarded rather than producing
incomplete classification data. Classification results for each system were
stored as a comma-separated file containing the pose index, RMSD value, and
assigned class.

\subsubsection{Identification of Scoring Function Failures}
\label{sec:scoring-failures}

A \emph{scoring function failure} occurs when the pose closest to the
crystallographic binding mode, i.e.\ the pose with the lowest RMSD to the
native structure, does not receive the most favourable (most negative)
\texttt{minimizedAffinity} score from Smina. In such cases, the scoring
function would direct a practitioner toward a geometrically inferior pose,
making these systems precisely the scenarios in which physics-based evaluation
beyond the docking score is most needed. Identifying scoring-function failures
is therefore central to defining our benchmark: they constitute the primary
evaluation target for the agentic ranking pipeline.

For each system that passed the pose classification step
(Section~\ref{sec:pose-classification}) and contained at least one near-native
pose (RMSD $< 2.0$~\AA{}), we extracted the \texttt{minimizedAffinity}
property stored in the Smina-produced docked SDF file for every pose using
RDKit~\cite{rdkit2025}. Two quantities were then compared:

\begin{enumerate}
    \item the index of the pose with the lowest RMSD to the native structure
    (the \emph{best-RMSD pose}), and
    \item the index of the pose with the most negative
    \texttt{minimizedAffinity} score (the \emph{best-scored pose}).
\end{enumerate}

\noindent If the two indices coincided, Smina was considered to have
\emph{succeeded}: its scoring function assigned the best energy to the
near-native pose. If they differed, the system was classified as a
\emph{scoring function failure}.

Of the 11{,}257 systems for which pose classifications were available,
8{,}597 contained at least one near-native pose and were therefore eligible
for scoring-function evaluation. Among these, Smina correctly identified
the near-native pose as the top-ranked candidate in 4{,}350 systems
(50.6\%), while 4{,}247 systems (49.4\%) exhibited scoring-function failures.
The near-parity between successes and failures underscores that empirical
scoring functions, despite their efficiency, misrank the best-RMSD pose in
roughly half of all cases, a finding consistent with the broader literature
on the limitations of knowledge-based and empirical docking
scores~\cite{warren2006critical}.

This binary partitioning of the dataset into \emph{Smina-success} and
\emph{Smina-failure} systems provides the foundation for the benchmark
design described in Section~\ref{sec:final-selection}: the final evaluation
set is drawn from both partitions so that the agentic pipeline can be
assessed on cases where the conventional score already identifies the correct
pose (controls) as well as on cases where it does not (target systems).

\subsubsection{Final System Selection}
\label{sec:final-selection}

From the 8{,}597 systems eligible for scoring-function evaluation
(Section~\ref{sec:scoring-failures}), we selected a compact benchmark of
ten protein-ligand complexes, five in which Smina correctly identifies
the near-native pose as the top-ranked candidate (\emph{success}
controls) and five in which it does not (\emph{failure} targets). This
balanced design enables a direct comparison of the agentic pipeline's
behaviour on systems where the conventional scoring function already
succeeds (testing whether the agent preserves a correct answer) versus
systems where it fails (testing whether the agent can recover the
correct pose through multi-criteria reasoning).

\paragraph{Protein metadata retrieval.}
To guide diversification, we retrieved protein names and functional
classifications for all candidate systems via the RCSB PDB
REST~API\@. For each system, the polymer entity endpoint
(\texttt{/rest/v1/core/polymer\_entity/\{id\}/1}) provided the
protein description, and the entry endpoint
(\texttt{/rest/v1/core/entry/\{id\}}) provided the
\texttt{pdbx\_keywords} classification, which we adopted as the
protein family label.

\paragraph{Selection criteria.}
Three objectives governed system selection:

\begin{enumerate}
    \item \textbf{Protein family diversity.}  The ten systems were
          chosen to span as many distinct protein families as
          possible, reducing the risk that benchmark performance
          reflects family-specific biases rather than generalisable
          ranking ability. The final set covers five families:
          hydrolase, nuclear receptor, transferase, lyase, and
          isomerase.

    \item \textbf{Failure severity diversity.}  Among the five
          Smina-failure systems, we required a range of scoring
          displacements, from near-misses where the near-native
          pose is ranked second or third by affinity, through
          moderate displacements (ranked seventh), to severe failures
          where the near-native pose falls to the 19th or 20th
          position. This gradient tests whether the agent can
          recover the correct pose regardless of how deeply the
          scoring function has buried it.

    \item \textbf{Structural quality.}  All systems originate from
          the PDBbind~2016 dataset
          (Section~\ref{sec:source-dataset}), and every system
          passed the pose classification pipeline
          (Section~\ref{sec:pose-classification}), ensuring that its
          docked poses include at least one near-native conformation
          (RMSD~$< 2.0$~\AA{}) so that a ground-truth best pose
          exists.
\end{enumerate}

\paragraph{Selection procedure.}
The selection was implemented in a four-stage automated pipeline. Starting from a manually curated seed list of twelve
pharmacologically notable protein-ligand systems, the pipeline
cross-referenced each system against the Smina scoring analysis,
partitioned the matches into successes and failures, and then expanded
each partition to exactly five systems by drawing additional candidates
from the full PDBbind pool. Additional success systems were chosen to
maximise protein family diversity relative to the systems already
selected. Additional failure systems were stratified by scoring
displacement severity, close (affinity rank~2-3), medium
(rank~7-11), and severe (rank~18+), and, within each stratum, the
system with the most distinct protein family was preferred. The
resulting ten systems are summarised in Table~\ref{tab:selected-systems}.

\begin{table}[htbp]
\centering
\small
\begin{tabular}{@{}llllcrr@{}}
\toprule
PDB ID & Protein & Family & Smina & Poses & Best-pose RMSD (\AA) & Affinity rank \\
\midrule
\multicolumn{7}{l}{\emph{Smina-success systems (controls)}} \\[2pt]
185L & T4 lysozyme             & Hydrolase        & \checkmark & 9  & 0.43 & 1  \\
1ERR & Estrogen receptor       & Nuclear receptor & \checkmark & 9  & 0.82 & 1  \\
2HYY & Tyrosine-protein kinase ABL1 & Transferase & \checkmark & 5  & 0.85 & 1  \\
2P16 & Coagulation factor Xa   & Hydrolase        & \checkmark & 20 & 1.24 & 1  \\
3HS4 & Carbonic anhydrase 2    & Lyase            & \checkmark & 20 & 0.96 & 1  \\
\addlinespace
\multicolumn{7}{l}{\emph{Smina-failure systems (targets)}} \\[2pt]
3OXC & HIV-1 protease          & Hydrolase        & \texttimes & 19 & 0.63 & 2  \\
4JFL & Peptidyl-prolyl isomerase FKBP5 & Isomerase & \texttimes & 20 & 1.76 & 3  \\
4MLT & Carbonic anhydrase 2    & Lyase            & \texttimes & 20 & 1.16 & 7  \\
2HA6 & Acetylcholinesterase    & Hydrolase        & \texttimes & 20 & 1.89 & 19 \\
4ZLS & HIV-1 protease          & Hydrolase        & \texttimes & 20 & 1.97 & 20 \\
\bottomrule
\end{tabular}
\caption{
    \textbf{The ten protein-ligand systems selected for benchmark
    evaluation.}  Systems are grouped by Smina scoring outcome
    (success or failure). \emph{Poses} is the number of docked
    conformations retained after curation.
    \emph{Best-pose RMSD} is the heavy-atom symmetry-corrected RMSD
    (\AA{}) of the near-native pose closest to the crystallographic
    binding mode.
    \emph{Affinity rank} is the 1-based position of that
    near-native pose when all poses are sorted by Smina
    \texttt{minimizedAffinity} (most negative first); a rank of~1
    indicates that Smina correctly assigned the best score to the
    near-native pose. The five failure systems span a range of
    scoring displacements from near-miss (rank~2) to severe
    (rank~20), testing the agent's ability to recover the correct
    pose at varying levels of scoring-function error. The five
    protein families represented, hydrolase, nuclear receptor,
    transferase, lyase, and isomerase, ensure structural and
    functional diversity within the benchmark.
}
\label{tab:selected-systems}
\end{table}

The ten systems collectively contain 162 docked poses (range: 5-20 per
system). Among the five success controls, the near-native pose RMSD
ranges from 0.43 to 1.24~\AA{}, confirming that Smina's best-scored pose
is indeed geometrically close to the crystal structure. Among the five
failure targets, near-native RMSDs range from 0.63 to 1.97~\AA{}, all
well within the 2.0~\AA{} binder threshold, yet Smina ranks these poses
between 2nd and 20th by affinity score, demonstrating that the scoring
failure is not caused by ambiguous or borderline RMSD values but by a
genuine inability of the empirical scoring function to assign the best
energy to a clearly near-native conformation.

\subsection{Evaluation Metrics}
\label{sec:evaluation-metrics}

We frame the evaluation as a \emph{best-pose identification} task:
given a set of $N_i$ docked poses for system~$i$, the agent must
select the single pose that most closely reproduces the
crystallographic binding mode. This formulation directly tests
whether the agentic pipeline can recover the biologically relevant
conformation from among the docking engine's candidates, the core
capability required for practical use in structure-based drug design.
Beyond measuring identification accuracy, we additionally analyse the
agent's decision-making process to assess whether its self-reported
reasoning faithfully reflects the underlying metric evidence
(Section~\ref{sec:decision-attribution}).

\subsubsection{Ground-Truth Definition}
\label{sec:ground-truth}

For each of the ten benchmark systems
(Section~\ref{sec:final-selection}), the ground-truth best pose is
defined as the docked conformation with the lowest heavy-atom
symmetry-corrected RMSD to the native structure, as determined during
the pose classification step (Section~\ref{sec:pose-classification}).
Formally, for a system with $N_i$ candidate poses
$\{p_1, \ldots, p_{N_i}\}$, the ground-truth pose index is

\begin{equation}
\label{eq:ground-truth}
g_i = \arg\min_{j \in \{1, \ldots, N_i\}}
      \mathrm{RMSD}(p_j,\; p_{\mathrm{native}})\,,
\end{equation}

\noindent where $p_{\mathrm{native}}$ is the crystallographic ligand
conformation re-processed through Smina in scoring-only mode to ensure
consistent atom ordering
(Section~\ref{sec:source-dataset}). All ten ground-truth poses satisfy
the 2.0~\AA{} near-native threshold established in
Section~\ref{sec:pose-classification}, with RMSDs ranging from 0.43
to 1.97~\AA{} (Table~\ref{tab:selected-systems}). Four of the ten
systems contain additional poses below this threshold (two systems
with two near-native poses each and two with three), but only the
single lowest-RMSD pose per system serves as the ground truth. The
RMSD deviation metric (Section~\ref{sec:rmsd-metric}) distinguishes
cases where the agent selects one of these secondary near-native
conformations from cases where it selects a geometrically distant
pose.

\paragraph{Native pose exclusion.}
The evaluation is performed in a \emph{without-native} configuration:
the candidate pool presented to the agent consists exclusively of the
Smina-generated docked poses. The native crystallographic
conformation is \emph{not} included among the candidates, ensuring
that the task tests the agent's ability to discriminate among
computationally generated geometries rather than to trivially
recognise an experimentally determined structure. The native pose is
used solely for RMSD-based ground-truth assignment and is never
exposed to the agent during analysis.

\subsubsection{Best-Pose Identification Accuracy}
\label{sec:top1-accuracy}

The primary evaluation metric is the \emph{best-pose identification
accuracy} (top-1 accuracy): the fraction of benchmark systems for
which the agent selects the ground-truth best pose as its top-ranked
candidate~\cite{su2019comparative}. For $m$ systems, the accuracy is

\begin{equation}
\label{eq:accuracy}
\mathrm{Acc} = \frac{1}{m}
  \sum_{i=1}^{m} \mathbf{1}\!\bigl[\hat{g}_i = g_i\bigr]\,,
\end{equation}

\noindent where $\hat{g}_i$ is the pose index selected by the agent
for system~$i$, $g_i$ is the ground-truth best pose index
(Equation~\ref{eq:ground-truth}), and $\mathbf{1}[\cdot]$ is the
indicator function, which returns~1 when the agent's top-ranked pose
is the lowest-RMSD conformation and~0 otherwise. The accuracy is
therefore the fraction of systems for which the agent identifies the
correct pose. This binary, per-system metric is deliberately
strict: the agent receives credit only for identifying the exact
correct pose. Alternative metrics that reward "close" selections
(e.g.\ selecting the second-lowest RMSD pose) were considered but
rejected, as they would obscure the practically important distinction
between selecting and failing to select the near-native conformation.
In the four systems that contain more than one near-native pose, this
strictness means that selecting a secondary near-native conformation
counts as a misidentification, even though the agent recovers a
biologically plausible binding mode. The RMSD deviation
(Section~\ref{sec:rmsd-metric}) captures these near-misses by
reporting a small~$\Delta_{\mathrm{RMSD}}$.

Because the accuracy is a proportion estimated from only $m = 10$
systems, the point estimate alone conveys limited information about the
agent's underlying success probability. We therefore report a 95\%
Wilson score confidence interval~\cite{wilson1927probable}, which
bounds the range of success probabilities consistent with the observed
number of correct identifications under a binomial model. The Wilson
interval is preferred over the normal approximation because it
provides better coverage for small samples and proportions near 0
or~1~\cite{brown2001interval}:

\begin{equation}
\label{eq:wilson}
\mathrm{CI}_{95}
= \frac{
    \hat{p} + \tfrac{z^{2}}{2m}
    \pm z\,\sqrt{
      \frac{\hat{p}(1-\hat{p})}{m}
      + \frac{z^{2}}{4m^{2}}
    }
  }{
    1 + \frac{z^{2}}{m}
  }\,,
\end{equation}

\noindent where $\hat{p} = \mathrm{Acc}$ is the observed accuracy and
$z = 1.96$ for 95\% confidence.

\subsubsection{RMSD of Selected Pose}
\label{sec:rmsd-metric}

The top-1 accuracy is a binary metric that does not capture the
\emph{severity} of misidentification. We therefore report a secondary
metric: the heavy-atom RMSD of the agent's selected pose to the native
structure. For a correctly identified system, this equals the
ground-truth best-pose RMSD; for a misidentified system, it quantifies
how far the agent's choice deviates from the crystallographic binding
mode.

The RMSD deviation between the selected and ground-truth poses is
computed as

\begin{equation}
\label{eq:rmsd-deviation}
\Delta_{\mathrm{RMSD},i} =
  \mathrm{RMSD}(\hat{p}_i,\; p_{\mathrm{native}})
  - \mathrm{RMSD}(p_{g_i},\; p_{\mathrm{native}})\,,
\end{equation}

\noindent where $\hat{p}_i$ is the agent's selected pose and $p_{g_i}$
is the ground-truth best pose. A value of
$\Delta_{\mathrm{RMSD}} = 0$ indicates a correct identification; positive
values indicate how much additional structural deviation the agent's
choice introduces relative to the best available pose. This metric
provides practical insight: a misidentification with
$\Delta_{\mathrm{RMSD}} = 0.5$~\AA{} is far less consequential than one
with $\Delta_{\mathrm{RMSD}} = 4.0$~\AA{}, even though both count
equally as failures under the binary accuracy metric.

\subsubsection{Baseline Comparisons}
\label{sec:baselines}

Two baselines contextualise the agent's performance.

\paragraph{Smina scoring baseline.}
Smina's conventional scoring function provides a natural reference
point: for each system, the pose with the most favourable (most
negative) \texttt{minimizedAffinity} score is taken as Smina's
top-ranked candidate; when this coincides with the ground-truth best
pose, Smina is deemed successful. The ten benchmark systems were
selected to include five Smina-success and five Smina-failure cases
(Section~\ref{sec:final-selection}), yielding a Smina baseline
accuracy of 50\% by construction. This balanced design ensures that
the agent is evaluated on both the "easy" regime (where the scoring
function already identifies the correct pose) and the "hard" regime
(where it does not), allowing direct comparison of the agentic
pipeline's accuracy against the conventional rescoring baseline on
each partition and on the full benchmark.

\paragraph{Random selection baseline.}
Under uniformly random selection, the probability of correctly
identifying the best pose in a system with~$N_i$ candidates is
$1 / N_i$. The expected random baseline accuracy across the ten
benchmark systems is therefore

\begin{equation}
\label{eq:random-baseline}
\mathrm{Acc}_{\mathrm{rand}} = \frac{1}{m}
  \sum_{i=1}^{m} \frac{1}{N_i}\,,
\end{equation}

\noindent where $N_i$ ranges from 5 to 20 across the benchmark
(Table~\ref{tab:selected-systems}). With $m = 10$ systems containing
a total of 162 poses, this yields an expected random accuracy of
approximately 7.7\%. The random baseline establishes that any
accuracy above this level reflects genuine discriminative ability
rather than chance.

\subsubsection{Partition-Stratified Accuracy}
\label{sec:stratified-accuracy}

Because the benchmark is balanced by design between five Smina-success
and five Smina-failure systems
(Section~\ref{sec:final-selection}), we additionally report the
agent's accuracy on each partition separately. Let
$\mathcal{S}^{+}$ and $\mathcal{S}^{-}$ denote the sets of
Smina-success and Smina-failure systems, respectively. The
partition-stratified accuracies are

\begin{equation}
\label{eq:acc-success}
\mathrm{Acc}^{+} = \frac{1}{|\mathcal{S}^{+}|}
  \sum_{i \in \mathcal{S}^{+}}
  \mathbf{1}\!\bigl[\hat{g}_i = g_i\bigr]\,,
\qquad
\mathrm{Acc}^{-} = \frac{1}{|\mathcal{S}^{-}|}
  \sum_{i \in \mathcal{S}^{-}}
  \mathbf{1}\!\bigl[\hat{g}_i = g_i\bigr]\,.
\end{equation}

\noindent $\mathrm{Acc}^{+}$ measures whether the agent preserves
correct identifications that the conventional scoring function already
makes, i.e.\ whether it avoids introducing regressions on "easy"
systems. $\mathrm{Acc}^{-}$ measures whether the agent can recover
the near-native pose on systems where the scoring function fails,
which is the primary capability that motivates the agentic approach.
The contrast between the two partition accuracies reveals whether any
overall improvement is driven by genuine recovery of scoring-function
failures or merely by retaining existing successes. Because each
partition contains only five systems, the stratified accuracies are
reported as descriptive summaries rather than as the basis for formal
hypothesis testing; with $|\mathcal{S}^{+}| = |\mathcal{S}^{-}| = 5$,
the achievable accuracy values are limited to the set
$\{0, 20, 40, 60, 80, 100\}\%$, and any associated confidence
intervals are necessarily wide.

\subsubsection{Statistical Significance Testing}
\label{sec:significance}

To establish that the agent's accuracy exceeds the random baseline
with statistical confidence, we apply a one-sided exact binomial
test. Under the null hypothesis that the agent selects the correct
pose independently for each system with probability equal to the
random baseline $p_0 = \mathrm{Acc}_{\mathrm{rand}}$
(Equation~\ref{eq:random-baseline}), the probability of observing $k$
or more successes out of $m$ systems is

\begin{equation}
\label{eq:binomial-test}
p\text{-value} = P(X \geq k \mid m,\, p_0)
= \sum_{j=k}^{m} \binom{m}{j}\, p_0^{\,j}\,(1 - p_0)^{m-j}\,,
\end{equation}

\noindent where $X \sim \mathrm{Binomial}(m,\, p_0)$. We reject the
null hypothesis at the $\alpha = 0.05$ significance level if the
$p$-value falls below~0.05.

The exact binomial test is preferred over asymptotic alternatives
(e.g.\ the $z$-test for proportions) because the sample size
($m = 10$) is too small for the normal approximation to be reliable,
and the null success probability ($p_0 \approx 0.077$) lies far from
0.5, exacerbating the approximation error. The test treats each
system as an independent Bernoulli trial, a reasonable assumption
given that the ten systems involve different proteins, ligands, and
pose sets with no shared structural or chemical features.

\subsubsection{Decision Attribution Analysis}
\label{sec:decision-attribution}

Beyond measuring \emph{whether} the agent identifies the correct pose,
we analyse \emph{how} it arrives at its decision. For each system the
agent produces (i)~a final pose selection backed by free-text
reasoning and (ii)~explicit self-reported percentage weights
$\{w_{i,t}\}_{t=1}^{T}$ that quantify the importance it assigned to
each of $T$~tool categories in making its choice
(Section~\ref{sec:agent-output}). We exploit this structured output
to perform two complementary analyses.

\paragraph{Metric separation.}
For each system~$i$ and tool category~$t$, we define the
\emph{metric separation} of the selected pose as the signed,
standardised distance between the selected pose's metric value and the
candidate-set mean, oriented so that positive values indicate a
favourable direction:

\begin{equation}
\label{eq:metric-separation}
\delta_{i,t} = s_t \cdot
  \frac{v_{i,t,\hat{g}_i} - \bar{v}_{i,t}}
       {\sigma_{i,t}}\,,
\end{equation}

\noindent where $v_{i,t,j}$ is the scalar metric value produced by
tool~$t$ for pose~$j$ of system~$i$,
$\bar{v}_{i,t}$ and $\sigma_{i,t}$ are the mean and standard
deviation of that metric across all $N_i$~candidate poses, and
$s_t \in \{+1, -1\}$ is a sign factor that ensures higher
$\delta_{i,t}$ corresponds to a more favourable outcome
($s_t = +1$ for metrics where larger is better, such as interaction
quality score and burial ratio;
$s_t = -1$ for metrics where smaller is better, such as clash energy,
conformational strain, and polar penalty score).
A large positive $\delta_{i,t}$ indicates that the selected pose
stands out favourably on tool~$t$; a value near zero indicates that
tool~$t$ does not differentiate the selected pose from the rest of the
candidate set.

\paragraph{Reasoning faithfulness.}
We assess whether the agent's self-reported tool weights align with
the objective metric separations by comparing the two rank orderings.
For each system~$i$, let $\mathbf{w}_i = (w_{i,1}, \ldots, w_{i,T})$
be the vector of stated weights and
$\boldsymbol{\delta}_i = (|\delta_{i,1}|, \ldots,
|\delta_{i,T}|)$ be the vector of absolute metric separations.
Concordance between the two is measured by the Spearman rank
correlation $\rho_i$ between $\mathbf{w}_i$ and
$\boldsymbol{\delta}_i$:

\begin{equation}
\label{eq:faithfulness}
\rho_i = \mathrm{Spearman}\bigl(
  \mathrm{rank}(\mathbf{w}_i),\;
  \mathrm{rank}(\boldsymbol{\delta}_i)
\bigr)\,.
\end{equation}

\noindent A high positive $\rho_i$ indicates faithful reasoning: the
agent assigns greater weight to tools on which the selected pose
genuinely excels relative to the candidate set. A weak or negative
$\rho_i$ suggests that the stated rationale may not reflect the actual
metric basis of the decision, a potential concern given that
chain-of-thought explanations from large language models are not
guaranteed to be faithful to the model's actual decision
process~\cite{turpin2023language,lanham2023measuring}. We report both the
per-system $\rho_i$ values and the median across systems.

\paragraph{Outcome-stratified attribution.}
To identify whether tool reliance patterns differ between correct and
incorrect decisions, we partition the systems by outcome and compare
the weight distributions. For each tool category~$t$, we compute the
mean stated weight across correctly identified systems
($\bar{w}_t^{\,\checkmark}$) and incorrectly identified systems
($\bar{w}_t^{\,\times}$):

\begin{equation}
\label{eq:outcome-weights}
\bar{w}_t^{\,\checkmark}
  = \frac{1}{|\mathcal{C}|}
    \sum_{i \in \mathcal{C}} w_{i,t}\,,
\qquad
\bar{w}_t^{\,\times}
  = \frac{1}{|\mathcal{I}|}
    \sum_{i \in \mathcal{I}} w_{i,t}\,,
\end{equation}

\noindent where $\mathcal{C}$ and $\mathcal{I}$ denote the sets of
systems with correct and incorrect identifications, respectively.
Systematic differences between $\bar{w}_t^{\,\checkmark}$ and
$\bar{w}_t^{\,\times}$ across tool categories may reveal failure
modes, for example, whether incorrect decisions are associated with
over-reliance on interaction quantity at the expense of interaction
quality, or with insufficient attention to steric clashes.

Because the analysis is based on $m = 10$~systems, the attribution
results are reported descriptively (e.g.\ as heatmaps of tool weights
and metric separations stratified by outcome) rather than as the basis
for formal inferential claims. The primary goal is to generate
actionable hypotheses about the agent's reasoning patterns that can
guide future prompt engineering and tool design.

\section{Results}

\subsection{Benchmark Summary}
\label{sec:results-benchmark-summary}

The benchmark comprises ten protein-ligand systems (162 docked poses,
5 to 20 per system) drawn from PDBbind~2016 and balanced between five
Smina-success controls and five Smina-failure targets
(Table~\ref{tab:selected-systems};
Section~\ref{sec:final-selection}). All ground-truth RMSDs fall
within the 2.0~\AA{} near-native threshold (range: 0.43 to
1.97~\AA{}; Section~\ref{sec:ground-truth}), and the evaluation uses
a \emph{without-native} configuration in which the crystallographic
pose is excluded from the candidate pool. This balanced design fixes
the Smina scoring baseline at 50.0\% by construction; uniformly
random selection yields an expected accuracy of 7.7\%
(Equation~\ref{eq:random-baseline}).

\subsection{Best-Pose Identification Accuracy}
\label{sec:results-best-pose-accuracy}

Across the ten benchmark systems, the agentic pipeline correctly
identified the ground-truth best pose in 5 of 10 cases, yielding a
best-pose identification accuracy of 50.0\%
(Equation~\ref{eq:accuracy}). The associated 95\% Wilson score
confidence interval is [23.7\%, 76.3\%]
(Equation~\ref{eq:wilson}), reflecting the substantial uncertainty
inherent in a ten-system evaluation.

An exact one-sided binomial test
(Equation~\ref{eq:binomial-test}) against the averaged random selection
baseline of 7.7\% (Equation~\ref{eq:random-baseline}) yields
$p < 0.001$, indicating that the agent's accuracy is significantly
above chance at the $\alpha = 0.05$ level. Because per-system pose
counts vary (5 to 20), the exact null distribution under uniformly
random selection is a Poisson-binomial with system-specific probabilities
$1/N_{i}$ rather than a single-parameter binomial with
$p_{0} = 0.077$; the binomial approximation adopted here inflates the
variance of the null and is therefore conservative for a one-sided
test, so the conclusion that the agent beats indiscriminate guessing
is robust to this refinement. The observed accuracy exceeds the
averaged random baseline by a factor of 6.5.

Compared with the Smina scoring baseline of 50.0\%
(Section~\ref{sec:results-benchmark-summary}), the agent achieves
numerically identical overall accuracy. However, the composition of
the agent's successes and failures differs from Smina's: the agent
correctly identified the best pose in 4 of 5 Smina-success systems
and 1 of 5 Smina-failure systems, whereas Smina, by construction of
the benchmark, succeeds on all 5 Smina-success systems and fails on
all 5 Smina-failure systems. This difference in error distribution
indicates that the agent partially recovers from scoring-function
failures (1 recovery on the Smina-failure partition) while
introducing one regression on the Smina-success partition; the net
effect is zero change in aggregate accuracy. The partition-stratified
analysis in Section~\ref{sec:results-partition-stratified} examines
these trade-offs in detail.

Table~\ref{tab:accuracy-summary} summarises the accuracy metrics.
The agent's performance is situated between the random baseline
(7.7\%) and the upper bound of the Wilson confidence interval
(76.3\%), with the point estimate coinciding with the Smina baseline.
The wide confidence interval underscores the need for larger-scale
validation, while the highly significant binomial test confirms that
the pipeline's selections are not attributable to chance.

\begin{table}[htbp]
\centering
\begin{tabular}{lc}
\hline
\textbf{Metric} & \textbf{Value} \\
\hline
Agent accuracy        & 50.0\% (5/10) \\
Wilson 95\% CI        & [23.7\%, 76.3\%] \\
Smina baseline        & 50.0\% (5/10) \\
Random baseline       & 7.7\% \\
$p$-value (vs.\ random) & $< 0.001$ \\
Agent / random ratio  & 6.5$\times$ \\
\hline
\end{tabular}
\caption{Best-pose identification accuracy of the agentic pipeline
compared with the Smina scoring and random selection baselines across
ten benchmark systems. The Wilson score 95\% confidence interval
(Equation~\ref{eq:wilson}) quantifies uncertainty from the small
sample. The $p$-value is from a one-sided exact binomial test
(Equation~\ref{eq:binomial-test}) against the random baseline.}
\label{tab:accuracy-summary}
\end{table}

\begin{figure}[htbp]
\centering
\includegraphics[width=0.6\textwidth]{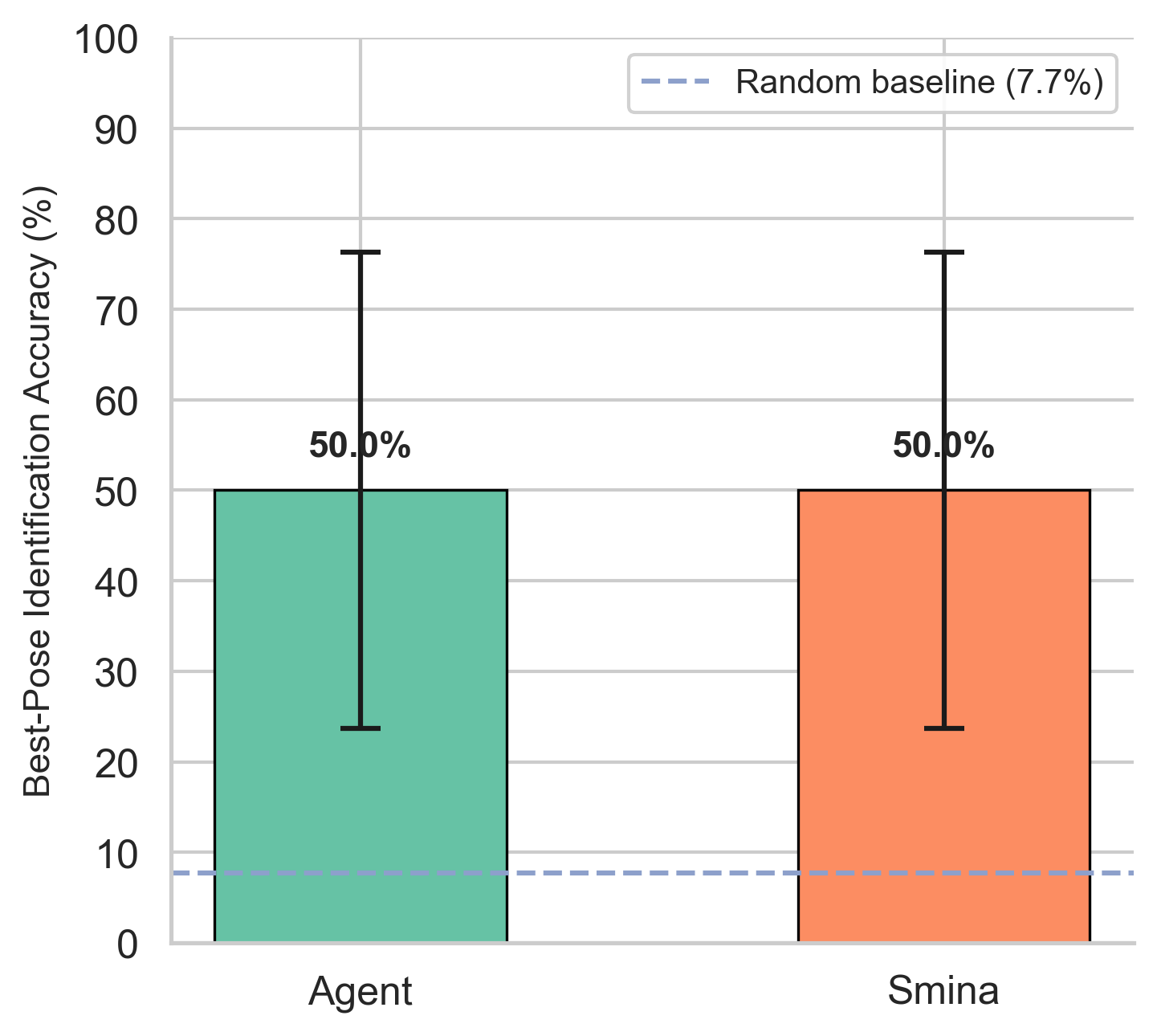}
\caption{Best-pose identification accuracy for the agentic pipeline
and the Smina scoring baseline, with 95\% Wilson score confidence
intervals (error bars). The dashed line indicates the expected
accuracy under uniformly random selection (7.7\%;
Equation~\ref{eq:random-baseline}). Both methods achieve 50.0\%
accuracy ($n = 10$ systems), though their per-system agreement
patterns differ (see Section~\ref{sec:results-partition-stratified}).}
\label{fig:best-pose-accuracy}
\end{figure}

Figure~\ref{fig:best-pose-accuracy} compares the agent and Smina
accuracies against the random selection baseline. Both methods
achieve the same point estimate, but the overlapping Wilson confidence
intervals highlight the limited statistical power of a ten-system
benchmark to discriminate between methods of similar performance.

Although the agent and Smina achieve the same overall accuracy, the
binary pass/fail metric does not capture how close the agent's
incorrect selections are to the true near-native conformation. The
following section examines the RMSD of the agent's selected poses to
quantify the severity of misidentifications.

\subsection{RMSD of Selected Poses}
\label{sec:results-rmsd}

The binary accuracy metric of Section~\ref{sec:results-best-pose-accuracy}
treats every misidentification equally, regardless of whether the agent
selected a pose 0.5~\AA{} or 5~\AA{} away from the ground truth. The RMSD
deviation $\Delta_{\mathrm{RMSD},i}$ (Equation~\ref{eq:rmsd-deviation}) provides
a continuous, structurally grounded complement: it measures how much
additional displacement the agent's choice introduces relative to the best
available docked conformation. Table~\ref{tab:rmsd-selected-poses}
reports the per-system results.

\begin{table}[htbp]
\centering
\begin{tabular}{llllcccl}
\hline
\textbf{System} & \textbf{Selected} & \textbf{GT pose} & \textbf{Correct?}
  & \textbf{Sel.\ RMSD (\AA{})}
  & \textbf{GT RMSD (\AA{})}
  & $\boldsymbol{\Delta_{\mathrm{RMSD}}}$ \textbf{(\AA{})}
  & \textbf{Severity} \\
\hline
185L & pose\_01 & pose\_01 & \checkmark & 0.43 & 0.43 & 0.00 & Correct \\
1ERR & pose\_01 & pose\_01 & \checkmark & 0.82 & 0.82 & 0.00 & Correct \\
2HYY & pose\_01 & pose\_01 & \checkmark & 0.85 & 0.85 & 0.00 & Correct \\
3OXC & pose\_02 & pose\_02 & \checkmark & 0.63 & 0.63 & 0.00 & Correct \\
2P16 & pose\_01 & pose\_01 & \checkmark & 1.24 & 1.24 & 0.00 & Correct \\
\hline
4JFL & pose\_01 & pose\_03 & $\times$   & 3.27 & 1.76 & 1.51 & Moderate \\
4MLT & pose\_01 & pose\_07 & $\times$   & 3.13 & 1.16 & 1.97 & Moderate \\
2HA6 & pose\_03 & pose\_19 & $\times$   & 4.72 & 1.89 & 2.82 & Severe \\
4ZLS & pose\_08 & pose\_20 & $\times$   & 5.92 & 1.97 & 3.95 & Severe \\
3HS4 & pose\_02 & pose\_01 & $\times$   & 5.28 & 0.96 & 4.32 & Severe \\
\hline
\end{tabular}
\caption{Per-system RMSD of the agent's selected pose to the crystallographic
binding mode, the ground-truth best-pose RMSD, the RMSD deviation
$\Delta_{\mathrm{RMSD}}$ (Equation~\ref{eq:rmsd-deviation}), and the resulting
severity category. Correct identifications ($\Delta_{\mathrm{RMSD}} = 0$) are
marked with a tick; misidentifications are categorised as moderate
($1.0 \leq \Delta_{\mathrm{RMSD}} < 2.0$~\AA{}) or severe
($\Delta_{\mathrm{RMSD}} \geq 2.0$~\AA{}). All RMSD values are
symmetry-corrected heavy-atom RMSDs computed against the crystallographic
ligand conformation.}
\label{tab:rmsd-selected-poses}
\end{table}

Across all ten benchmark systems, the mean RMSD of the agent's selected
pose to the crystallographic binding mode is 2.63~\AA{} (range
0.43--5.92~\AA{}; median 2.18~\AA{}). Restricting to the five
correctly identified systems, all selected poses lie well within the
2.0~\AA{} near-native threshold, with a maximum RMSD of 1.24~\AA{}
(2P16). Across the five \emph{misidentified} systems, the mean
$\Delta_{\mathrm{RMSD}}$ is 2.92~\AA{} (median 2.82~\AA{}; maximum
4.32~\AA{}), and the mean selected-pose RMSD is 4.46~\AA{}, indicating
that the misidentifications correspond to poses in substantially
different orientations rather than near-misses close to the ground truth.

Applying the severity thresholds defined in the Methods, none of the five
misidentifications falls in the mild category
($0 < \Delta_{\mathrm{RMSD}} < 1.0$~\AA{}). Two systems exhibit moderate
deviation: 4JFL ($\Delta_{\mathrm{RMSD}} = 1.51$~\AA{}) and 4MLT
($\Delta_{\mathrm{RMSD}} = 1.97$~\AA{}). In both cases the selected pose
remains plausibly near the correct binding pocket, though with meaningful
geometric displacement. The remaining three misidentifications are
severe ($\Delta_{\mathrm{RMSD}} \geq 2.0$~\AA{}): 2HA6
($\Delta_{\mathrm{RMSD}} = 2.82$~\AA{}), 4ZLS
($\Delta_{\mathrm{RMSD}} = 3.95$~\AA{}), and 3HS4
($\Delta_{\mathrm{RMSD}} = 4.32$~\AA{}). These three systems represent
cases where the agent selected a pose in a substantially different
orientation from the ground-truth binding mode, and the severity of
the error warrants detailed investigation in the Representative System
Analysis (Section~\ref{sec:results-representative}).

From a practical standpoint, a pose with RMSD below 2.0~\AA{} to the
crystal structure is generally considered near-native and useful for
lead optimisation. Of the ten agent-selected poses, exactly five
(50.0\%) satisfy this criterion, the five correctly identified
systems (Table~\ref{tab:rmsd-selected-poses}). The five misidentified
poses all have RMSD values above 3.0~\AA{}, meaning that in every
failure case the agent selected a pose outside the near-native
threshold and therefore of limited direct utility for structure-based
drug design. No misidentification falls in the mild category
($\Delta_{\mathrm{RMSD}} < 1.0$~\AA{}), but this observation must be
interpreted against the available pose landscape. In two of the five
systems (4ZLS, 4MLT), the docked-pose pool contains no conformation
with $\Delta_{\mathrm{RMSD}} < 1.0$~\AA{} relative to the ground
truth, so a mild error was geometrically impossible regardless of the
agent's reasoning. In the remaining three systems (3HS4, 4JFL,
2HA6), poses with $\Delta_{\mathrm{RMSD}} < 1.0$~\AA{} were
available (two, two, and six candidates, respectively), yet the agent
bypassed all of them in favour of a substantially more distant
conformation.

\begin{figure}[htbp]
\centering
\includegraphics[width=\textwidth]{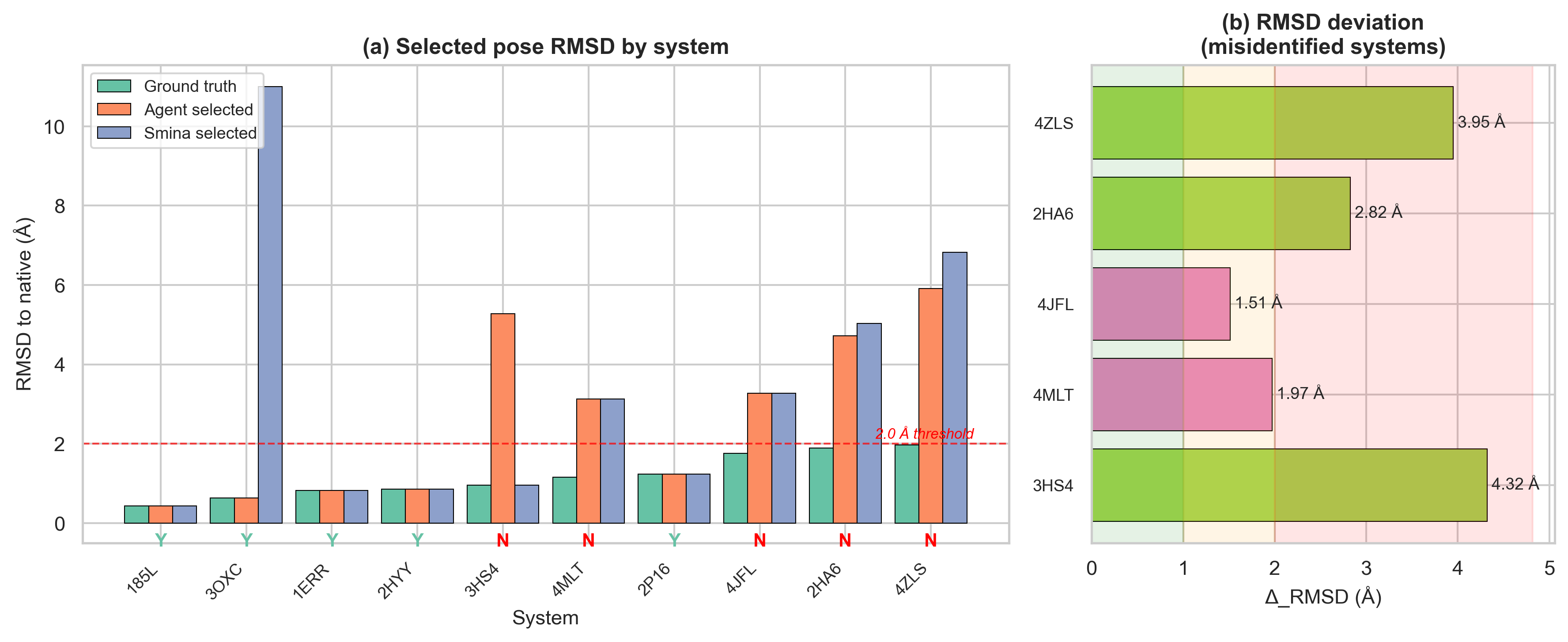}
\caption{(a) Heavy-atom symmetry-corrected RMSD to the crystallographic
binding mode for the agent-selected pose (orange), the ground-truth best
pose (teal), and the Smina-selected pose (green) across all ten benchmark
systems. The dashed red line marks the 2.0~\AA{} near-native threshold.
Tick labels below bars indicate correct~(Y) or incorrect~(N) agent
identifications.  (b) RMSD deviation $\Delta_{\mathrm{RMSD}}$
(Equation~\ref{eq:rmsd-deviation}) for the five misidentified systems,
with shading indicating mild (green; none observed), moderate (orange;
$1.0 \leq \Delta_{\mathrm{RMSD}} < 2.0$~\AA{}), and severe (red;
$\Delta_{\mathrm{RMSD}} \geq 2.0$~\AA{}) categories.}
\label{fig:rmsd-selected-poses}
\end{figure}

Figure~\ref{fig:rmsd-selected-poses} displays the per-system selected-pose
RMSD alongside the ground-truth RMSD and the $\Delta_{\mathrm{RMSD}}$
of each misidentified system.
The partition-stratified analysis in
Section~\ref{sec:results-partition-stratified} examines whether error
severity is concentrated in Smina-failure systems or distributed across
both partitions.

\subsection{Partition-Stratified Accuracy}
\label{sec:results-partition-stratified}

Having characterised the structural quality of the agent's selections via RMSD,
we now examine whether the overall 50.0\% accuracy is uniformly distributed
across the benchmark or whether it masks an asymmetry between systems where
Smina already succeeds and those where it fails. This partition-stratified
analysis (Equation~\ref{eq:acc-success};
Section~\ref{sec:stratified-accuracy}) is the primary diagnostic for the
added value of the agentic approach: does the pipeline merely replicate the
scoring function on easy cases, or does it genuinely recover near-native poses
that the scoring function misses?

On the five Smina-success systems (185L, 1ERR, 2HYY, 2P16, 3HS4), the agent
correctly identified the ground-truth best pose in four cases, yielding
$\mathrm{Acc}^{+} = 80.0\%$ (4/5; Equation~\ref{eq:acc-success}). The sole
regression occurred in 3HS4, where the agent selected pose\_02 instead of the
ground-truth pose\_01 despite Smina ranking the correct pose first. The
$\mathrm{Acc}^{+}$ of 80.0\% substantially exceeds the partition-specific random
baseline of 10.4\% (factor of 7.7$\times$), confirming that the agent retains
strong discriminative ability on systems that the scoring function handles well.

On the five Smina-failure systems (2HA6, 3OXC, 4JFL, 4MLT, 4ZLS), the agent
successfully recovered the ground-truth best pose in one case, yielding
$\mathrm{Acc}^{-} = 20.0\%$ (1/5; Equation~\ref{eq:acc-success}). The single
recovery was system 3OXC, where the agent correctly identified pose\_02 as the
near-native conformation despite Smina ranking it second. The four remaining
Smina-failure systems (2HA6, 4JFL, 4MLT, 4ZLS) were not recovered; the
agent's selections in these cases correspond to severe or moderate RMSD
deviations (Table~\ref{tab:rmsd-selected-poses}). At 20.0\%, $\mathrm{Acc}^{-}$
remains substantially above the partition-specific random baseline of 5.1\%
(factor of 4.0$\times$), indicating that the agent's ranking is not random, but
the recovery rate is limited.

Table~\ref{tab:partition-category} cross-classifies each system by the
joint Smina and agent outcome. Because the single regression (3HS4) and
the single recovery (3OXC) cancel, the overall accuracy remains unchanged
relative to Smina.

\begin{table}[htbp]
\centering
\begin{tabular}{llcl}
\hline
\textbf{Partition} & \textbf{Category} & \textbf{Systems} & \textbf{System IDs} \\
\hline
Smina-success & A: both correct          & 4 & 185L, 1ERR, 2HYY, 2P16 \\
              & B: regression            & 1 & 3HS4                    \\
\hline
Smina-failure & C: recovery              & 1 & 3OXC                    \\
              & D: both incorrect        & 4 & 2HA6, 4JFL, 4MLT, 4ZLS  \\
\hline
\end{tabular}

\vspace{0.8em}

\begin{tabular}{lccc}
\hline
\textbf{Partition} & \textbf{Agent accuracy} & \textbf{Smina accuracy} & \textbf{Random baseline} \\
\hline
Smina-success ($\mathrm{Acc}^{+}$) & 80.0\% (4/5) & 100.0\% (5/5) & 10.4\% \\
Smina-failure ($\mathrm{Acc}^{-}$) & 20.0\% (1/5) &   0.0\% (0/5) &  5.1\% \\
Overall                             & 50.0\% (5/10) & 50.0\% (5/10) &  7.7\% \\
\hline
\end{tabular}
\caption{Partition-stratified accuracy of the agentic pipeline across the ten
benchmark systems. Systems are classified by the joint outcome of the Smina
and agent decisions: Category~A (both correct), B (Smina correct, agent
incorrect, regression), C (Smina incorrect, agent correct, recovery), and D
(both incorrect). Partition-specific random baselines use the mean of $1/N_i$
across the five systems in each partition, where $N_i$ is the number of
candidate poses for system $i$.}
\label{tab:partition-category}
\end{table}

Taken together, the contrast between $\mathrm{Acc}^{+}$ and
$\mathrm{Acc}^{-}$ suggests that the multi-metric evaluation currently acts
conservatively, often arriving at conclusions similar to the scoring function
rather than overriding it on the basis of physics-based signals. Because each
partition comprises only five systems, the estimates are fragile. A single
additional recovery would shift $\mathrm{Acc}^{-}$ from 20.0\% to 40.0\%, so
these partition-level values should be read as descriptive rather than
statistically definitive.

\begin{figure}[htbp]
\centering
\includegraphics[width=0.6\textwidth]{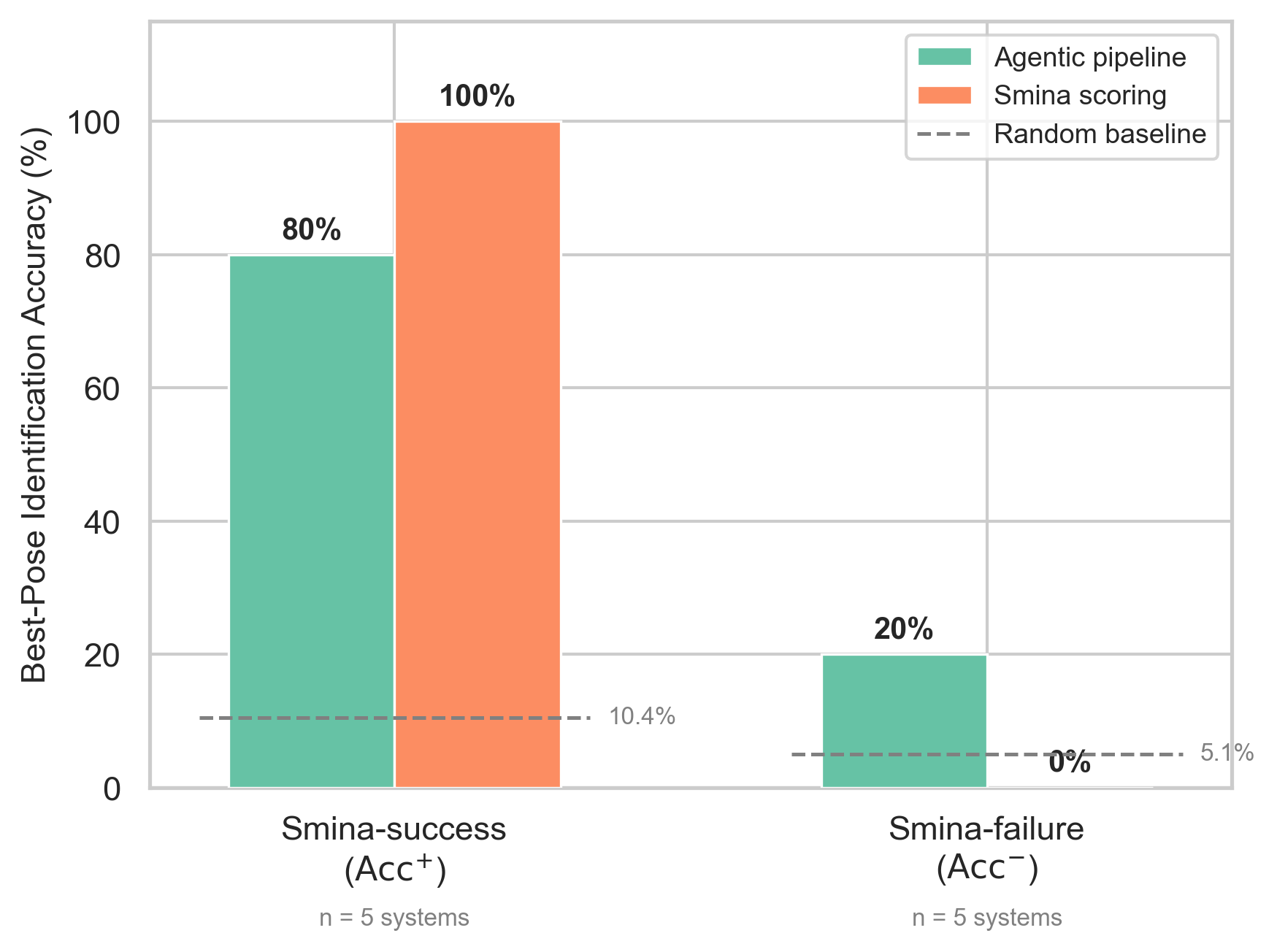}
\caption{Partition-stratified best-pose identification accuracy for the
agentic pipeline (teal) and Smina scoring (orange) on the Smina-success
($\mathrm{Acc}^{+}$) and Smina-failure ($\mathrm{Acc}^{-}$) partitions of
the ten-system benchmark. Dashed horizontal segments indicate the
partition-specific random baseline (mean of $1/N_i$ across the five systems
in each partition). Each partition contains $n = 5$ systems; the achievable
accuracy values are $\{0, 20, 40, 60, 80, 100\}\%$.}
\label{fig:partition-stratified-accuracy}
\end{figure}

Figure~\ref{fig:partition-stratified-accuracy} shows the grouped bar chart
comparing the agentic pipeline and Smina scoring function across both
partitions, together with the partition-specific random baselines.
The Representative System Analysis in
Section~\ref{sec:results-representative} examines four concrete systems, one
from each cell of the Smina-outcome $\times$ Agent-outcome matrix (3OXC, 2P16,
3HS4, and 4JFL), providing a qualitative explanation for the metric patterns
that drove both successes and failures.

\subsection{Representative System Analysis}
\label{sec:results-representative}

\begin{figure}[htbp]
\centering
\begin{subfigure}[t]{0.48\textwidth}
\centering
\includegraphics[width=\textwidth]{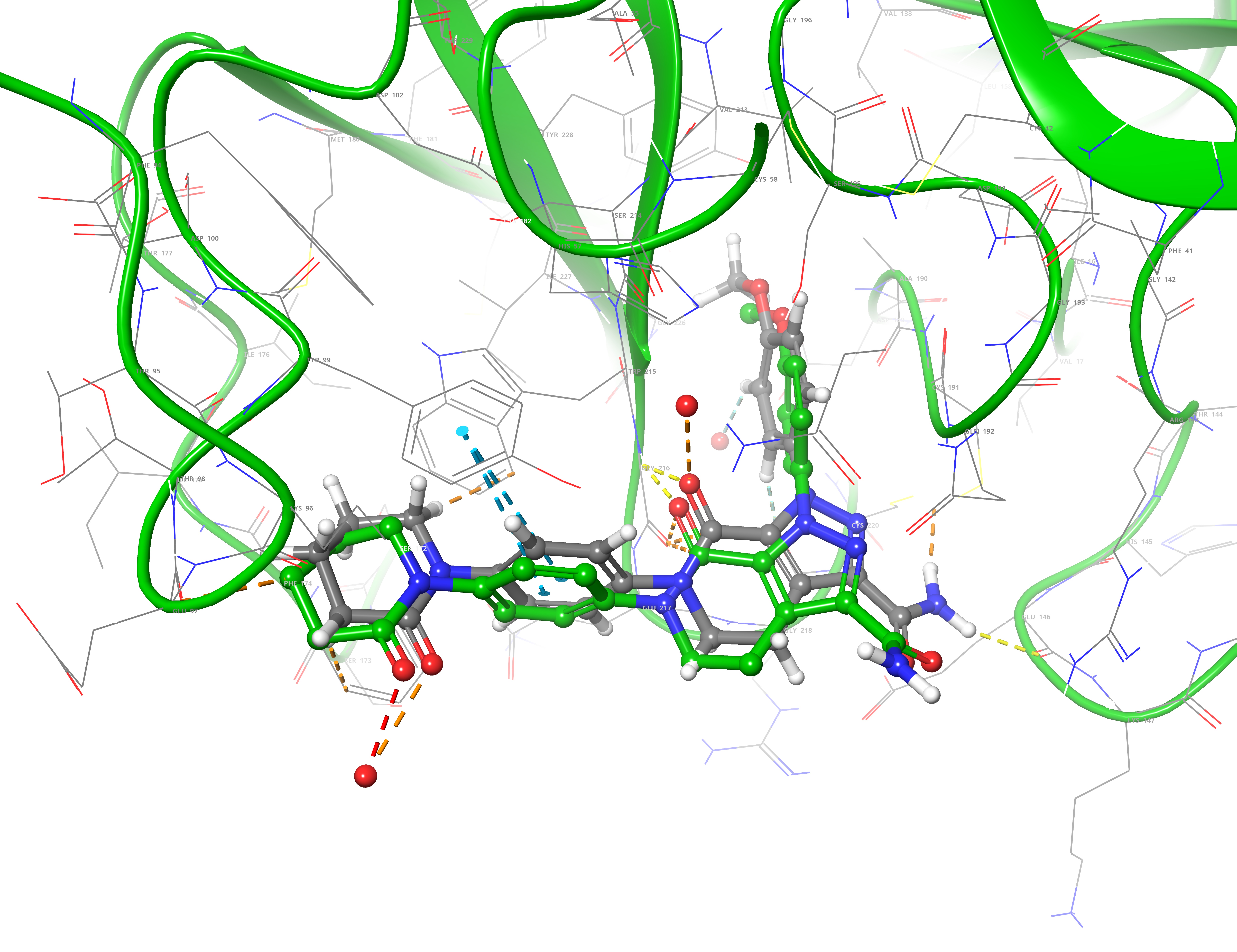}
\caption{2P16 (Category~A, agreement). The agent's selected pose\_01
(green; RMSD 1.24~\AA{}) closely overlaps the crystallographic ligand
(grey), confirming correct near-native identification.}
\label{fig:pymol-2p16}
\end{subfigure}
\hfill
\begin{subfigure}[t]{0.48\textwidth}
\centering
\includegraphics[width=\textwidth]{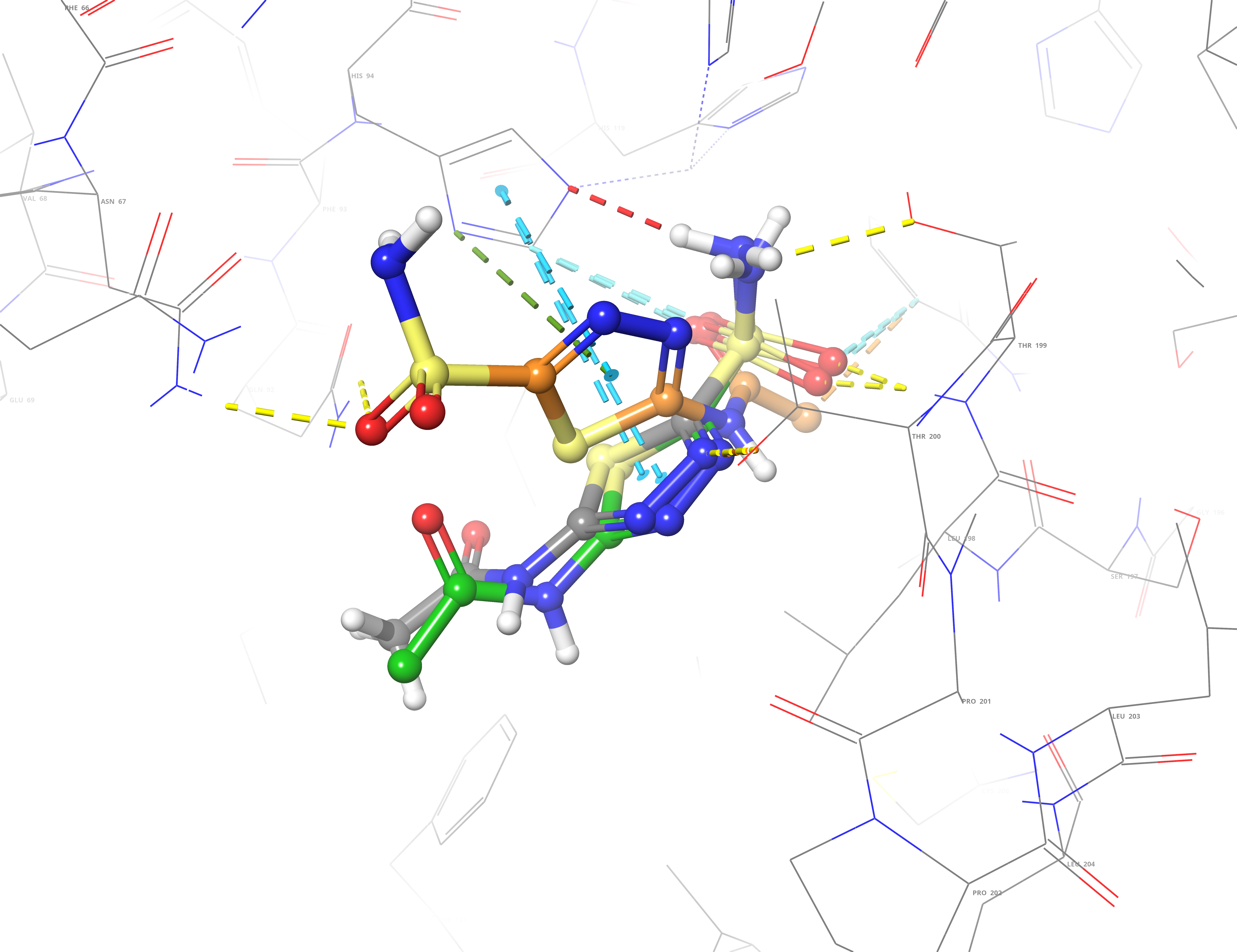}
\caption{3HS4 (Category~B, regression). The agent's selected pose\_02
(orange; RMSD 5.28~\AA{}) is spatially displaced from both the
crystallographic ligand (grey) and the correct pose\_01 (green; RMSD
0.96~\AA{}).}
\label{fig:pymol-3hs4}
\end{subfigure}

\vspace{1em}

\begin{subfigure}[t]{0.48\textwidth}
\centering
\includegraphics[width=\textwidth]{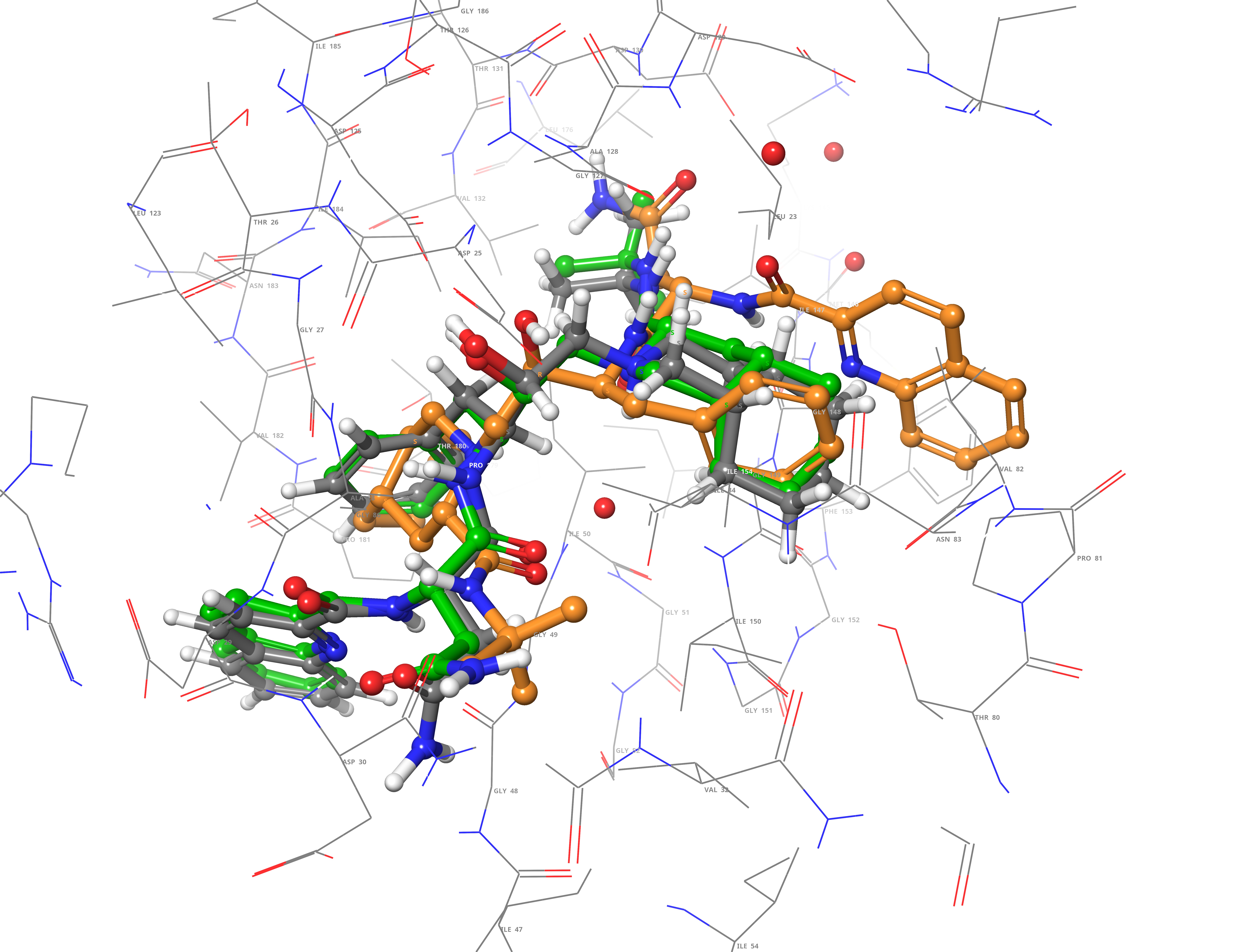}
\caption{3OXC (Category~C, recovery). The agent's selected pose\_02
(green; RMSD 0.63~\AA{}) overlaps the crystallographic ligand (grey),
while Smina's top-ranked pose\_01 (orange; RMSD 11.00~\AA{}) adopts
an inverted orientation within the HIV-1 protease binding channel,
yielding a large heavy-atom RMSD despite remaining within the active
site.}
\label{fig:pymol-3oxc}
\end{subfigure}
\hfill
\begin{subfigure}[t]{0.48\textwidth}
\centering
\includegraphics[width=\textwidth]{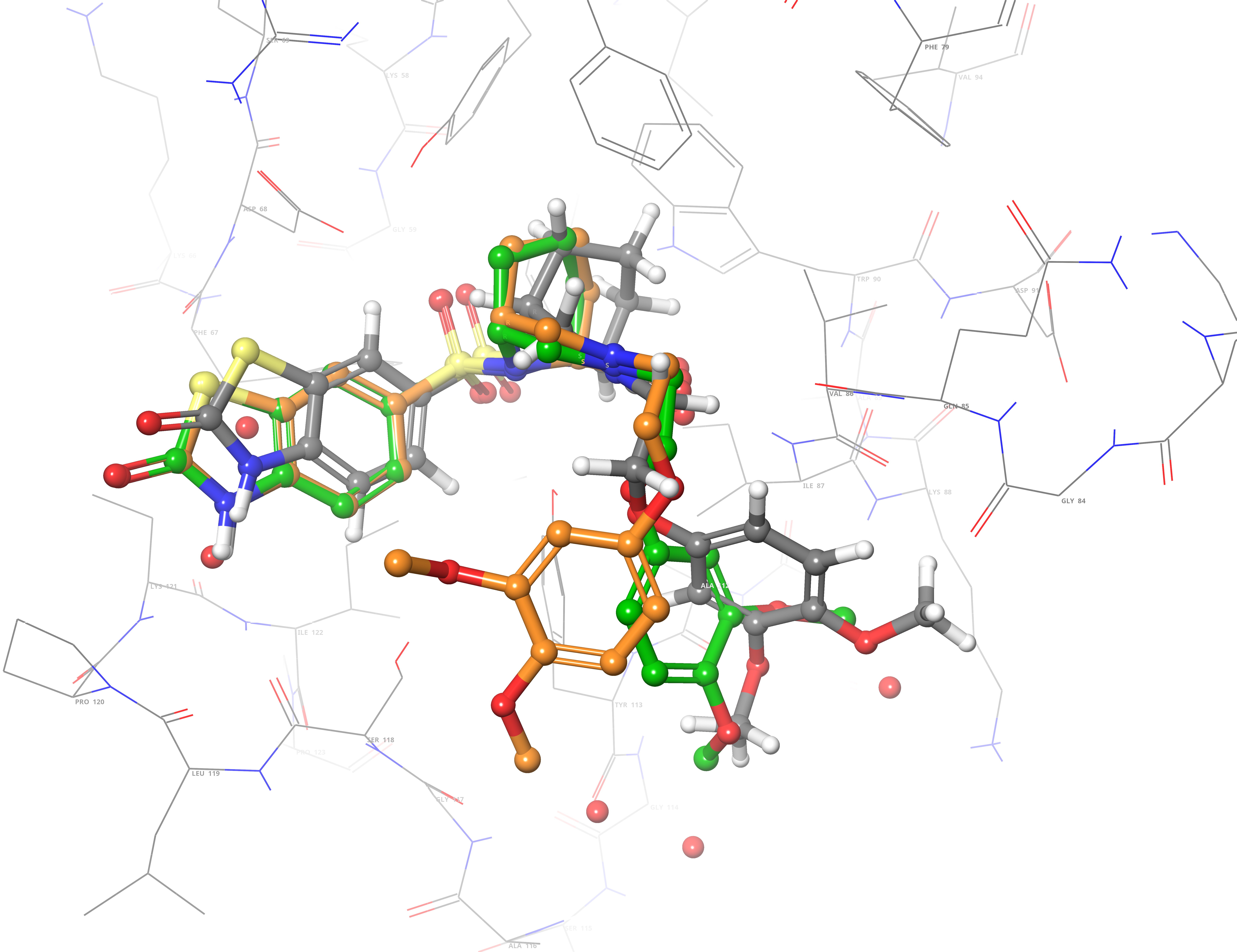}
\caption{4JFL (Category~D, instructive failure). The agent's selected
pose\_01 (orange; RMSD 3.27~\AA{}) and the correct pose\_03 (green;
RMSD 1.76~\AA{}) both occupy the FKBP51 binding pocket, with the
displacement concentrated in the solvent-exposed tail of the molecule.}
\label{fig:pymol-4jfl}
\end{subfigure}
\caption{Binding-site views of the four representative systems from the
Smina-outcome $\times$ Agent-outcome classification matrix. In each panel the
protein is shown with binding-site residues in standard element colours.
The crystallographic reference ligand is rendered in grey; the
agent-selected or ground-truth near-native pose is shown in green;
incorrectly ranked poses are shown in orange. All panels were generated
with Maestro~\cite{maestro}.}
\label{fig:pymol-representative}
\end{figure}

The preceding aggregate analyses establish the pipeline's overall accuracy and
error severity. To illustrate how the agent arrives at its decisions in
practice, we examine four representative systems, one from each cell of the
Smina-outcome $\times$ Agent-outcome classification matrix
(Table~\ref{tab:partition-category}). For each system, we present the agent's
selection, the key metric comparisons, and verbatim excerpts from the
reasoning trace that demonstrate the auditability of the agentic approach.
The cases are presented in category order (A--D), progressing from agreement
through regression and recovery to joint failure.
Figure~\ref{fig:pymol-representative} provides structural context for the
four systems, showing how each agent-selected pose compares spatially to the
crystallographic reference ligand within the protein binding site.

\paragraph{Category~A: Agreement: 2P16 (Smina success, agent success).}

System 2P16 ($\beta$-trypsin, 20 candidate poses) exemplifies a challenging
case in which Smina and the agent agree on the same correct answer, pose\_01
(RMSD 1.24~\AA{}), but the agent provides a richer, auditable justification
(Figure~\ref{fig:pymol-2p16}).
The ensemble was dominated by docking artifacts: 9 of 20 poses (45.0\%) were
classified as \textsc{surface} binders, and no pose achieved \textsc{deep\_pocket}
classification (best burial 55.6\%).
Table~\ref{tab:case-2p16} shows the metric profile of the selected pose and
its closest competitors.

\begin{table}[htbp]
\centering
\begin{tabular}{lccc}
\hline
\textbf{Metric} & \textbf{pose\_01 (selected)} & \textbf{pose\_03} & \textbf{pose\_04} \\
\hline
TQS                          & 3.72  & 4.50  & 6.32 \\
Burial ratio                 & 55.6\% & 52.8\% & 50.0\% \\
Binding mode quality         & 0.56  & ---   & --- \\
Strain (kcal/mol)            & 77.30 & 77.54 & 96.54 \\
Steric clashes               & 1     & 0     & 1 \\
Polar penalty                & 1.00  & 1.50  & 1.50 \\
H-bonds                      & 2 (avg quality 0.59) & --- & --- \\
\hline
\end{tabular}
\caption{Metric comparison for the agent's selected pose (pose\_01) and the
two closest competitors in system 2P16. All three poses are classified as
\textsc{partial\_pocket}; no ensemble member achieved \textsc{deep\_pocket}
status.  pose\_04 had the highest total quality score (TQS; Equation~\ref{eq:total-quality}) in the ensemble but was penalised for
extreme conformational strain.}
\label{tab:case-2p16}
\end{table}

\begin{figure}[htbp]
\centering
\begin{subfigure}[t]{0.48\textwidth}
\centering
\includegraphics[width=\textwidth]{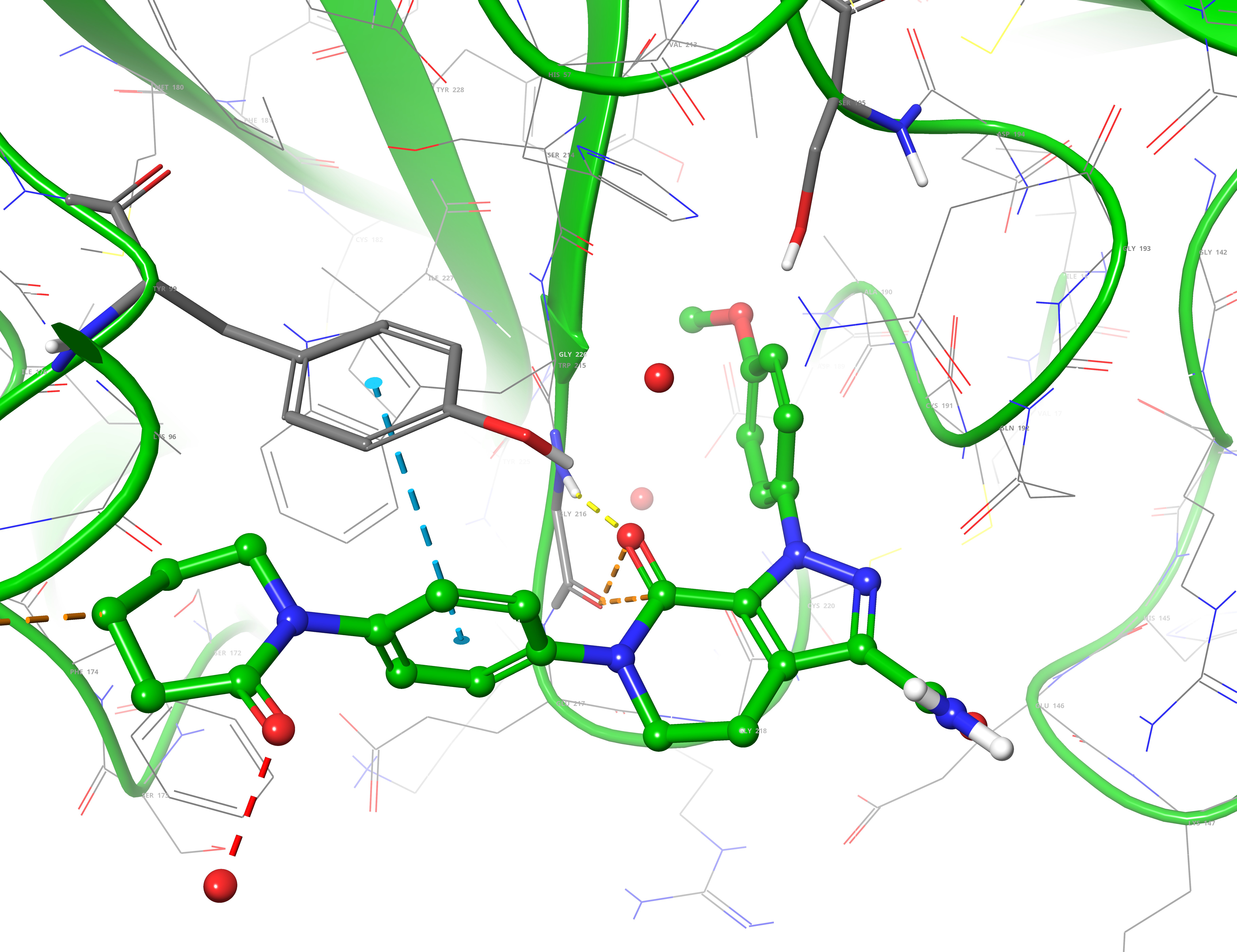}
\caption{Interaction network of the agent's selected pose\_01 (green) with
binding-site residues, including the hydrogen bond to GLY216A and the
$\pi$-stack to TYR99A. The crystallographic ligand is shown in grey.}
\label{fig:maestro-2p16-interactions}
\end{subfigure}
\hfill
\begin{subfigure}[t]{0.48\textwidth}
\centering
\includegraphics[width=\textwidth]{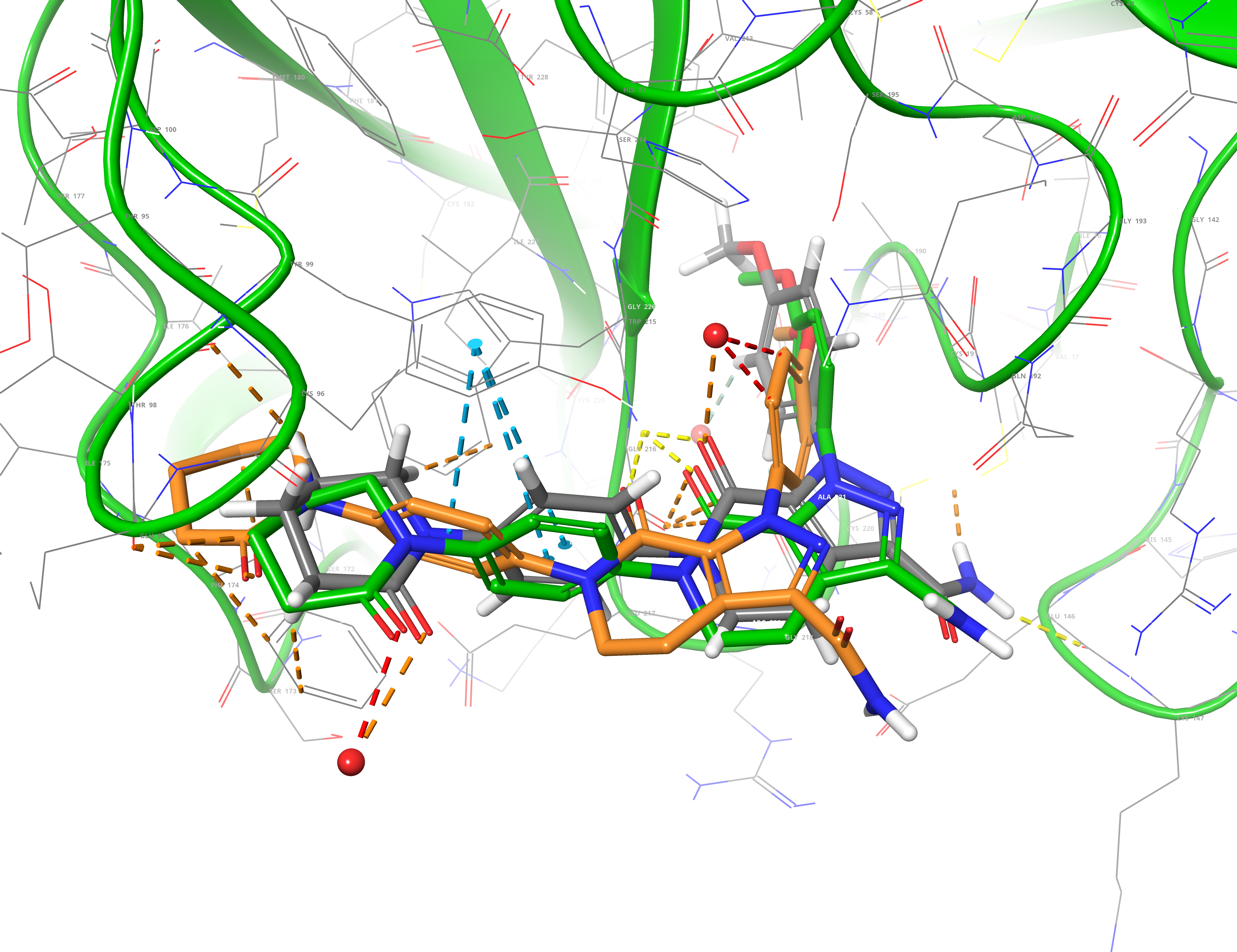}
\caption{Overlay of the selected pose\_01 (green; RMSD 1.24~\AA{}) and
the rejected competitor pose\_04 (orange; RMSD 3.28~\AA{}). Despite
pose\_04's highest-in-set TQS of 6.32, the agent rejected it for
${\sim}$20~kcal/mol additional conformational strain.}
\label{fig:maestro-2p16-vs-pose04}
\end{subfigure}
\caption{Structural details for system 2P16.  (a)~Key protein--ligand
interactions for the agent's selected pose.
(b)~Spatial comparison of the selected pose against the rejected
highest-scoring competitor. The crystallographic ligand is shown in grey
in both panels.}
\label{fig:maestro-2p16-detail}
\end{figure}

\subparagraph{Metric overview and binding mode gating.}
The agent's strategy centred on using the binding mode classification tool as a
gating filter. As the reasoning trace states:

\begin{quote}
\textit{``I first eliminated SURFACE-mode poses from top consideration.
pose\_01 is one of the few poses that engages the pocket best among this
set, with the highest burial in the dataset (55.6\% buried), classified as
PARTIAL\_POCKET with the top binding-mode quality score (0.56). In a dataset
where no pose reaches DEEP\_POCKET ({>}70\% burial), this deeper-than-peers
engagement is a decisive advantage.''}
\end{quote}

\noindent Notably, the agent rejected pose\_04
(Figure~\ref{fig:maestro-2p16-vs-pose04}) despite its highest-in-set TQS
of 6.32, reasoning that
\textit{``the interaction quality advantage does not plausibly offset
${\sim}$20~kcal/mol extra conformational cost in the absence of superior burial
or polar management.''}
The agent further highlighted that pose\_01's chemically diversified
interaction network (Figure~\ref{fig:maestro-2p16-interactions}), comprising
an excellent H-bond to GLY216A
(2.81~\AA{}, 162.2\textdegree), a $\pi$-stack to TYR99A, and a water bridge
to SER195A, was
\textit{``somewhat undercounted by the reported total quality''} because the
TQS does not include $\pi$-stacking and water bridge contributions.

\subparagraph{Deliberative process and adaptive weighting.}
Importantly, the reasoning trace reveals that the agent did not converge on
pose\_01 immediately. The Chain of Thought records an intermediate stage in
which the agent leaned towards a different candidate:
\textit{``pose\_04 has the best total quality at 6.32 but comes with extremely
high strain. Finally, it looks like pose\_03 might be the most balanced
choice based on overall metrics and lower strain values compared to
pose\_04.''}
It was only after systematically applying the binding mode filter and
re-evaluating burial rankings that pose\_01 emerged as the top selection.
This iterative deliberation underscores that the agent's final answer is not
the output of a single scoring pass but the product of an explicit
hypothesis-revision cycle. Crucially, the agent also adapted its metric
weights to the characteristics of this specific ensemble rather than relying on
a fixed formula:

\begin{quote}
\textit{``The set contains many SURFACE poses; therefore, pocket engagement
must be a strong discriminator (binding mode + burial = 45\% combined).''}
\end{quote}

\noindent By allocating 45\% of its composite weight to pocket engagement
(25\% binding mode + 20\% burial), the agent ensured that the dominant failure
mode of this ensemble, surface binding, would be the primary axis of
discrimination. Its treatment of strain was similarly context-sensitive:
\textit{``Strain is used to break ties among similarly pocketed, similarly
interacting poses; however, all poses show very high strain, so differences
of a few kcal/mol are not decisive unless very large
(${\geq}$15--20~kcal/mol) and uncompensated.''}
This explicit articulation of a quantitative threshold, rather than a blind
penalty, illustrates the kind of reasoned, auditable judgement that a
single-score function cannot provide.

\subparagraph{Self-aware limitations and scientific judgement.}
Beyond ranking, the agent exercised self-critical scientific judgement by
flagging the limitations of its own metrics. In the Confidence Assessment, it
noted:
\textit{``All poses exhibit very high strain energies; while I used strain as
a tiebreaker, the absolute MMFF-derived values may be inflated or
system-dependent. Relative differences were considered rather than
absolutes.''}
This awareness of force-field artefacts, distinguishing relative from absolute
energetics, mirrors the reasoning that a computational chemist would apply
when interpreting MMFF strain values. The agent further questioned the
upstream docking protocol itself:
\textit{``No pose attains DEEP\_POCKET classification. If the true site is
deep, docking may have partially failed, which lowers overall confidence.''}
Such a hypothesis is notable because it shifts responsibility from the ranking
step to the pose-generation step, illustrating that the agent can reason about
the provenance of its inputs rather than treating them as given. Even at the
individual-pose level, the agent exercised quality-over-quantity reasoning.
When evaluating competitor pose\_09, which reported five hydrogen bonds, the
agent observed that
\textit{``H-bond geometry is only marginal on average (0.45) despite 5
reported H-bonds, suggesting overcounting of weakly supportive
interactions.''}
The Key Findings section reinforced this multi-metric tension across the full
ensemble:
\textit{``High conformational strain across the board (66--97~kcal/mol). Only
small relative differences exist among pocketed poses; two SURFACE poses have
the lowest strain but are disqualified by binding mode.''}
Taken together, these passages demonstrate that the agent does not merely
aggregate metrics; it interprets them with domain-appropriate caveats,
identifies the conditions under which each metric is, or is not, decisive,
and propagates uncertainty into its confidence estimate.

\subparagraph{Confidence assessment, synthesis, and suggested experiments.}
The agent reported a confidence level of ``Medium'' and explicitly identified
the additional data that would raise it, including
\textit{``local minimization with side-chain repacking and explicit-water
placement for the top poses to reassess strain, clashes, and H-bonding''}
as well as
\textit{``experimental footprinting or site-directed mutagenesis (e.g.,
GLY216A, TYR99A, PHE174A) to confirm the contact pattern.''}
The latter suggestion reveals embedded domain knowledge: the agent nominated
the very residues it identified as key interaction partners, connecting its
computational findings to a testable experimental hypothesis.
Its final synthesis encapsulated the multi-metric reasoning in a single
passage:

\begin{quote}
\textit{``pose\_01's combination of the best burial, a coherent and chemically
sensible interaction network (including one excellent H-bond and a stabilising
$\pi$ contact to TYR99A), low clash burden, and manageable polar penalty
outweighs its only moderate total quality score and the generally high strain
that affects nearly all poses here.''}
\end{quote}

\noindent This sentence is precisely the kind of structured, multi-factor
justification that a single docking score cannot express: it names the
decisive features (burial, interaction coherence), acknowledges the
weaknesses (moderate TQS, high strain), and contextualises the weaknesses as
ensemble-wide rather than pose-specific.

\noindent This case demonstrates the explainability advantage of the agentic approach:
whereas Smina merely assigns an affinity score and this relies on the pre-defined terms and weights of the scoring function, the agent's trace provides a
complete audit of which physical features favoured the selected pose, which
competitors were considered, and why they were rejected. The remaining three
Category~A systems (185L, 1ERR, 2HYY) were cleaner cases where the best pose
was separated from competitors by wider metric margins and did not require
extensive tiebreaking logic.

\paragraph{Category~B: Regression: 3HS4 (Smina success, agent failure).}

System 3HS4 (carbonic anhydrase~II, 20 candidate poses) is the sole regression
case in the benchmark: Smina correctly ranked pose\_01 (RMSD 0.96~\AA{})
first, but the agent selected pose\_02 instead, with \emph{high} confidence
(Figure~\ref{fig:pymol-3hs4}).
This case reveals a fundamental limitation: the correct pose can score poorly
on computed metrics relative to a competing conformation.
The structural consequences are substantial: the agent's selected pose\_02 has
an RMSD of 5.28~\AA{} to the crystallographic binding mode, compared with
0.96~\AA{} for the ground-truth pose\_01
($\Delta_{\mathrm{RMSD}} = 4.32$~\AA{}, classified as \emph{severe} in
Table~\ref{tab:rmsd-selected-poses}), the largest RMSD deviation in the
entire benchmark.
Table~\ref{tab:case-3hs4} contrasts the two poses.

\begin{table}[htbp]
\centering
\begin{tabular}{lcc}
\hline
\textbf{Metric} & \textbf{pose\_02 (selected)} & \textbf{pose\_01 (correct)} \\
\hline
TQS                          & 6.52 & 3.23 \\
H-bonds (avg quality)        & 5 (0.62) & 4 (0.63) \\
Hydrophobic contacts (total) & 4 (3.40) & 1 (0.70) \\
$\pi$-stacking               & 1 (HIS94A) & 0 \\
Burial ratio                 & 93.8\% & 75.0\% \\
Binding mode quality         & 1.00 & 0.83 \\
Strain (kcal/mol)            & 47.08 & 46.45 \\
Steric clashes               & 0 & 2 \\
Polar penalty                & 1.00 & 0.50 \\
Ligand efficiency             & 0.769 (excellent) & 0.462 (excellent) \\
\hline
\end{tabular}
\caption{Metric comparison for the agent's selected pose (pose\_02) and the
ground-truth best pose (pose\_01) in system 3HS4.  pose\_02 dominates on
nearly every metric the agent evaluates; pose\_01 does not appear in the
agent's top-7 ranking.}
\label{tab:case-3hs4}
\end{table}

\subparagraph{Why the agent was wrong with high confidence.}
The metrics in Table~\ref{tab:case-3hs4} explain both why the agent selected
pose\_02 and why the failure carries an important lesson.  pose\_02 dominated
on every axis the agent evaluates: it achieved the highest TQS in the
20-pose ensemble (6.52 vs.\ 3.23 for pose\_01, a 2.0-fold difference), the
deepest burial (93.8\% vs.\ 75.0\%), zero steric clashes (vs.\ 2), a
$\pi$-stacking contact with HIS94A absent from pose\_01, and the best ligand
efficiency in the set (0.769). The agent stated:

\begin{quote}
\textit{``The selected pose, pose\_02, is the only pose that simultaneously
achieves the highest interaction quality (Total Quality Score 6.52), a true
DEEP\_POCKET binding mode with exceptional burial (93.8\% of ligand atoms
buried; classification quality 1.00), [and] zero steric clashes.
[\ldots] This three-way advantage (quality, deep burial, geometric
cleanliness) is decisive under the stated weighting scheme.''}
\end{quote}

\noindent Critically, the ground-truth pose\_01 does not appear in the agent's
reasoning trace or top-7 ranking at all. Its modest TQS of 3.23 and lower
burial placed it below the agent's attention threshold. The sole metrics
where pose\_01 was superior, slightly lower strain (46.45 vs.\ 47.08~kcal/mol,
a difference of only 0.63~kcal/mol) and a marginally better polar penalty
(0.50 vs.\ 1.00), were heavily outweighed under the agent's 32\%/32\%
interaction-plus-burial weighting scheme.

\subparagraph{Adaptive weighting and thermodynamic commensuration.}
The weight-setting process itself reveals ensemble-aware adaptation. In its
key findings, the agent characterised the 3HS4 landscape as unusually clean:
\textit{``Most poses classify as DEEP\_POCKET with good-to-excellent burial;
a few are PARTIAL\_POCKET (pose\_13, pose\_17, pose\_20). No clear SURFACE
artifacts were present, which is good.''}  Unlike the 2P16 ensemble, where
9 of 20 poses were \textsc{surface} binders, the 3HS4 pool was predominantly
well-docked, leaving the binding mode classifier with limited discriminatory
power. The decisive burden therefore fell on interaction quality and
secondary penalty metrics, and the agent adapted its strain treatment
accordingly:
\textit{``Given the uniformly high strain across the dataset, strain was
used as a tie-breaker and assigned a smaller but non-negligible weight,''}
while
\textit{``clashes and polar penalties were weighted more than strain because
they directly reflect physical plausibility and thermodynamic penalties not
otherwise captured by the quality score.''}
This weighting inversion, elevating clashes and polar penalties above
strain, reflects a pharmacologically sound judgement: in an ensemble where
all conformations carry similarly inflated strain values
(44--53~kcal/mol), differential clash and polar profiles carry more
information about relative binding plausibility. The reasoning chain further
reveals an attempt to ground abstract penalty scores in approximate
thermodynamic costs:

\begin{quote}
\textit{``Each unit may represent a cost of about 2--3~kcal per unsatisfied
polar atom.  pose\_05's 0 unsatisfied polar atoms would save it around
4--6~kcal compared to pose\_02. However, pose\_02 has a much better total
quality and burial. Despite the polar penalties, pose\_05's high strain
makes it less favorable.''}
\end{quote}

\noindent This conversion, estimating ${\sim}$2--3~kcal/mol per buried
unsatisfied polar atom, demonstrates a form of thermodynamic commensuration
rarely performed by automated scoring tools. The agent placed heterogeneous
metrics on an approximate common energy scale, weighed competing advantages,
and concluded that pose\_02's interaction and burial superiority outweighed
the estimated 4--6~kcal/mol polar desolvation cost. Yet it maintained
appropriate epistemic caution about its own framework, noting that
\textit{``it's not helpful to equate TQS values directly to
energy''}, acknowledging that interaction quality is a geometric surrogate,
not a free energy estimate.

\subparagraph{Confidence assessment and self-identified limitations.}
The agent assigned \emph{high} confidence to its selection of pose\_02 despite
explicitly acknowledging several methodological limitations. In its key
findings, the agent attributed the uniformly elevated strain values to a
specific force-field limitation:

\begin{quote}
\textit{``Conformational strain is consistently very high for all poses.
This is atypical in absolute terms and likely reflects limitations of the
MMFF reference or unminimized bound states.''}
\end{quote}

\noindent By naming MMFF specifically, the agent attributed the anomalous
strain landscape (44--53~kcal/mol across all 20 poses) to a known limitation
of the reference force field rather than treating these values as physically
meaningful absolute penalties, a methodologically appropriate response that
justified the decision to use strain only as a relative tie-breaker. The
uncertainty section of the confidence assessment reiterated this reasoning:
\textit{``Conformational strain values are uniformly very high across the
dataset (44--53~kcal/mol), suggesting force field or reference conformer
artifacts. I treated strain as a relative tie-breaker rather than an absolute
discriminator.''}  The agent further flagged ranking fragility, cautioning
that \textit{``water-mediated networks and protonation states can subtly
influence both H-bond detection and polar satisfaction; explicit hydration or
pKa refinement might reshuffle close runners-up slightly.''}  This
acknowledgement identifies precise physical factors, explicit water
bridges and protonation-dependent hydrogen bonding, that neither the
agent's tool suite nor Smina's empirical scoring function captures, and
which may partly explain why pose\_01, the crystallographic answer, was
invisible to the metric landscape of both approaches. The
juxtaposition of high confidence with candid uncertainty disclosure is itself
informative: the agent was confident in its ranking \emph{given the available
metrics}, while recognising that those metrics may not encompass all
determinants of binding. The failure is therefore not one of reasoning
coherence but of tool-suite coverage.

\subparagraph{Cross-system contrast: context-dependent tool roles.}
The comparison between 3HS4 and the Category~A system 2P16 illustrates how
the same tool suite plays fundamentally different roles across ensembles. In
2P16, the binding mode classifier was the decisive instrument: it eliminated
9 of 20 poses (45.0\%) as \textsc{surface} binders, collapsing the candidate
pool before interaction quality was assessed. In 3HS4, by contrast, only
three \textsc{partial\_pocket} outliers required deprioritisation. As the
agent stated:
\textit{``Partial\_pocket binders (pose\_13, pose\_17, pose\_20) were
deprioritized per the classifier; even when polar penalties were favorable,
their shallower engagement reduces confidence relative to true deep pocket
binders.''}
With the binding mode filter providing only marginal discrimination, the
agent relied instead on interaction quality, clash profiles, and polar
penalties to separate the remaining 17 \textsc{deep\_pocket} conformations.
This shift manifested in how secondary metrics were deployed: for example,
the agent flagged that
\textit{``pose\_16 shows an unusually high clash severity (0.099), a red flag
despite decent interaction quality (3.31) and burial (75\%)''}, using clash
\emph{severity}, not merely count, as a discriminator within an otherwise
homogeneous ensemble. The same tools thus served as hard gates in one
context and fine-grained discriminators in another, illustrating the
adaptive, context-dependent nature of the agent's deliberative framework.

\subparagraph{Smina scoring-term decomposition.}
Decomposing Smina's scoring function into its five component
terms~\citep{koes2013smina} clarifies which physical dimensions separate
pose\_01 from pose\_02 and, by extension, what the agent's tool suite
fails to capture. Table~\ref{tab:smina-terms-3hs4} reports the
per-term raw values and their weighted contributions for both poses.

\begin{table}[htbp]
\centering
\begin{tabular}{lrrrrrr}
\hline
\textbf{Term} & \textbf{Weight} & \textbf{p01 raw} & \textbf{p01 wtd}
  & \textbf{p02 raw} & \textbf{p02 wtd} & \textbf{$\Delta$~wtd} \\
\hline
gauss$_1$ (shape)       & $-0.036$ & 59.64 & $-2.12$ & 64.97 & $-2.31$ & $-0.19$ \\
gauss$_2$ (volume)      & $-0.005$ & 804.56 & $-4.15$ & 904.37 & $-4.66$ & $-0.52$ \\
repulsion               &  $0.840$ &   9.78 &  $8.22$ &   4.35 &  $3.66$ & $-4.56$ \\
hydrophobic             & $-0.035$ &   2.89 & $-0.10$ &  10.22 & $-0.36$ & $-0.26$ \\
non-dir.\ H-bond        & $-0.587$ &   7.92 & $-4.65$ &   6.44 & $-3.79$ & $+0.87$ \\
\hline
\textbf{Total}          &          &        & $-2.81$ &        & $-7.46$ & $-4.65$ \\
\hline
\end{tabular}
\caption{Smina scoring-term decomposition for the ground-truth pose
(pose\_01) and the agent's selected pose (pose\_02) in system~3HS4,
obtained by single-point rescoring (\texttt{--score\_only}) of the
docked coordinates. Each raw term value is multiplied by a learned
weight to produce its contribution; negative weighted values are
favourable.  $\Delta$~=~pose\_02~$-$~pose\_01.}
\label{tab:smina-terms-3hs4}
\end{table}

In four of five terms, pose\_02 is favourable: it achieves better shape
complementarity (gauss$_1$, gauss$_2$), 3.5-fold stronger hydrophobic
contacts, and substantially lower steric repulsion (4.35 vs.\ 9.78
raw), which alone accounts for a weighted advantage of~4.56~kcal/mol.
These four dimensions map directly onto the agent's own metrics (burial
ratio, hydrophobic contacts, and clash detection), and both systems agree
that pose\_02 is superior on all of them.
The sole term favouring pose\_01 is the non-directional hydrogen-bond
function, which sums pairwise donor--acceptor proximity without angular
cutoffs: 7.92 vs.\ 6.44 raw (weighted advantage~0.87~kcal/mol for
pose\_01). This result exposes a specific discrepancy with the agent's
tools. PLIP detects \emph{more} hydrogen bonds for pose\_02 than for
pose\_01 (5~vs.~4, Table~\ref{tab:case-3hs4}), because PLIP applies
strict geometric criteria (distance \emph{and} angle thresholds) that
count individual well-formed contacts. Smina's distance-only function
captures a broader measure of polar complementarity: the cumulative
proximity of all donor and acceptor atoms to their protein partners,
including weak or geometrically suboptimal contacts that PLIP does not
register. The crystallographic binding mode, having been refined against
electron density, places polar atoms at distances that maximise this
cumulative proximity even when individual contact angles fall outside
PLIP's detection thresholds.

The polar penalty tool partially captured this signal: pose\_01 has one
unsatisfied buried polar atom (penalty~0.50) while pose\_02 has two
(penalty~1.00). But the polar penalty operates only as a deficit
counter, it penalises unsatisfied polar atoms without rewarding the
degree of polar complementarity, and its contribution was overwhelmed
by the 2.0-fold TQS advantage and 18.8~percentage-point burial
differential that favoured pose\_02 under the agent's weighting scheme.
The Smina decomposition therefore identifies a concrete tool-suite gap:
a continuous measure of polar complementarity that rewards
well-satisfied polar environments, not merely penalises unsatisfied
ones. Such a metric would have narrowed the scoring differential
between pose\_01 and pose\_02 and potentially brought the correct pose
within the agent's attention threshold.

\paragraph{Category~C: Recovery: 3OXC (Smina failure, agent success).}

System 3OXC (HIV-1 protease, 19 candidate poses) represents the
single recovery in the benchmark (Figure~\ref{fig:pymol-3oxc}). Smina selected pose\_01
(RMSD 11.00~\AA{}) as its top-ranked pose by affinity, placing the
correct near-native conformation, pose\_02 (RMSD 0.63~\AA{}), second.
The agent reversed this ranking and promoted pose\_02 to first. The
decisive factor was conformational strain used as a tiebreaker between two
deep-pocket binders with similar interaction profiles.
Table~\ref{tab:case-3oxc} compares the agent's selected pose against
Smina's top-ranked pose and the agent's closest competitor in its own
deliberation (pose\_06, RMSD 9.51~\AA{}).

\begin{table}[htbp]
\centering
\begin{tabular}{lccc}
\hline
\textbf{Metric} & \textbf{pose\_02 (agent)} & \textbf{pose\_01 (Smina)} & \textbf{pose\_06 (runner-up)} \\
\hline
TQS                          & 16.81 & 13.57 & 17.56 \\
Total interactions           & 24    & 20    & 25 \\
H-bonds (avg quality)        & 6 (0.88) & 6 (0.88) & 6 (0.74) \\
Hydrophobic contacts (avg)   & 14 (0.81) & 10 (0.79) & 15 (0.84) \\
Burial ratio                 & 80.0\% & 76.4\% & 83.6\% \\
Binding mode quality         & 0.89 & 0.85 & 0.93 \\
Strain (kcal/mol)            & 46.06 & 45.01 & 66.42 \\
Steric clashes               & 2 (negligible) & 2 (negligible) & 2 (small) \\
Polar penalty                & 2.00 & 1.00 & 1.00 \\
\hline
\end{tabular}
\caption{Metric comparison for system 3OXC across the agent's selected
pose (pose\_02, RMSD 0.63~\AA{}), Smina's top-ranked pose (pose\_01,
RMSD 11.00~\AA{}), and the agent's runner-up (pose\_06,
RMSD 9.51~\AA{}). All three are classified as \textsc{deep\_pocket} by
the binding mode classifier.}
\label{tab:case-3oxc}
\end{table}

\begin{figure}[htbp]
\centering
\includegraphics[width=0.55\textwidth]{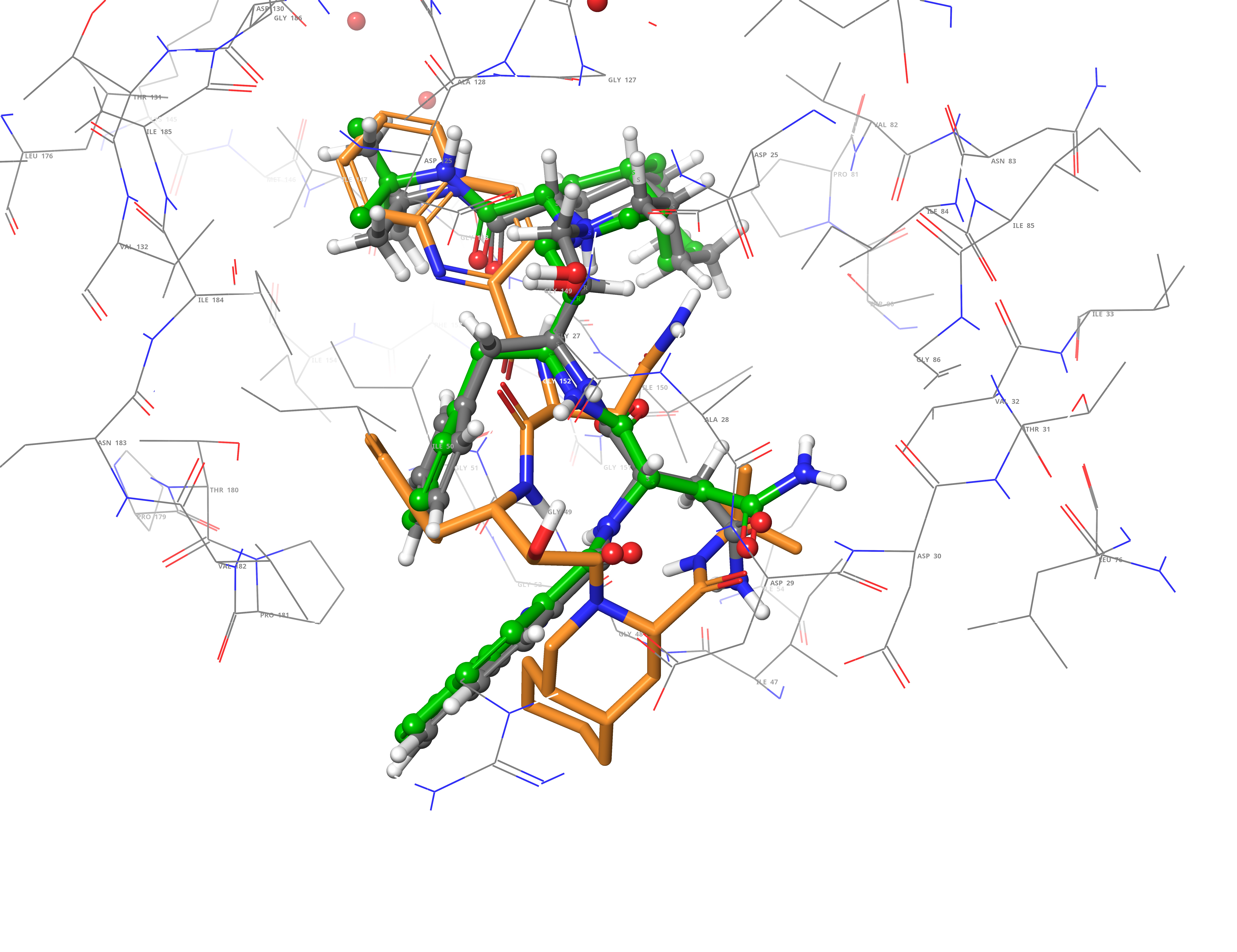}
\caption{System 3OXC: overlay of the agent's selected pose\_02 (green;
RMSD 0.63~\AA{}) and the rejected runner-up pose\_06 (orange; RMSD
9.51~\AA{}). Both achieved \textsc{deep\_pocket} classification with
comparable interaction profiles (TQS 16.81 vs.\ 17.56), but the agent
selected pose\_02 on the basis of a 20~kcal/mol conformational strain
advantage (46.06 vs.\ 66.42~kcal/mol). The crystallographic ligand is
shown in grey.}
\label{fig:maestro-3oxc-vs-pose06}
\end{figure}

\subparagraph{Reasoning trace and strain-based tiebreaker.}
The agent's reasoning trace reveals a clear decision pathway. Both pose\_02
and pose\_06 achieved \textsc{deep\_pocket} classification with high burial
ratios (80.0\% vs.\ 83.6\%) and near-identical interaction counts, making
binding mode unable to discriminate between them. The TQS slightly favoured
pose\_06 (17.56 vs.\ 16.81;
Figure~\ref{fig:maestro-3oxc-vs-pose06}), but the agent identified a
20~kcal/mol conformational strain differential as the decisive
tiebreaker. In the
agent's own words:

\begin{quote}
\textit{``With interactions being close (pose\_02 16.81 vs pose\_06 17.56),
the 20+ kcal/mol strain disadvantage for pose\_06 is too large to ignore.
Given both are deep-pocketed, strain was used as a critical tie-breaker.
Therefore, pose\_02 outranks pose\_06.''}
\end{quote}

\subparagraph{Deliberative process and decision hierarchy.}
The selection rationale above conveys the final verdict, but the
Chain of Thought reveals how that verdict crystallised during the
deliberation. Early in the reasoning, the agent diagnosed an ensemble-level
anomaly and set the analytical frame:

\begin{quote}
\textit{``I need to interpret the high strain across the dataset, which might
suggest a force-field mismatch for this ligand and protonation. Even though
relative differences between poses matter, I should penalize high strain as a
tiebreaker among similarly strong candidates.''}
\end{quote}

\noindent With this principle established, the agent worked through the top
contenders with explicit conditional reasoning:
\textit{``pose\_06 shows top quality but suffers from high strain of 66.42.
For the best selection, lower strain is preferred, though high strain might be
acceptable if interactions are notably better.  [\ldots] Despite a worse polar
penalty (2.0 vs 1.0), the substantial 20~kcal strain difference favors
pose\_02.''}  The strain tiebreaker was therefore formulated as a general
policy during the reasoning process and then applied to the specific
comparison, rather than appearing as a post-hoc rationalisation in the final
output. In parallel, the agent enforced binding mode classification as a
categorical gate. When evaluating pose\_11 (TQS 17.24, comparable to the top
candidates), it concluded:
\textit{``Per best practice, a DEEP\_POCKET pose with slightly inferior
metrics outranks a PARTIAL\_POCKET pose with slightly superior interaction
quality. Thus, pose\_11 cannot be \#1.''}  This decision
hierarchy, \textsc{deep\_pocket} status as a hard filter, interaction quality
as the primary discriminator, strain as the final tiebreaker, directly
mirrors the hierarchical decision framework encoded in the system prompt
(Section~\ref{sec:agent-architecture}). The agent further demonstrated
critical evaluation of its own tool outputs: in the Chain of Thought, it
noted that \textit{``salt bridge distances over 5.00~\AA{} likely aren't
realistic,''} and in the metric breakdown for pose\_02 it reinforced this
scepticism, observing that \textit{``The salt bridges are distant (quality
low) but are not needed for ranking given the already excellent
H-bond/hydrophobic profile.''}  Rather than naively incorporating all detected
interactions, the agent discounted geometrically implausible contacts, a
behaviour that mirrors expert structural biologist judgement.

\subparagraph{Self-aware uncertainty and tool-output anomaly detection.}
Equally revealing is the agent's handling of uncertainty and tool-output
anomalies. In the Key Findings section, it attributed the systematically
elevated strain energies to a specific computational cause:

\begin{quote}
\textit{``Conformational strain is consistently high across all poses
(41--71~kcal/mol), which points to systematic issues (e.g., MMFF94
parameterization or ligand protonation). Use strain as a relative
differentiator only.''}
\end{quote}

\noindent By naming the MMFF94 force field explicitly and recommending
relative rather than absolute interpretation, the agent demonstrated
awareness that its own tool's output was partially artefactual, yet still
informative for ranking. Despite these acknowledged limitations, the agent did not
retreat into indecision; instead, it explicitly identified the conditions
under which its ranking might be revised:
\textit{``A few poses (e.g., pose\_06) have very strong quality but extreme
strain; if future refinement reduces their strain substantially without
degrading interactions, rankings could change.''}  The agent thus assigned a
``Medium'' confidence level that was calibrated to the evidence: high enough
to recommend a specific pose, yet transparent about the conditions that could
alter the outcome.

\subparagraph{Cross-metric trade-offs and emergent correlations.}
The agent's reasoning further demonstrates cross-metric trade-off analysis
rather than siloed evaluation of individual metrics. When justifying its
tolerance of pose\_02's higher polar penalty (4~unsatisfied polars, score 2.00
vs.\ 1.00 for pose\_06), the agent acknowledged the weakness explicitly:

\begin{quote}
\textit{``I accepted pose\_02's moderate polar penalty (4 unsatisfied polars)
because the strong H-bond geometry (+ excellent hydrophobics) suggests net
favorable binding; the alternatively lower-polar-penalty poses either had
weaker interaction quality or worse burial or were partial-pocket binders.''}
\end{quote}

\noindent This passage shows the agent reasoning across all metric axes
simultaneously: a deficit on one axis (polar satisfaction) is tolerated only
because every alternative that performs better on that axis pays a worse price
elsewhere. An ensemble-level insight reinforced this contextualisation:
\textit{``Polar penalties range from minimal (1--2 unsatisfied polars) to
concerning ($\geq$5), and large polar penalties usually correlate with lower
H-bond quality.''}  The agent thus discovered an emergent cross-metric
correlation, poor polar satisfaction tends to co-occur with weaker hydrogen
bonding, directly from the data, and used it to contextualise pose\_02's
moderate penalty as an acceptable outlier given its strong hydrogen-bond
profile (average quality 0.88). The same multi-axis logic governed the
comparison with Smina's preferred pose\_01:
\textit{``Although pose\_01 has slightly better strain and better polar
satisfaction, pose\_02's substantially stronger interaction network ($\Delta$
quality ${\sim}$+3.24; +4 total interactions; more and stronger hydrophobics)
outweighed those advantages under the chosen weights.''}  By quantifying the
delta across multiple axes simultaneously, the agent arrived at a
transparent, auditable justification for overriding the scoring-function
ranking, precisely the kind of interpretable decision trace that a
single-score approach cannot provide.

\noindent This case provides the strongest evidence for the paper's central
thesis: a multi-metric evaluation that preserves individual physical
contributions can recover near-native poses that a single-score function
misses. Smina's unified affinity score conflated the favourable interactions
of pose\_01 (its top-ranked pose, TQS 13.57) with its lower strain profile;
the agent's decomposed evaluation elevated pose\_02 on the basis of a
superior interaction network while simultaneously penalising the
higher-quality competitor (pose\_06) for its thermodynamically unfavourable
strain.

\paragraph{Category~D: Instructive failure: 4JFL (Smina failure, agent
failure).}

System 4JFL (peptidyl-prolyl isomerase FKBP51, 20 candidate poses) is the most
instructive of the four Category~D systems because the agent explicitly
evaluated and rejected the correct pose (pose\_03, RMSD 1.76~\AA{}),
providing a fully transparent account of where the reasoning diverged from
ground truth (Figure~\ref{fig:pymol-4jfl}). The agent's selected pose\_01 has an RMSD of 3.27~\AA{} to the
crystallographic binding mode
($\Delta_{\mathrm{RMSD}} = 1.51$~\AA{}, classified as \emph{moderate} in
Table~\ref{tab:rmsd-selected-poses}), placing it near the boundary of a
plausible alternative binding mode rather than a catastrophic misprediction.
No pose in the ensemble achieved \textsc{deep\_pocket}
classification (best burial 64.9\%, pose\_02), and 4 of 20 were classified as
\textsc{surface} binders. Table~\ref{tab:case-4jfl} compares the agent's
selection (pose\_01) with the correct pose (pose\_03).

\begin{table}[htbp]
\centering
\begin{tabular}{lcc}
\hline
\textbf{Metric} & \textbf{pose\_01 (selected)} & \textbf{pose\_03 (correct)} \\
\hline
TQS                          & 8.70 & 7.91 \\
H-bonds (avg quality)        & 5 (0.67) & 4 (0.78) \\
Hydrophobic contacts (avg)   & 7 (0.76) & 6 (0.80) \\
Burial ratio                 & 56.8\% & 43.2\% \\
Binding mode quality         & 0.45 & 0.35 \\
Strain (kcal/mol)            & 55.08 & 67.71 \\
Steric clashes               & 0 & 1 \\
Polar penalty                & 0.50 & 1.00 \\
Ligand efficiency             & 0.389 & 0.361 \\
\hline
\end{tabular}
\caption{Metric comparison for the agent's selected pose (pose\_01) and the
ground-truth best pose (pose\_03) in system 4JFL. Both poses are classified
as \textsc{partial\_pocket}.  pose\_01 leads on most metrics; pose\_03
has superior hydrogen-bond geometry.}
\label{tab:case-4jfl}
\end{table}

\subparagraph{Explicit rejection and reasoning divergence.}
The agent's reasoning trace reveals the exact point of divergence. After
evaluating all 20 poses, it compared pose\_01 and pose\_03 on each metric
axis and concluded:

\begin{quote}
\textit{``[pose\_03 has] lower burial (43.2\% vs 56.8\% for pose\_01), higher
strain (67.71 kcal/mol, worse by ${\sim}$12.6~kcal/mol), higher polar penalty
(1.00 vs 0.50), and 1 minor clash. [\ldots] The much higher strain and worse
burial/polar profile outweighed its slightly superior H-bond geometry.
pose\_01 remains preferable.''}
\end{quote}

\noindent The agent acknowledged pose\_03's excellent hydrogen-bond geometry
(average quality 0.78 vs.\ 0.67) but assigned insufficient weight to this
advantage. Under the agent's weighting scheme (interaction quality 35\%,
binding mode and burial 25\%, strain 20\%, polar penalty 15\%, clashes 5\%),
the metrics favouring pose\_01, higher TQS, deeper burial, lower
strain, and fewer clashes, collectively outweighed the single metric
favouring pose\_03 (H-bond geometry). This is, in part, a genuine trade-off: the agent applied its
weighting scheme consistently. Yet three structural observations
suggest that the trade-off is less clear-cut than the metric table
implies.

\begin{figure}[htbp]
\centering
\begin{subfigure}[t]{0.48\textwidth}
\centering
\includegraphics[width=\textwidth]{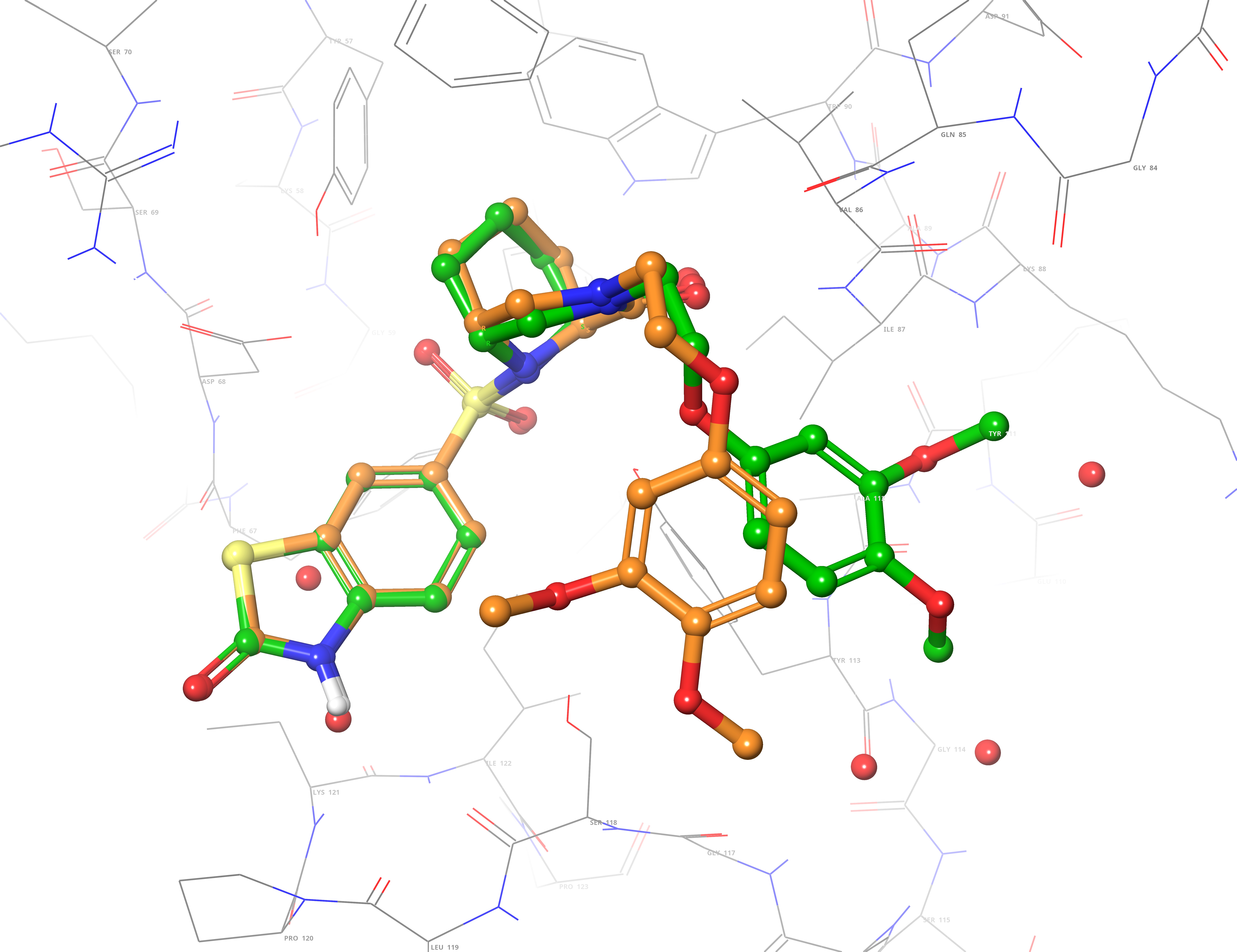}
\caption{Close-up of the region where the agent's selected pose\_01
(orange; RMSD 3.27~\AA{}) and the correct pose\_03 (green; RMSD
1.76~\AA{}) diverge. Both poses occupy the binding-site core
similarly; the displacement concentrates in the solvent-exposed tail of
the molecule.}
\label{fig:maestro-4jfl-tail}
\end{subfigure}
\hfill
\begin{subfigure}[t]{0.48\textwidth}
\centering
\includegraphics[width=\textwidth]{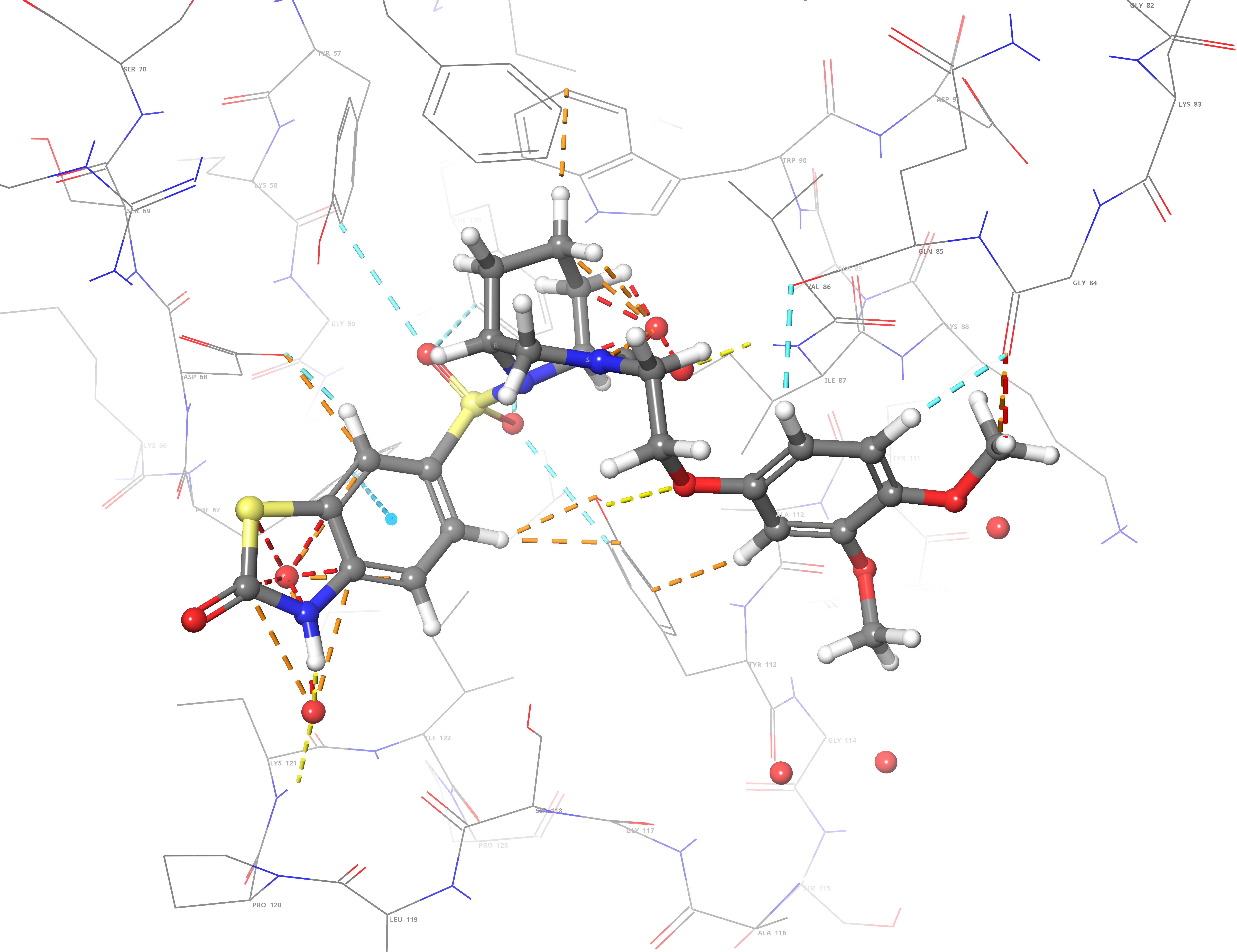}
\caption{Crystallographic water molecules in the 4JFL binding site
mediate hydrogen bonds between the ligand and protein that the
pipeline's polar penalty tool does not credit, contributing to
pose\_03's artificially elevated polar penalty.}
\label{fig:maestro-4jfl-water}
\end{subfigure}
\caption{Structural details for system 4JFL illustrating two of the three
factors behind the agent's incorrect selection.  (a)~The RMSD displacement
between the two poses concentrates in the solvent-exposed tail rather than
in the binding-site core.  (b)~Ordered crystallographic water molecules
near the ligand mediate hydrogen bonds that the pipeline's polar penalty
tool cannot account for. The crystallographic ligand is shown in grey in
both panels.}
\label{fig:maestro-4jfl-detail}
\end{figure}

First, the 13.6-percentage-point gap in burial ratio (56.8\% vs.\
43.2\%) overstates the actual difference in pocket engagement. The
SASA data show that pose\_03 has a \emph{lower} total
solvent-accessible surface area
(182.4~\AA{}$^{2}$ vs.\ 194.1~\AA{}$^{2}$) and a lower average SASA
per atom (4.93~\AA{}$^{2}$ vs.\ 5.25~\AA{}$^{2}$) than pose\_01; the
burial \emph{ratio} reverses because it applies a binary
1.0~\AA{}$^{2}$ threshold, so five atoms that fall just above the
cutoff in pose\_03 count as ``exposed'' while the corresponding atoms
in pose\_01 fall just below it and count as ``buried.''  This
sensitivity to the threshold is a known limitation of the current SASA
implementation (Section~\ref{sec:sasa}), which does not decompose
protein-contributed burial from intramolecular self-occlusion. Visual
inspection of the two poses confirms that they occupy the pocket
similarly (Figure~\ref{fig:maestro-4jfl-tail}); the RMSD difference
between them (3.27~\AA{} vs.\ 1.76~\AA{}) concentrates in the
solvent-exposed tail of the ligand rather than in the binding-site core.

Second, the 4JFL crystal structure contains ordered water molecules
that mediate hydrogen bonds between the ligand and the protein
(Figure~\ref{fig:maestro-4jfl-water}). The
pipeline's polar penalty tool evaluates only direct protein--ligand
contacts within 3.5~\AA{} (Section~\ref{sec:polar-penalty}) and does
not credit water-mediated satisfaction of buried polar atoms. Visual
inspection of the X-ray structure confirms that at least some of
pose\_03's nominally unsatisfied polar atoms are in fact satisfied
through crystallographic water bridges, so its true polar penalty is
lower than reported, narrowing the gap with pose\_01 on this axis as
well.

Third, the agent's weighting scheme undervalues H-bond geometry
quality relative to burial ratio when no pose achieves
\textsc{deep\_pocket} status. Increasing the weight of H-bond
geometry in shallow-site ensembles would address this component of the
failure.

These three factors, an artifact-inflated burial ratio, an unmodelled
water-mediated interaction network, and a weighting heuristic biased
toward burial depth, converge to explain the 4JFL failure more
completely than any single cause.

\subparagraph{Ensemble diagnosis and quality-over-quantity framework.}
Beyond the final metric comparisons, the chain-of-thought trace reveals a
deliberative process in which the agent first diagnosed the ensemble's
distributional properties before applying any ranking logic. Having
characterised the ensemble, the agent adapted its evaluation strategy to its
specific characteristics:

\begin{quote}
\textit{``pose\_02 has the highest burial but weaker interaction quality.
I think the scoring should weigh interaction quality heavily, along with
binding mode classification and strain. It's essential to highlight that
this dataset primarily consists of partial\_pocket modes.''}
\end{quote}

\noindent This passage is noteworthy because the agent explicitly recognised
that, in the absence of any \textsc{deep\_pocket} pose, interaction quality
must serve as the primary discriminator rather than burial depth. The agent
also identified internal counterexamples that stress-tested this framework:
\textit{``Notably, pose\_15 has the best strain and polar penalty but a low
interaction quality of 2.70 with only five total interactions.  [\ldots]
Interesting trade-offs are at play here.''}  By flagging pose\_15 as a case
where favourable strain (${\sim}$49~kcal/mol) and zero polar penalty
coincided with poor interaction quality (TQS of~2.70), the agent rehearsed
the multi-dimensional trade-off logic it would later apply to the critical
pose\_01~vs.\ pose\_03 comparison. The culmination of this deliberative arc
was a self-labelled decision philosophy, \textit{``In the
quality-over-quantity framework, pose\_01's interaction geometry and network
coherence decisively beat alternatives''}, a framework that, while
internally consistent, ultimately led to the incorrect selection.

\subparagraph{Cross-metric insights and metric dominance hierarchy.}
The agent's ensemble-level observations further reveal its capacity for
cross-metric reasoning that extends beyond pairwise pose comparisons.
Commenting on the relationship between polar satisfaction and interaction
richness across all 20~poses, it noted:
\textit{``Polar penalties range from excellent (pose\_15, 0.00) to concerning
(pose\_08, 2.00), but poses with zero polar penalty tended to have weak
interactions/poor quality.''}  This anti-correlation reflects a
pharmacologically meaningful pattern: poses that form extensive contacts with
the protein interior are precisely those most likely to bury unsatisfied polar
atoms, whereas surface-exposed poses trivially satisfy all polar groups by
remaining solvent-accessible. The agent's acknowledgement of this trade-off
demonstrates genuine ensemble-level reasoning rather than purely local
comparisons. When arbitrating among its top candidates, the agent codified a
clear metric dominance hierarchy. On the role of conformational strain, it
stated:

\begin{quote}
\textit{``Strain was used as a tie-breaker among strong interaction networks
because all poses show unusually high strain; large absolute differences
(e.g., 55 vs 68~kcal/mol) were considered, but strain alone could not
overcome clearly superior interaction quality plus better burial.''}
\end{quote}

\noindent It further quantified the energetic basis for its polar penalty
treatment, \textit{``Polar penalties were factored to penalize buried
unsatisfied polars (${\sim}$2--3~kcal/mol per atom) but given less weight
than primary interaction quality''}, and articulated a principled stance on
the burial~vs.\ interaction trade-off exemplified by pose\_02:
\textit{``Interaction quality is weighted more than burial when both are in
acceptable ranges. The interaction deficit (both in count and quality) and
higher polar penalty keep pose\_02 behind pose\_01 despite its superior
burial.''}  The explicit subordination of strain and burial to interaction
quality when both remain within acceptable ranges reveals a coherent decision
architecture, one in which each metric occupies a defined role rather than
contributing interchangeably, even though this hierarchy led to an incorrect
selection for this particular system.

\subparagraph{Confidence assessment and constructive failure.}
Perhaps most strikingly, the agent's confidence assessment section reveals an
unusual degree of epistemic self-awareness for an automated system. It
assigned only ``Medium'' confidence to its selection and articulated specific
reasons:

\begin{quote}
\textit{``All poses show very high conformational strain (49--81~kcal/mol).
This raises concern about the docking protocol or protonation/tautomeric
states. Strain limits absolute confidence.''}
\end{quote}

\noindent The agent further proposed two alternative hypotheses for
the absence of deeply buried poses: \textit{``No DEEP\_POCKET binders
({>}70\% burial) were observed; several are SURFACE. The binding site may
be shallow, or docking constraints may have led to surface-biased
solutions.''}  It also identified a methodological limitation of its own tool
suite, \textit{``Water networks and protein side-chain repositioning were
not exhaustively modeled; marginal differences among top poses could change
with explicit solvent/side-chain repacking''}, thereby acknowledging that the
static, implicit-solvent analysis underpinning its ranking may not capture
solvation effects that could reorder top-tier candidates. Finally, the agent
proposed a concrete protocol-level intervention: \textit{``Consider re-running
docking with alternative parameters or a different pocket definition if
experimental evidence suggests a deeper site. The uniformly high strain
numbers are a red flag; refinement or re-docking is warranted.''}  This
cascade, from flagging systemic anomalies, through generating alternative
hypotheses, to recommending protocol modifications, transforms what would
otherwise be an uninformative wrong answer into a constructive diagnostic
report. The 4JFL failure thus illustrates a key property of the agentic
approach: even when the system arrives at the wrong answer, it makes its
reasoning fully traceable, providing not
only a complete account of its reasoning but also identifying the conditions
under which its confidence is limited and what protocol-level changes might
improve the outcome.

This case demonstrates that transparent failure can be more valuable than
opaque success: by providing a complete decision trace, the agent identifies
the precise weighting limitation, insufficient emphasis on H-bond geometry in
shallow binding sites, that a future system-prompt revision can target
directly.

Across the four case studies, several cross-cutting patterns emerge. First,
the binding mode classification tool acted as an effective gating filter in
2P16, eliminating 9 of 20 surface artifacts that would otherwise have
confounded the ranking. Second, the 3HS4 regression reveals a ceiling on
what computed metrics can achieve: when the correct binding mode lacks
favourable computed properties relative to a competitor, no weighting scheme
applied to the current tool outputs will recover it. Third, the
conformational strain tool proved decisive in 3OXC, where a 20~kcal/mol
strain differential correctly tipped the ranking in favour of the near-native
pose that Smina had missed, the strongest evidence for the multi-metric
thesis. Fourth, the 4JFL failure traces to three converging factors: a
burial ratio inflated by the binary SASA threshold, unmodelled
water-mediated hydrogen bonds that artificially penalise the correct
pose, and a weighting heuristic that undervalues H-bond geometry in
shallow binding sites. Each factor is independently actionable.

A second cross-cutting observation concerns adaptive weighting. In every
system, the agent tailored its metric weights to the ensemble's
distributional properties rather than applying a fixed formula: pocket
engagement dominated in surface-rich ensembles (2P16, 45\% combined weight),
interaction quality dominated in uniformly well-docked ensembles (3OXC,
35\%), and clashes and strain were weighted equally when absolute strain values were
uninformative (3HS4, both 12\%). This context-dependent deployment of the same tool
suite, sometimes as a hard gate, sometimes as a fine-grained
discriminator, distinguishes the agentic approach from fixed-weight
rescoring schemes.

A third pattern is the agent's consistent self-critical behaviour. In all
four systems, the agent attributed elevated strain values to MMFF94
force-field limitations, recommended relative rather than absolute
interpretation, and proposed concrete follow-up protocols (restrained
minimisation of the protein-ligand complex with side-chain repacking,
explicit hydration, or re-docking with alternative parameters).
These epistemic disclaimers and experimental suggestions transform each
output from a bare ranking into a diagnostic report whose transparency is
maintained regardless of whether the final answer is correct.

These patterns are consistent with the aggregate accuracy contrast reported
in Table~\ref{tab:partition-category}: the high $\mathrm{Acc}^{+}$ (80\%)
reflects the agent's ability to preserve correct identifications through
robust multi-metric reasoning (exemplified by 2P16), while the low
$\mathrm{Acc}^{-}$ (20\%) reflects the tool-suite coverage ceiling exposed
by 3HS4 and the weighting limitation exposed by 4JFL.
The formal Decision Attribution Analysis in the following section
(Section~\ref{sec:decision-attribution}) quantifies these patterns across all
ten systems.

\subsection{Decision Attribution Analysis}
\label{sec:results-decision-attribution}

Having established what the agent decided in the preceding sections,
we now analyse \emph{why} it decided that way using the three-part
attribution framework defined in
Section~\ref{sec:decision-attribution}. For each system the agent
produces a pose selection together with self-reported percentage
weights and a free-text chain-of-thought. The analyses below exploit
both the structured weights and the objective per-pose metric values
to assess which tools drove each selection, whether the agent's stated
reasoning faithfully reflects the underlying metric evidence, and
whether weight-allocation patterns differ systematically between
correct and incorrect decisions.

\begin{figure}[htbp]
\centering
\includegraphics[width=\textwidth]{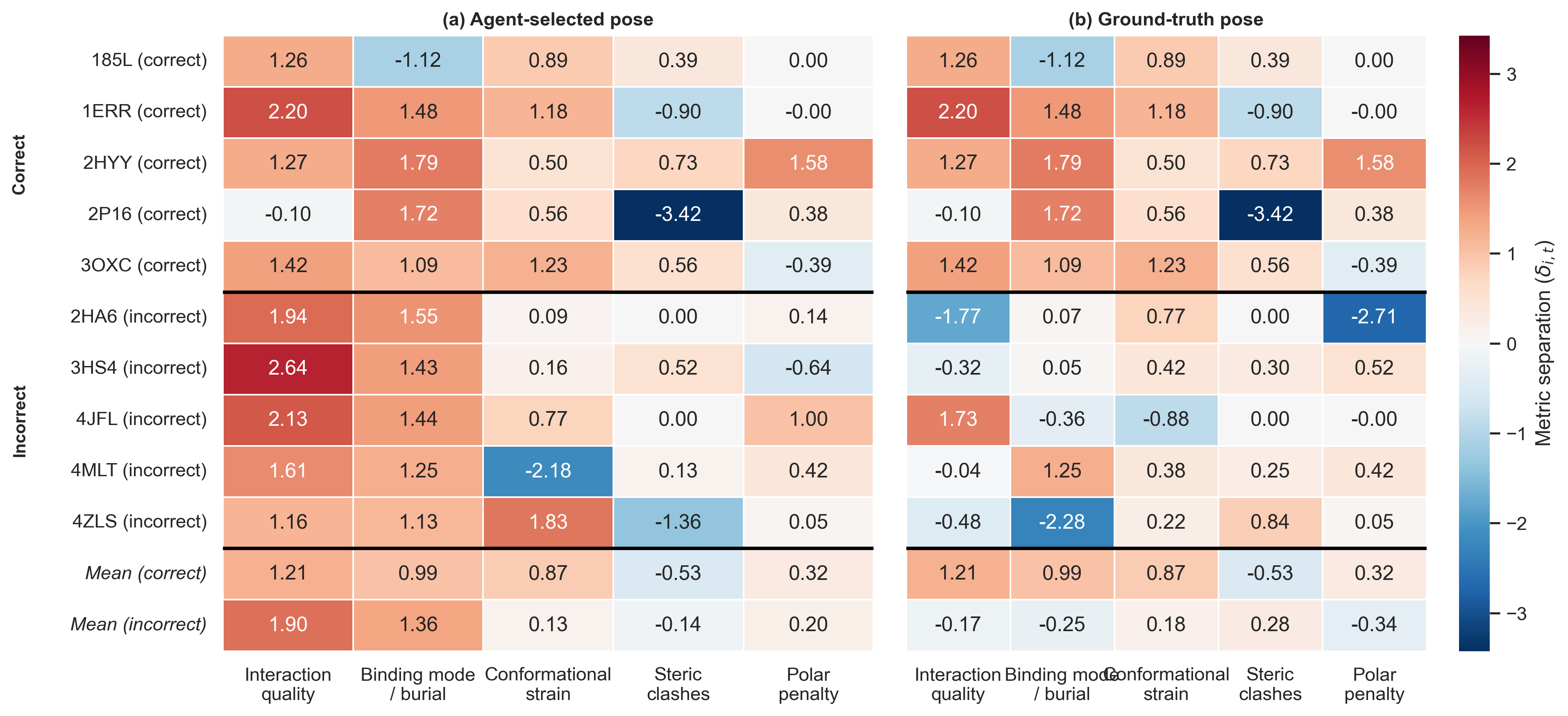}
\caption{Metric separation ($\delta_{i,t}$;
Equation~\ref{eq:metric-separation}) for each system (rows) and tool
category (columns). Panel~(a): agent-selected pose;
panel~(b): ground-truth (lowest-RMSD) pose. The two panels are
identical for correctly identified systems and differ only for failure
cases. Red cells indicate that the target pose is more favourable
than the candidate-pool mean; blue cells indicate a less favourable
selection. Per-partition means appear below the second black line.}
\label{fig:metric-separation-heatmap}
\end{figure}

\paragraph{Metric separation.}
Figure~\ref{fig:metric-separation-heatmap} reports the metric separation
$\delta_{i,t}$ (Equation~\ref{eq:metric-separation}) for each
system--tool combination as a colour-coded heatmap, with per-partition
means in the bottom rows. Panel~(a) shows the separations computed
for the agent-selected pose; panel~(b) shows the same metric for the
ground-truth (lowest-RMSD) pose. For the five correctly identified
systems the two panels are identical; differences appear only in the
five failure cases. Positive values indicate that the target pose is
more favourable than the candidate-pool mean on that metric; negative
values indicate a less favourable selection.

Two tool categories, interaction quality (IQ) and binding
mode/burial (BM/B), show positive separations in the large majority
of systems (9/10 and 9/10, respectively), indicating that the agent
almost universally selects poses that score above the candidate-pool
mean on these dimensions. The two exceptions, 2P16, where the
selected pose has a near-zero IQ separation ($\delta = -0.10$), and
185L, where burial is below the pool mean ($\delta = -1.12$), are
cases where the agent prioritised other decisive metrics (binding
mode gating and interaction quality, respectively). The mean IQ separation is higher for
incorrect selections ($\bar{\delta} = +1.90$) than for correct ones
($\bar{\delta} = +1.21$), suggesting that incorrect decisions tend to
be driven by an \emph{exceptional} TQS advantage, a high interaction
quality score alone is not sufficient to guarantee a correct choice.

The most discriminating tool category is \textbf{conformational
strain}. Correctly identified systems exhibit a mean strain
separation of $+0.87$ (the selected pose is nearly one standard
deviation below the pool mean in strain energy), whereas incorrectly
identified systems average only $+0.13$. Two failure cases
illustrate the extremes: in 4MLT the selected pose has a markedly
\emph{unfavourable} strain separation ($\delta = -2.18$, meaning the
selected pose is strained relative to the pool), while in 2HA6 the
strain dimension provided essentially no discriminative signal
($\delta = +0.09$). These patterns align with the qualitative
observations from the Representative System Analysis
(Section~\ref{sec:results-representative}) and suggest that
strain-based filtering is the principal determinant of decision
correctness.

Steric clashes and polar penalty exhibit near-zero mean separations
(clash: $-0.53$ and $-0.14$ for correct and incorrect, respectively;
polar: $+0.32$ and $+0.20$), with individual systems showing large
excursions in both directions. The polar penalty tool produces
$|\delta| < 0.3$ in four of ten systems, indicating that it rarely
differentiates the selected pose from the pool. The clash metric is
dominated by scale effects in systems where most poses register zero
clashes: a single pose with even a small clash severity value produces
an extreme z-score (e.g., 2P16: $\delta = -3.42$), making
cross-system comparison of this metric less informative than the
others.

Comparing panels~(a) and~(b) for the five failure cases reveals why
the agent errs. The ground-truth poses in these systems are
metric-indistinguishable from the candidate pool on the two dimensions
the agent weights most heavily: mean IQ separation is $-0.17$ and mean
BM/B separation is $-0.25$, both effectively at or below the pool
average. By contrast, the agent-selected poses in the same systems
show strong positive separations on both dimensions (mean IQ
$= +1.90$, BM/B $= +1.36$). The correct pose simply does not stand
out on interaction quality or burial, so a selection strategy that
prioritises these metrics will systematically miss it. Strain
separation tells the opposite story: ground-truth poses in failure
cases average $+0.18$, comparable to the $+0.13$ of agent-selected
poses, confirming that neither pose is markedly strained and that
strain alone cannot rescue the decision. One exception is 4JFL, where
the ground-truth pose retains a high IQ separation ($+1.73$) yet the
agent selected a different pose; the failure here traces to an
unfavourable binding mode/burial score ($-0.36$) on the correct pose,
which the agent penalised despite favourable interaction quality.

\paragraph{Reasoning faithfulness.}
Table~\ref{tab:faithfulness} reports the Spearman rank correlation
$\rho_i$ (Equation~\ref{eq:faithfulness}) between each system's
self-reported weight vector and the absolute metric separation vector.

\begin{table}[htbp]
\centering
\begin{tabular}{lcrl}
\hline
\textbf{System} & \textbf{Outcome} & $\boldsymbol{\rho_i}$ & \textbf{$p$-value} \\
\hline
185L & \checkmark & $+0.975$ & $0.005$ \\
1ERR & \checkmark & $+0.975$ & $0.005$ \\
2HYY & \checkmark & $+0.462$ & $0.434$ \\
2P16 & \checkmark & $-0.154$ & $0.805$ \\
3OXC & \checkmark & $+0.800$ & $0.104$ \\
\hline
2HA6 & $\times$ & $+0.975$ & $0.005$ \\
3HS4 & $\times$ & $+0.866$ & $0.058$ \\
4JFL & $\times$ & $+0.900$ & $0.037$ \\
4MLT & $\times$ & $-0.051$ & $0.935$ \\
4ZLS & $\times$ & $+0.359$ & $0.553$ \\
\hline
\multicolumn{2}{l}{\textit{Median (all)}} & $+0.833$ & --- \\
\multicolumn{2}{l}{\textit{Median (correct)}} & $+0.800$ & --- \\
\multicolumn{2}{l}{\textit{Median (incorrect)}} & $+0.866$ & --- \\
\hline
\end{tabular}
\caption{Reasoning faithfulness per system, measured as the Spearman
rank correlation $\rho_i$ between the agent's self-reported tool
weights $\mathbf{w}_i$ and the absolute metric separations
$|\boldsymbol{\delta}_i|$ (Equation~\ref{eq:faithfulness}). Higher
$\rho_i$ indicates that the agent assigns more weight to tools on
which the selected pose genuinely stands out.}
\label{tab:faithfulness}
\end{table}

The median faithfulness across all ten systems is $\rho = +0.83$,
indicating that the agent's self-reported weight rankings are
\emph{substantially concordant} with the objective metric-separation
rankings. Six of ten systems reach $\rho \geq +0.80$, and three
achieve $\rho = +0.975$ (the maximum attainable Spearman coefficient
for $T = 5$ categories). These results indicate that the
chain-of-thought reasoning is largely faithful: the agent assigns the
most weight to tools on which the selected pose genuinely excels
relative to the candidate pool.

Two systems show weak or negative faithfulness: 2P16
($\rho = -0.15$) and 4MLT ($\rho = -0.05$). In 2P16, the agent's
dominant weight was assigned to binding mode/burial (45\%), which
indeed produced the largest positive $\delta$ ($+1.72$), but the
second-highest weight (interaction quality, 30\%) corresponded to a
near-zero separation ($-0.10$), depressing the overall rank
correlation. In 4MLT, the near-uniform strain values across all 20
poses (range: $18.78$--$18.81$~kcal~mol$^{-1}$) resulted in an
extreme strain $\delta$ ($-2.18$) driven entirely by tiny absolute
differences, providing little genuine discriminative power. The
agent assigned only 5\% weight to strain, the lowest in the
benchmark, suggesting some awareness of the metric's limited
informativeness, yet the formal $\rho_i$ remains low because the
absolute $|\delta|$ ranking is dominated by this inflated z-score.

Descriptively, the median $\rho$ is similar across the two outcome
groups: $+0.80$ for correct systems and $+0.87$ for incorrect systems.
A Mann--Whitney $U$ test against these per-group rank distributions
yields $U = 13.0$, $p = 1.00$; this result is reported descriptively
because with $n_{1} = n_{2} = 5$ the test has minimal power to support
either rejection or equivalence. The reasoning alignment measured
here is therefore compatible with either a shared alignment profile
across outcomes or a modest difference that the present sample cannot
resolve. Taken at face value, the similarity is consistent with the
interpretation that incorrect decisions do not arise from a visible
realignment between stated and observed weightings, the agent relies
on the same ordering of metrics that it reports, but rather that the
metric evidence itself is insufficient to discriminate the correct pose
from high-scoring decoys. This descriptive measure does not, on its
own, distinguish faithful introspection from internally coherent
post-hoc rationalisation~\citep{turpin2023language, lanham2023measuring}.

\begin{figure}[htbp]
\centering
\includegraphics[width=0.55\textwidth]{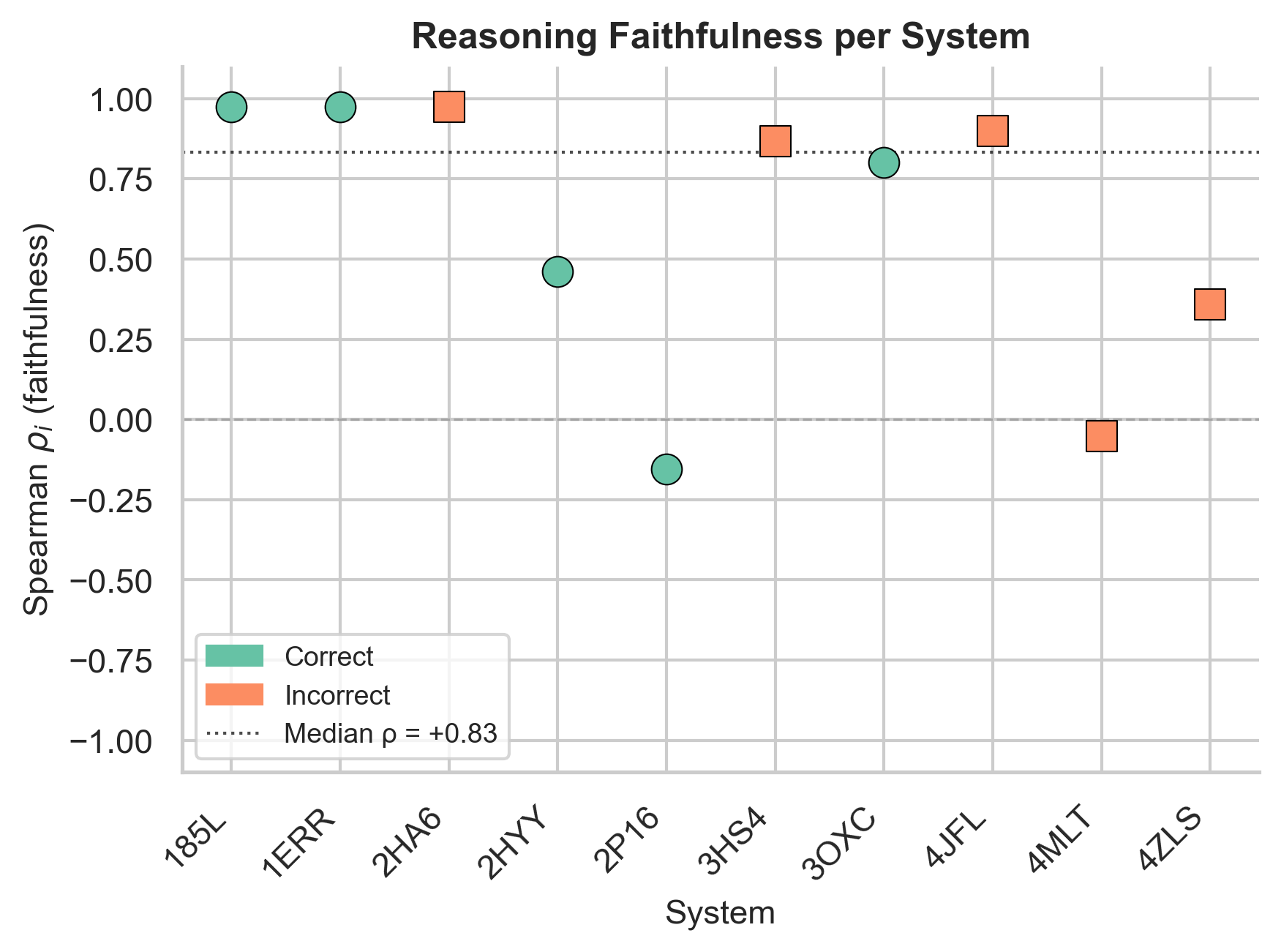}
\caption{Reasoning faithfulness ($\rho_i$;
Equation~\ref{eq:faithfulness}) for each system, coloured by outcome
(teal = correct, orange = incorrect). The dotted horizontal line
marks the median across all systems ($\rho = +0.83$). Faithfulness
is not systematically related to decision correctness.}
\label{fig:reasoning-faithfulness}
\end{figure}

Figure~\ref{fig:reasoning-faithfulness} visualises the per-system
$\rho_i$ values, confirming the absence of a systematic
correct--incorrect separation.

\paragraph{Outcome-stratified attribution.}
Table~\ref{tab:outcome-weights} compares the mean stated tool weights
for correctly and incorrectly identified systems
(Equation~\ref{eq:outcome-weights}).

\begin{table}[htbp]
\centering
\begin{tabular}{lrrr}
\hline
\textbf{Tool category} & $\bar{w}_t^{\,\checkmark}$ & $\bar{w}_t^{\,\times}$ & $\Delta$ \\
\hline
Interaction quality      & 36.0\% & 34.4\% & $+1.6$ \\
Binding mode / burial    & 30.0\% & 28.4\% & $+1.6$ \\
Conformational strain    & 16.0\% & 15.4\% & $+0.6$ \\
Steric clashes           &  8.0\% &  8.4\% & $-0.4$ \\
Polar penalty            & 10.0\% & 13.4\% & $-3.4$ \\
\hline
\end{tabular}
\caption{Outcome-stratified mean tool weights
(Equation~\ref{eq:outcome-weights}).
$\bar{w}_t^{\,\checkmark}$ and $\bar{w}_t^{\,\times}$ denote the
mean weight assigned to tool category~$t$ across the five correct and
five incorrect systems, respectively.  $\Delta$ is the difference
(correct $-$ incorrect).}
\label{tab:outcome-weights}
\end{table}

\begin{figure}[htbp]
\centering
\includegraphics[width=0.65\textwidth]{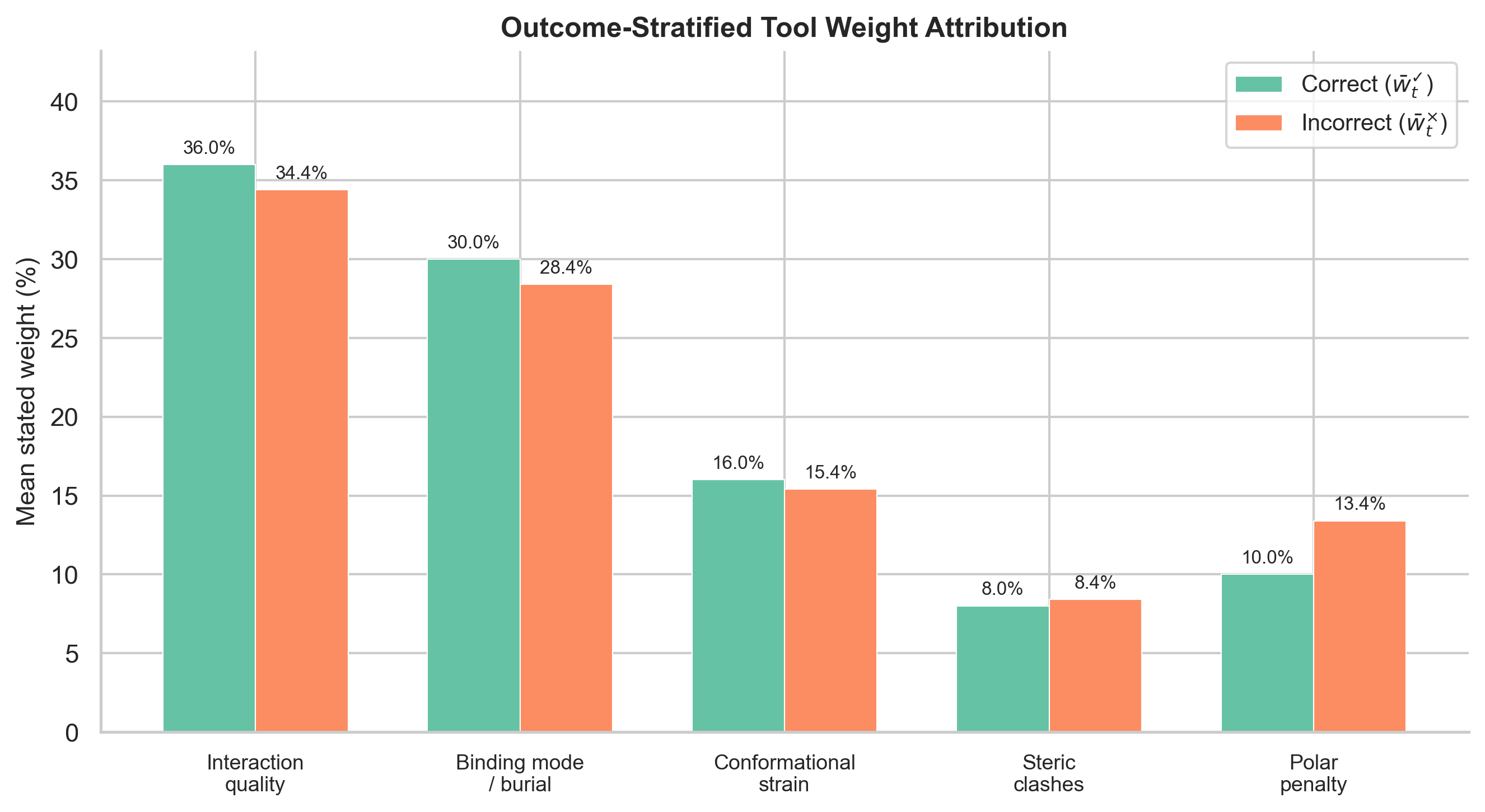}
\caption{Outcome-stratified tool weight comparison
(Equation~\ref{eq:outcome-weights}). Each pair of bars shows the mean
self-reported weight for correctly identified systems (teal) and
incorrectly identified systems (orange). Weight profiles are
remarkably similar, indicating that the failure mode lies in the
tool-suite's discriminative power rather than in the agent's attention
allocation.}
\label{fig:outcome-stratified-weights}
\end{figure}

Figure~\ref{fig:outcome-stratified-weights} provides a visual
comparison of the two weight profiles.

The weight profiles for correct and incorrect decisions are strikingly
similar: no tool category exhibits a mean difference exceeding
$|\Delta| = 3.5$~percentage points. The largest individual
difference is a $-3.4$ percentage-point lower allocation to polar
penalty in correct systems ($\bar{w}^{\,\checkmark} = 10.0\%$) than
in incorrect ones ($\bar{w}^{\,\times} = 13.4\%$), but this is
within the expected variability for $m = 5$~systems per group. The
interaction quality and binding mode/burial categories jointly receive
approximately 63--66\% of the total weight in both groups, confirming
that the agent treats these as the primary decision drivers regardless
of the outcome.

The near-identical weight profiles, combined with the faithful
reasoning demonstrated in Analysis~2, lead to a key insight: the
agent's failure mode is \emph{not} a misallocation of attention
across tool categories. Rather, incorrect decisions arise when the
metric evidence itself, particularly the strain dimension, fails to
separate the correct pose from high-scoring decoys. This directs
future improvement efforts toward enhancing individual tool outputs
(e.g., more accurate force fields, additional binding-quality
indicators) rather than toward prompt engineering of the agent's
weighting strategy.

\subsection{Supplementary Analyses}
\label{sec:supplementary_analyses}

\begin{figure}[htbp]
    \centering
    \includegraphics[width=0.55\textwidth]{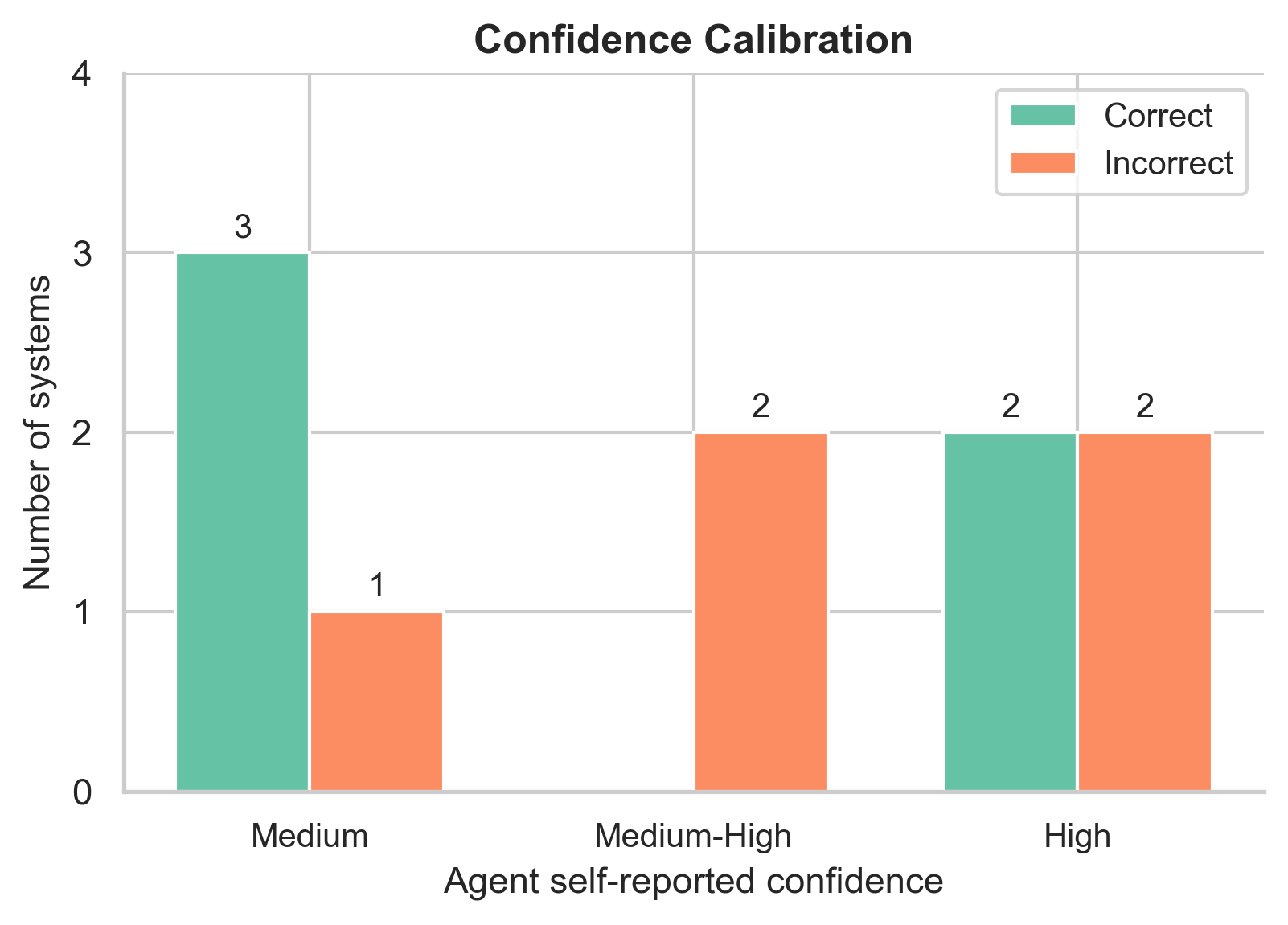}
    \caption{
        Confidence calibration.
        Bars show the number of correct (green) and incorrect (orange) systems
        at each self-reported confidence level.
        The descriptive ordering (high-confidence 50\% accuracy,
        medium-confidence 75\%) is opposite to what empirical
        calibration would predict, but cannot be resolved inferentially
        at $n = 10$ (Fisher exact test, $p = 1.000$; reported
        descriptively).
    }
    \label{fig:confidence_calibration}
\end{figure}

\paragraph{Confidence Calibration.}
Each output report includes a self-reported confidence level, \emph{Medium},
\emph{Medium-High}, or \emph{High}, reflecting the agent's assessment of its own
certainty.
Across the ten benchmark systems, the agent assigned High confidence to four
systems (185L, 1ERR, 3HS4, 4MLT), of which only two were correct (50\%
accuracy).
Medium-confidence systems (2HYY, 2P16, 3OXC, 4JFL) achieved 75\% accuracy
(3/4 correct), while the two Medium-High systems (2HA6, 4ZLS) were
both incorrect.
A Fisher exact test of the two-by-two table (High versus non-High
confidence, correct versus incorrect) returns OR~=~1.00,
$p$~=~1.000; this is reported descriptively, because at $n = 10$ systems
the test has insufficient power to support a confident statement
either way. Taken descriptively, the high-confidence subset reached
50\% accuracy and the medium-confidence subset 75\%, an ordering
opposite to the one that empirical calibration would predict. We
therefore refrain from concluding that the self-reported confidence
score is uncalibrated; we note instead that calibration cannot be
assessed at this sample size and that the observed ordering warrants
larger-scale evaluation
(Figure~\ref{fig:confidence_calibration}).

\paragraph{Ground-Truth Pose Metric Profile.}
To understand where correct ground-truth poses sit in metric space relative to
their ensembles, we computed the signed $z$-score of each ground-truth pose's
five tool-derived metrics within its ensemble, defining positive direction as
favourable (higher IQ and BM/B; lower CS, SC, and UP).
In the five correct systems, the ground-truth pose exhibits consistently
favourable metric separation: mean signed $z$-scores are
$+1.21$ (IQ), $+0.99$ (BM/B), and $+0.87$ (CS), indicating that
these ground-truth poses genuinely stand out as ensemble-best across
multiple tools.
By contrast, in the five incorrect systems the ground-truth pose is metric-indistinguishable
from the ensemble mean (IQ $z = -0.17$, BM/B $z = -0.25$,
CS $z = +0.18$), offering the agent no physics-based signal to favour
the native-like geometry.
This result suggests that failures are concentrated in systems where the
ground-truth pose has no metric advantage, making correct selection
fundamentally harder with the current tool suite.

\begin{figure}[htbp]
    \centering
    \includegraphics[width=0.55\textwidth]{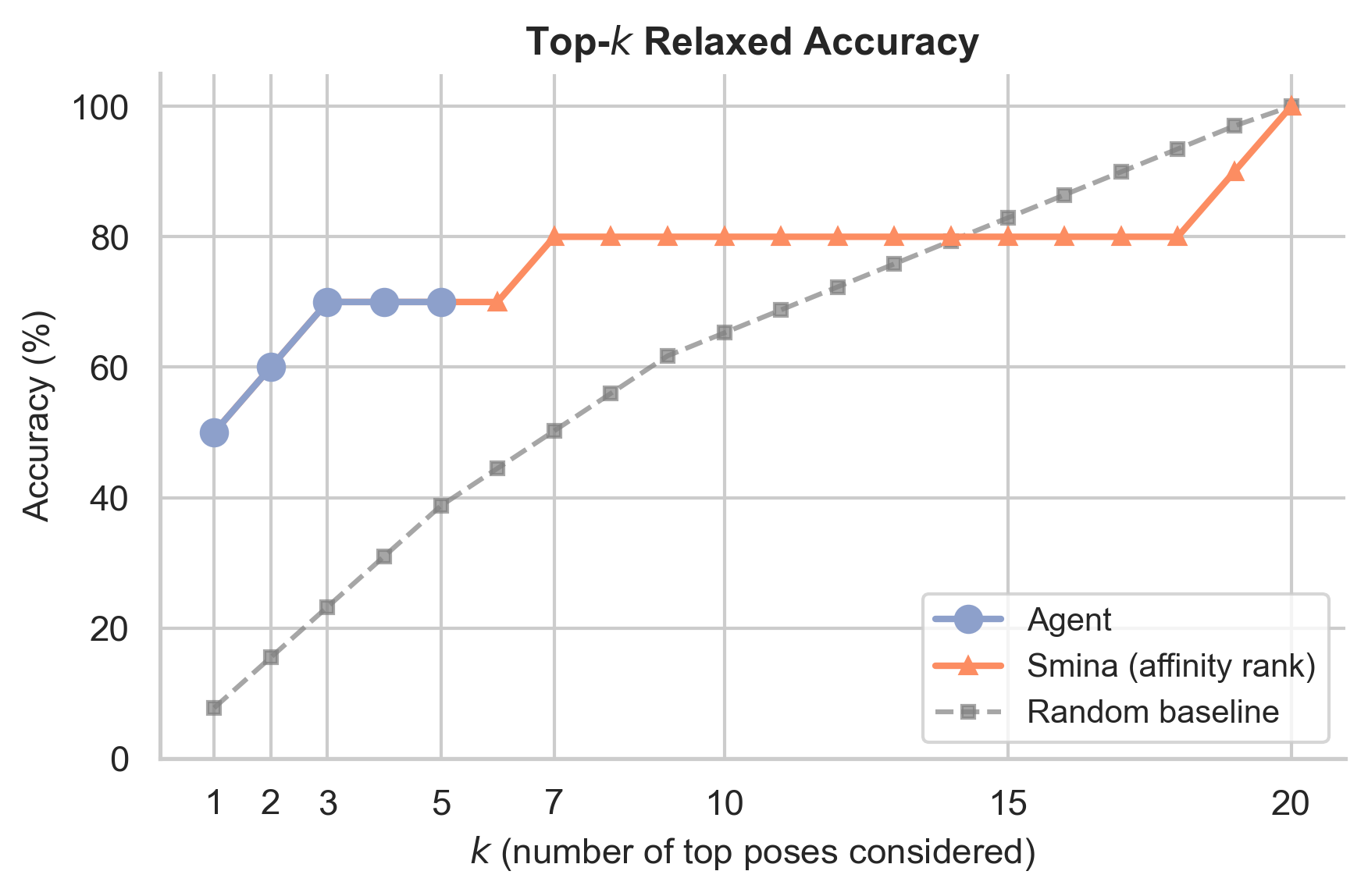}
    \caption{
        Top-$k$ relaxed accuracy.
        The agent's accuracy (circles, solid line) rises from 50\% at $k = 1$
        to 70\% at $k = 3$ and then plateaus; the plateau is a
        configuration artefact caused by the prompt instructing the
        agent to emit a short ranked shortlist (typically four to five
        poses) rather than a full ordering, not a property of its
        underlying ranking ability.
        Smina's affinity-based ranking (triangles, solid line) matches
        the agent through $k = 3$ and then continues to 100\% at
        $k = 20$ because it ranks every docked conformation.
        Head-to-head comparison is therefore meaningful only at matched
        ranks $k \leq 3$.
        The random-selection baseline (squares, dashed line) grows as
        $k / \bar{n}_{\mathrm{poses}}$.
    }
    \label{fig:topk_accuracy}
\end{figure}

\paragraph{Top-\texorpdfstring{$k$}{k} Relaxed Accuracy.}
The comparative analysis section of each report implicitly ranks poses
by decreasing preference. A structural caveat conditions the
interpretation of this analysis: the system prompt instructs the agent
to emit a short ranked shortlist (typically the top four to five poses)
rather than a complete ordering of the ensemble, whereas Smina
produces an affinity score for every pose and therefore a full
ranking. The agent's top-$k$ curve therefore plateaus where its
emitted shortlist ends (a configuration choice) and not at a ceiling of
its underlying ranking ability. Head-to-head comparison between agent
and Smina top-$k$ accuracies is meaningful only at matched ranks
$k \leq 3$, which is the depth covered by every run in the benchmark.
Relaxing the success criterion from top-1 to top-$k$ reveals a sharp
improvement from $k = 1$ to $k = 3$:
accuracy rises from 50\% (5/10) at $k = 1$, to 60\% (6/10) at $k = 2$
when 4MLT's ground truth (pose\_07) enters the shortlist, and 70\% (7/10)
at $k = 3$ when 4JFL's ground truth (pose\_03) is included.
No further gains occur at $k = 4$ or $k = 5$: the three remaining failures
(2HA6, 3HS4, 4ZLS) have ground-truth poses that were never emitted in
the agent's shortlist.
Smina's affinity-based ranking follows the same trajectory through
$k = 3$ (50\%, 60\%, 70\%), though the systems recovered at $k = 2$
differ: the agent recovers 4MLT while Smina recovers 3OXC.
Beyond $k = 5$, Smina continues to accumulate correct systems, reaching
80\% at $k = 7$ (4MLT), 90\% at $k = 19$ (2HA6), and 100\% at $k = 20$
(4ZLS), because it produces a complete ranking over all docked
conformations; the agent does not, so its top-$k$ ceiling at 70\%
reflects the truncated-emission configuration rather than a limit on
its ability to rank the remaining poses had it been prompted to do so.
Both methods substantially exceed the random-selection baseline
(7.7\% at $k = 1$, 38.7\% at $k = 5$) at every $k$ within the agent's
ranking depth, confirming that both rankings carry meaningful signal
even when the top-1 pick is incorrect
(Figure~\ref{fig:topk_accuracy}).

\paragraph{Single-Metric Accuracy Baselines.}
To quantify the added value of multi-metric integration, we tested five na\"ive
baselines that each select the pose optimising a single tool
(highest IQ, highest BM/B, lowest CS, lowest SC, or lowest UP).
No single-metric strategy exceeds 30\% accuracy:
IQ-only, BM/B-only, and SC-only each achieve 30\% (3/10),
UP-only achieves 20\% (2/10), and CS-only achieves only 10\% (1/10).
A majority-vote baseline, selecting whichever pose wins the most individual
metrics, also reaches only 30\% (3/10), because pose-level metric rankings
often conflict.
The agent's 50\% accuracy therefore represents a meaningful improvement
over any single-metric heuristic, consistent with the interpretation that
the LLM reasoning layer integrates complementary signals from
heterogeneous tools.

\begin{figure}[htbp]
    \centering
    \includegraphics[width=0.70\textwidth]{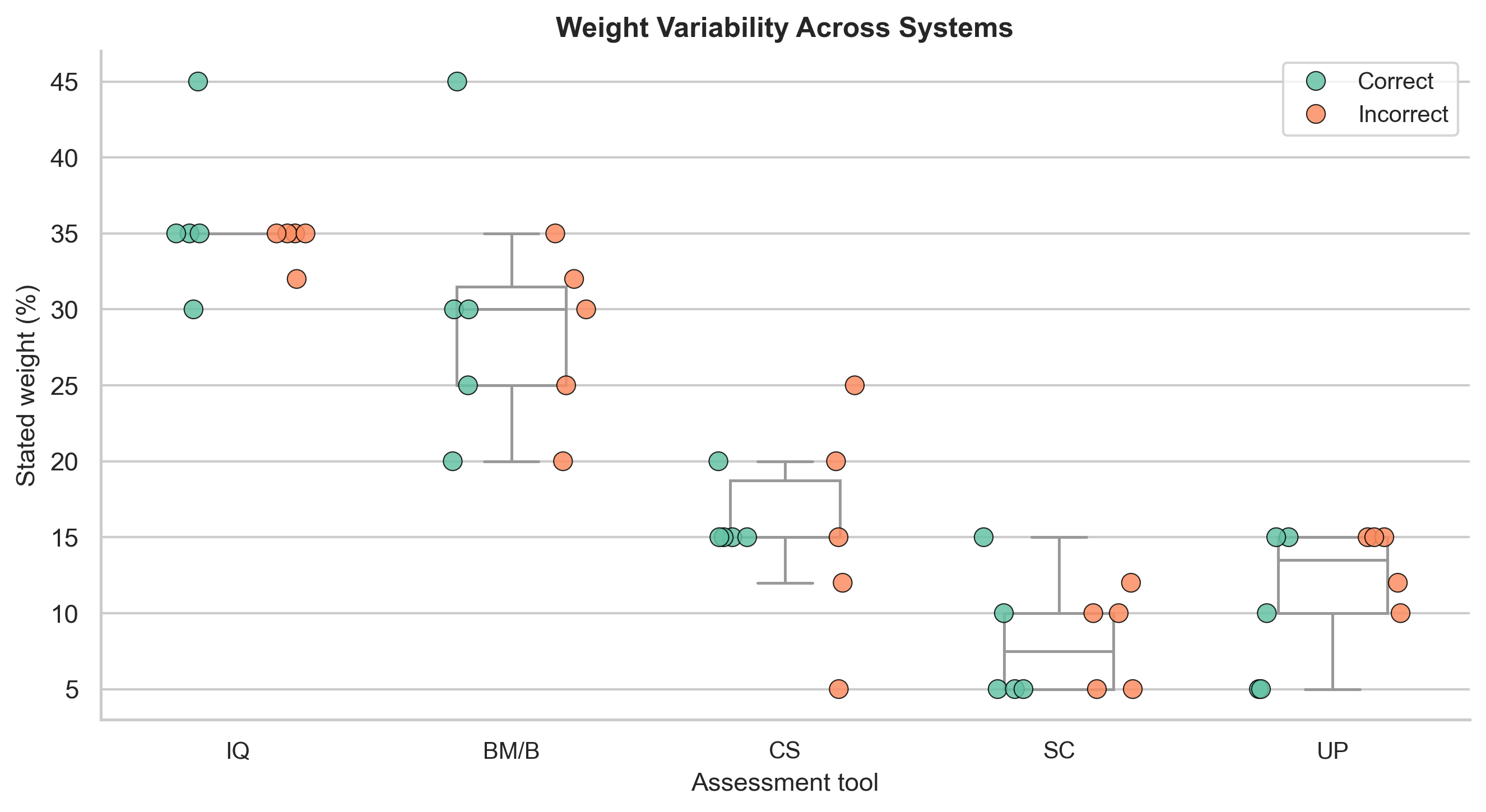}
    \caption{
        Distribution of self-reported tool weights across ten systems.
        Box plots show the median and interquartile range; individual points
        are coloured by selection outcome (green = correct, orange = incorrect).
        IQ receives the most consistent weighting; penalty tools show
        higher relative variability.
    }
    \label{fig:weight_variability}
\end{figure}

\paragraph{Weight Variability Across Systems.}
Figure~\ref{fig:weight_variability} visualises the distribution of
self-reported weights across all ten systems for each tool category.
Interaction Quality is the most stable weight (mean $35.2 \pm 3.9$\%,
CV~=~10.9\%), reflecting its consistent primacy in the agent's reasoning.
Binding-Mode/Burial is the next most stable (mean $29.2 \pm 7.4$\%,
CV~=~25.5\%) but shows the widest absolute range (20--45\%),
driven by 2P16 where the agent up-weighted burial to 45\% due to the
ligand's shallow-pocket binding landscape.
The three penalty-oriented tools—Strain, Clashes, and Polar—exhibit
higher relative variability (CV~=~33.8--44.8\%), consistent with
their context-dependent role as tiebreakers rather than primary drivers.
Notably, the distribution of weights does not differ systematically
between correct and incorrect systems within any tool category.

\paragraph{Ensemble Difficulty and Performance.}
To probe whether ensemble-level properties track selection difficulty,
we computed two complexity proxies for each system: ensemble size
($n_{\mathrm{poses}}$) and the coefficient of variation (CV) of
Interaction Quality scores across all poses.
Both proxies covary negatively with correctness in the descriptive sense.
Correct systems have a mean ensemble size of 12.4 poses versus 20.0 for
incorrect systems (point-biserial $r = -0.668$), and a mean
TQS CV of 20.1\% versus 35.7\% ($r = -0.739$).
The nominal $p$-values for these correlations ($p = 0.035$ and
$p = 0.015$, uncorrected) are reported descriptively and
should be interpreted as exploratory: with $n = 10$ systems and
multiple candidate descriptors examined, no correction for multiple
comparisons is applied, and neither correlation supports an
inferential claim.
Larger ensembles mechanically increase the search space, while higher
TQS variability may reflect more degenerate binding landscapes where
multiple poses present plausible interaction profiles.
These tentative associations motivate follow-up evaluation at larger
scale, where ensemble size and metric heterogeneity could be assessed
as difficulty indicators for confidence-aware selection strategies.

\paragraph{Chain-of-Thought Depth.}
We quantified the depth of the agent's reasoning by measuring the word count of
the Chain-of-Thought section preceding each final report.
The mean Chain-of-Thought comprised $873 \pm 101$ words across the ten systems
(range: 699--1027).
Correct systems tended to elicit slightly longer chains ($923 \pm 91$ words)
than incorrect systems ($824 \pm 92$ words), though this difference
did not reach significance (Mann-Whitney $U = 21.0$, $p = 0.095$).
More strikingly, when normalised by ensemble size, correct systems received
$99.1 \pm 65.2$ words per pose versus $41.2 \pm 4.6$ for incorrect systems,
suggesting that the agent allocates disproportionately more reasoning
effort per pose in smaller, ultimately easier ensembles.
Neither raw word count nor confidence level correlated significantly
with Chain-of-Thought length (Spearman $\rho = -0.494$, $p = 0.147$ for
$n_{\mathrm{poses}}$ and $\rho = -0.156$, $p = 0.668$ for confidence),
indicating that reasoning verbosity is relatively stable regardless
of the ensemble complexity or the agent's stated certainty.

\section{Discussion}

\subsection{The Value of Multi-Metric Decomposition over Composite Scoring}
\label{sec:disc-multi-metric}

\paragraph{Single-metric insufficiency.}
The single-metric accuracy baselines reported in
Section~\ref{sec:supplementary_analyses} reveal that no individual
physicochemical criterion is sufficient for near-native pose identification.
Selecting the pose that optimises a single tool (highest interaction quality,
deepest burial, lowest strain, fewest clashes, or fewest unsatisfied polar
atoms) never exceeds 30\% accuracy, and conformational strain alone achieves
only 10\%. A majority-vote heuristic that selects whichever pose wins the
plurality of individual metrics also caps at 30\%, because pose-level rankings
across tools frequently conflict: a deeply buried pose may harbour steric
clashes, and the candidate with the richest hydrogen-bond network may adopt an
energetically strained conformation. The agent's 50\% accuracy
(Section~\ref{sec:results-best-pose-accuracy}) therefore cannot be replicated
by any single-metric shortcut or their na\"ive combination; it requires the
kind of cross-metric integration that explicit reasoning over decomposed
observables enables. This finding echoes a longstanding theme in
scoring-function research~\citep{warren2006critical, su2019comparative}: the
physics of protein--ligand binding is intrinsically multi-dimensional, and
collapsing competing contributions (van der Waals contacts, electrostatic
complementarity, desolvation penalties, conformational strain) into a single
composite score inevitably discards the trade-off information needed to
adjudicate between closely ranked poses. Where traditional scoring functions
address this dimensionality reduction through fixed-weight summation or
consensus protocols~\citep{charifson1999consensus, wang2003comparative}, the
agentic approach preserves each observable as a separate input to an explicit
reasoning process, allowing the relative importance of each criterion to vary
with the physicochemical context of the ensemble under evaluation.

\paragraph{Context-dependent tool deployment.}
The representative system analyses
(Section~\ref{sec:results-representative}) illustrate how the agent
reconfigures its evaluation strategy in response to each ensemble's
distributional properties. In~2P16, where 9 of 20 poses were classified as
surface binders, the agent allocated 45\% of its composite weight to pocket
engagement (25\% binding mode, 20\% burial), effectively deploying binding
mode as a hard gate that eliminated nearly half the candidate pool before
interaction quality was assessed. In~3HS4, by contrast, 17 of 20 poses
occupied the deep-pocket category, leaving the binding mode classifier with
negligible discriminatory power; the agent responded by shifting emphasis to
interaction quality and elevating clashes and polar penalties above strain
(both 12\%), which better separated the remaining candidates. In~3OXC, all
top-ranked poses were deeply buried and presented similar interaction
profiles (TQS 16.81 vs.\ 17.56), so the agent promoted conformational strain
to a decisive tiebreaker, judging a 20~kcal/mol strain differential too
large to overlook despite a marginal quality-score disadvantage. A
fixed-weight rescoring function, by definition, applies the same coefficient
vector to every ensemble regardless of its composition; the agentic approach
adapts the weighting hierarchy to the ensemble at hand (gate, discriminator,
or tiebreaker) according to which criterion carries the most information in
each context. The weight variability analysis
(Figure~\ref{fig:weight_variability}) corroborates this pattern
quantitatively: interaction quality is the most stable weight across the ten
systems (mean $35.2 \pm 3.9$\%, CV~=~10.9\%), reflecting its consistent
primacy, whereas the three penalty-oriented tools (strain, clashes, and
unsatisfied polar atoms) exhibit substantially higher variability
(CV~=~33.8--44.8\%), consistent with their context-dependent role as
tiebreakers rather than primary drivers. The variability itself is therefore
a feature of the architecture, not noise: it reflects principled adaptation
of the decision hierarchy to the physicochemical landscape of each pose
ensemble.

\paragraph{Conformational strain as both the strongest discriminator and the principal failure bottleneck.}
The metric separation analysis
(Figure~\ref{fig:metric-separation-heatmap}) identifies conformational strain as
the most outcome-sensitive tool in the benchmark, but its influence is
double-edged. Correctly identified systems exhibit a mean strain separation
of $\delta = +0.87$, the selected pose lies nearly one standard deviation
below the ensemble mean in strain energy, whereas incorrectly identified
systems average only $\delta = +0.13$. When genuine strain differentiation
exists, as in~3OXC, where a 20~kcal/mol differential was large enough to
override a marginal interaction-quality disadvantage, strain drives the
agent toward the correct answer. When it does not, strain becomes the
principal failure bottleneck: in~4MLT the near-uniform strain landscape
across all 20 poses ($18.78$--$18.81$~kcal~mol$^{-1}$) inflated the
z-score of a negligible absolute difference into a markedly unfavourable
separation ($\delta = -2.18$), while in~2HA6 the strain dimension provided
essentially no discriminative signal ($\delta = +0.09$). The Decision
Attribution Analysis (Section~\ref{sec:results-decision-attribution})
reinforces this duality: the agent's failure mode is not a misallocation of
attention across tool categories, but arises specifically when the strain
evidence fails to separate the correct pose from high-scoring
decoys. Conformational strain is therefore both the strongest
single predictor of success and the dimension whose absence of signal most
reliably predicts failure. This pattern is consistent with the broader
literature on ligand strain: crystallographic surveys indicate that most
protein-bound ligands tolerate only moderate conformational penalties
(\textless 5~kcal/mol for $\sim$60\% of complexes)~\citep{perola2004conformational},
and strain-based filtering has been shown to improve enrichment in
large-library docking campaigns~\citep{gu2021strain}.
A caveat is warranted: absolute MMFF94 strain energies across our benchmark
systems are systematically elevated (44--97~kcal/mol), suggesting force-field
artefacts that inflate the raw values
(Section~\ref{sec:disc-limitations}). The agent identified this pattern in
every representative system and adopted a relative rather than absolute
interpretation of strain, a methodologically sound response that a rigid
scoring function, operating on the same inflated energies, could not execute.

\paragraph{The 3OXC recovery: anatomy of a successful scoring-function override.}
The 3OXC system, the sole Category~C case in the partition-stratified
analysis (Table~\ref{tab:partition-category}), provides the clearest
illustration of how decomposed multi-metric evaluation can recover a
near-native pose that a composite scoring function misses. Smina ranked the
near-native conformation (pose\_02, RMSD~$= 0.63$~\AA{}) second by affinity,
yet the agent correctly identified it from a pool of 19~candidates
(Section~\ref{sec:results-representative}). The decision turned on two
reasoning steps that a single composite score cannot express. First, having
established that the principal competitor, pose\_06, held marginal advantages
in interaction quality (TQS~17.56 vs~16.81) and burial (83.6\% vs~80.0\%),
the agent explicitly weighed these gains against a
20~kcal~mol$^{-1}$ conformational-strain penalty
(46.06 vs~66.42~kcal~mol$^{-1}$) and concluded that ``the strain
disadvantage is too large to ignore''
(Table~\ref{tab:case-3oxc}). Second, the agent discounted salt-bridge
contacts reported at distances exceeding 5~\AA{} as geometrically
implausible, a critical evaluation of its own tool outputs that prevented an
inflated interaction score from biasing the ranking. Together, these
judgements illustrate the core mechanism the framework was designed to enable:
individual physical contributions remain visible and separately evaluable, so
that a dominant penalty in one dimension can override marginal advantages in
others. This recovery must, however, be interpreted with appropriate
restraint: 3OXC is the only successful override among five
scoring-function failures (Acc$^{-} = 20.0\%$, 1/5;
Section~\ref{sec:results-partition-stratified}), with the remaining four
Smina-failure systems falling into Category~D. Although a 20\% recovery rate
substantially exceeds the random baseline of 5.1\%, it falls far short of
what routine deployment would require; the result therefore constitutes a
proof of principle, decomposed evaluation \emph{can} recover poses missed by
composite ranking, rather than a claim of systematic superiority.

\subsection{Diagnosing the Failure Mode: Tool-Suite Coverage, Not Reasoning Quality}
\label{sec:disc-failure-mode}

\paragraph{Ground-truth poses in failure systems are metric-indistinguishable from their ensembles.}
The ground-truth pose metric profile analysis
(Section~\ref{sec:supplementary_analyses}) provides the sharpest diagnostic
for understanding why the agent fails. In the five correctly identified
systems, the ground-truth conformation exhibits consistently favourable
z-scores across the three primary observables: $+1.21$ (interaction
quality), $+0.99$ (binding mode/burial), and $+0.87$ (conformational
strain), the native-like geometry genuinely stands out from its ensemble on
every dimension the tool suite evaluates. In the five incorrect systems, by
contrast, the ground-truth pose is metric-indistinguishable from the
ensemble mean (IQ $z = -0.17$, BM/B $z = -0.25$, CS $z = +0.18$), offering
no physics-based signal that any weighting of the current tools could
exploit. The 3HS4 regression, the sole Category~B case
(Table~\ref{tab:partition-category}), illustrates this ceiling most
starkly: the correct pose (pose\_01, RMSD~$= 0.96$~\AA{}) scored a TQS of
only 3.23, exhibited 75.0\% burial, and carried two steric clashes, whereas
the agent's selection (pose\_02) achieved a TQS of 6.52, 93.8\% burial, and
zero clashes (Table~\ref{tab:case-3hs4}). Under any reasonable weighting of
these observables, pose\_02 would be preferred; the ground-truth
conformation did not even appear in the agent's top-seven ranking, because
its modest interaction profile placed it below the attention threshold.
This pattern establishes a fundamental performance ceiling: the framework
cannot outperform the collective discriminative power of its tool suite.
When the native-like pose offers no metric advantage over competing
conformations, no reweighting of the current observables can recover it;
the bottleneck lies upstream of the reasoning layer, in the evidence on
which that reasoning acts.

\paragraph{Alignment between stated weights and observed metric separations.}
The faithfulness analysis reinforces this diagnosis. The median per-system
Spearman rank correlation between the agent's self-reported weight vectors
and the objective metric separations of the selected pose is $\rho = +0.83$
(Table~\ref{tab:faithfulness}), indicating substantial concordance between
stated reasoning priorities and the quantitative evidence that accompanies
them. An interpretive caveat is warranted: this correlation measures
alignment between what the agent says it weighted and what the selected
pose actually exhibits; it does not, on its own, distinguish faithful
introspection from internally coherent post-hoc rationalisation, a
failure mode documented for frontier models even in the presence of
plausible chain-of-thought
traces~\citep{turpin2023language, lanham2023measuring}. Descriptive
comparison between the outcome groups is consistent with
symmetry: the median $\rho$ is $+0.80$ for correctly identified systems
and $+0.87$ for incorrectly identified ones (Mann--Whitney $U = 13.0$,
$p = 1.00$; reported descriptively, since with $n_{1} = n_{2} = 5$ the
test has minimal power to support either rejection or equivalence).
The outcome-stratified weight analysis
(Table~\ref{tab:outcome-weights}) corroborates this symmetry: the mean
weight allocated to each tool category differs by at most
3.4~percentage points between the two groups, a margin well within the
descriptive variability expected for partitions of five systems. The
agent does not visibly reallocate attention across tool categories when
it fails; it applies the same evaluation framework, with comparable
emphasis on each physicochemical dimension, regardless of outcome. This combination (faithful reasoning,
consistent attention allocation, yet divergent outcomes) localises the
failure mode upstream of the reasoning layer, in the discriminative power of
the input metrics themselves. The finding contrasts with the broader
chain-of-thought faithfulness literature, where LLMs have been shown to
produce plausible but fabricated rationalisations for bias-driven
decisions~\citep{turpin2023language} and to exhibit \emph{decreasing}
faithfulness with increasing model
capability~\citep{lanham2023measuring}. In the present application, where
the evidence consists of structured numerical outputs presented under
explicit domain-specific interpretation guidelines, reasoning concordance is
substantially higher, suggesting that the structured, tool-mediated
architecture may itself act as a faithfulness safeguard, a design property
with implications for AI-assisted scientific workflows beyond molecular
docking.

\paragraph{Implications for system improvement.}
The preceding diagnosis directs improvement efforts to three concrete,
evidence-grounded targets. First, the tool suite requires expansion:
the 3HS4 failure arose because the ground-truth binding mode is preferred
for reasons the current observables capture only partially.
Smina's scoring-term decomposition
(Table~\ref{tab:smina-terms-3hs4}) pinpoints the gap: the sole term
favouring pose\_01 is non-directional hydrogen-bond proximity, a
continuous polar complementarity measure that the agent's tools lack.
PLIP's binary contact detection and the polar penalty's deficit-only
counting cannot represent this dimension, and its
0.87~kcal/mol contribution was overwhelmed by the agent's
interaction-plus-burial weighting
(Section~\ref{sec:results-representative}). Incorporating physics-based
rescoring methods that model solvation explicitly, such as
MM-GBSA~\citep{greenidge2014mmgbsa} or free-energy
perturbation, would introduce physical dimensions capable of separating
poses that the present five tools treat as equivalent. Second,
conformational strain, the strongest individual diagnostic for decision
correctness (Section~\ref{sec:disc-multi-metric}), is degraded by MMFF94
force-field artefacts that inflate absolute energies across all benchmark
systems; adopting a higher-accuracy force field or performing explicit
minimisation of the bound-state complex would sharpen the most informative
discriminator the framework currently possesses
(Section~\ref{sec:disc-limitations}). Third, the 4JFL analysis identified three converging factors behind
the failure: a burial ratio inflated by the binary SASA threshold
(Section~\ref{sec:sasa}), unmodelled water-mediated hydrogen bonds
that artificially penalise the correct pose
(Table~\ref{tab:case-4jfl}), and a weighting heuristic that
undervalues H-bond geometry quality when no pose achieves deep-pocket
classification. The first two factors require tool-suite
improvements (a continuous burial metric and water-bridge-aware polar
penalties); the third is directly addressable through system-prompt
refinement. Importantly, the
ability to trace each incorrect decision to a specific, actionable
deficiency (in the tool suite, the force field, or the weighting
heuristics) is itself a product of the decomposed, transparent
architecture. A conventional scoring function that achieved the same
overall accuracy would not reveal whether the bottleneck lies in the
feature representation, the functional form, or the training
data~\citep{warren2006critical}, and would therefore offer no roadmap for
targeted improvement; diagnosability is thus a practical advantage of the
framework independent of its numerical performance.

\subsection{Quality and Faithfulness of LLM Scientific Reasoning}
\label{sec:disc-llm-reasoning}

\paragraph{Adaptive weighting and thermodynamic commensuration.}
The adaptive weighting documented in
Section~\ref{sec:disc-multi-metric}, where the agent reconfigured its
evaluation hierarchy from system to system, also bears on the quality of
the underlying reasoning process. Beyond adjusting relative priorities,
the representative system analyses
(Section~\ref{sec:results-representative}) reveal two behaviours that
transcend simple metric comparison: approximate thermodynamic
commensuration and explicit hypothesis revision. In the 3HS4 analysis,
the agent estimated a cost of approximately 2--3~kcal/mol per buried
unsatisfied polar atom and weighed this against a competing candidate's
interaction and burial superiority, concluding that the estimated
4--6~kcal/mol polar desolvation disadvantage was outweighed by gains on
the primary axes (Table~\ref{tab:case-3hs4}). Placing heterogeneous
observables on an approximate common energy scale is a form of reasoning
typically performed only by expert computational chemists during manual
pose inspection~\citep{bissantz2010guide}. The system prompt provides
energy-scale heuristics, for example, an approximate 2--3~kcal/mol
desolvation cost per buried unsatisfied polar atom and an ``interaction
compensation principle'' allowing moderate strain when offset by strong
binding, but the agent's specific application, chaining these prompted
values into a connected multi-step trade-off argument and arriving at a
quantitative estimate of the net disadvantage, goes beyond what the
instructions literally prescribe. The agent
simultaneously maintained epistemic caution, noting that ``it's not
helpful to equate TQS values directly to energy'', a self-imposed
boundary between geometric surrogates and thermodynamic quantities. In
the 3OXC recovery, a complementary pattern appeared: the agent explicitly
quantified the inter-pose delta across multiple axes
($\Delta$~quality~${\sim}$+3.24; +4~total interactions) and used these
simultaneous comparisons to construct a transparent justification for
overriding the scoring-function ranking
(Table~\ref{tab:case-3oxc}). The 2P16 analysis further demonstrates
hypothesis revision: the reasoning trace records an intermediate stage in
which the agent favoured a different candidate before systematically
applying the binding-mode filter and revising its selection
(Table~\ref{tab:case-2p16}), iterative deliberation rather than a single
scoring pass. Together, these behaviours (commensuration, cross-axis
quantification, and hypothesis revision) suggest that LLMs, when
provided with structured numerical evidence and domain-specific
interpretation guidelines, can perform a form of multi-criterion
scientific reasoning that goes beyond pattern matching. A caveat is
warranted: the system prompt
(Section~\ref{sec:agent-architecture}) encodes much of the evaluation
framework, and the extent to which the observed behaviour reflects genuine
understanding versus pattern-following of prompt instructions cannot be
resolved from these results alone~\citep{wang2023towards}.

\paragraph{Self-critical reasoning and tool-output anomaly detection.}
A complementary facet of the reasoning quality emerges from the agent's
treatment of its own tool limitations. The individual flagging episodes
documented in the preceding subsections (MMFF94 force-field attribution
across all four representative systems
(Section~\ref{sec:disc-multi-metric}), salt-bridge geometry filtering in
3OXC, and identification of unmodelled water networks in 3HS4
(Section~\ref{sec:disc-failure-mode})) are instances of a systematic
self-critical pattern confirmed in the cross-system summary
(Section~\ref{sec:results-representative}). Each episode follows a common
structure: the agent detects a systematic anomaly in a tool's output,
attributes it to a specific methodological cause, recommends an adjusted
interpretation strategy, and proposes a concrete follow-up, mutagenesis
and footprinting in 2P16, explicit hydration and pKa refinement in 3HS4,
re-docking with alternative pocket definitions in 4JFL. In the last of
these cases, the self-critical reasoning extended beyond force-field
artefacts to the upstream docking protocol itself: the agent proposed that
the absence of any deeply buried conformation might reflect sampling bias
or pocket mis-definition rather than the ligand's true binding mode
(Table~\ref{tab:case-4jfl}), a hypothesis that shifts responsibility
from the ranking step to the pose-generation step. Collectively, these
behaviours transform each output from a bare ranking into a structured
diagnostic report that delivers not only a pose selection but also the
reasoning behind it, the conditions under which it may be unreliable, and
a roadmap for strengthening the evidence base. This diagnostic richness
is preserved regardless of whether the final selection is correct: the
3HS4 and 4JFL traces, despite producing incorrect rankings, correctly
identified the factors most likely to undermine the
result (unmodelled solvation and sampling limitations,
respectively) providing actionable guidance for targeted
improvement. Whether this reasoning-level
self-awareness translates into empirically calibrated confidence labels is
examined in the following paragraph.

\paragraph{Confidence calibration: a gap between self-assessment and empirical accuracy.}
The confidence calibration analysis
(Section~\ref{sec:supplementary_analyses}) shows that this granular
self-awareness does not aggregate into a reliable summary signal at
the present sample size. Of the four systems to which the agent
assigned high confidence, only two were correctly identified (50\%
accuracy); medium-confidence systems achieved 75\% (3/4 correct),
while both medium-high systems were incorrect
(Figure~\ref{fig:confidence_calibration}). Calibration cannot be
assessed inferentially at $n = 10$: a Fisher exact test of the
high-versus-non-high confidence table returns OR~=~1.00,
$p$~=~1.000, but with only ten systems and a $2 \times 2$ table the
test has essentially no power to support either rejection or
equivalence. Descriptively, the high-confidence subset achieved
50\% accuracy and the medium-confidence subset 75\%, an ordering
opposite to what empirical calibration would predict, and warrants
larger-scale evaluation rather than a definitive judgement on
calibration from the present benchmark. The observation is
qualitatively consistent with the broader literature on miscalibrated
verbalised confidence in LLMs~\citep{xiong2024llm_uncertainty}. The 3HS4 regression sharpens
this paradox: the agent assigned high confidence to its worst failure
($\Delta_{\mathrm{RMSD}} = 4.32$~\AA{}, the largest deviation in the
benchmark; Table~\ref{tab:rmsd-selected-poses}), precisely because the
selected pose dominated on every metric the tool suite provides
(Section~\ref{sec:disc-failure-mode}). Simultaneously, the reasoning
trace explicitly flagged force-field artefacts and unmodelled water
networks as potential confounders, the very factors that explain the
failure. The confidence label was thus well-calibrated to the metric
evidence but not to the outcome, because the metric evidence was itself
misleading. This dissociation between reasoning-level uncertainty,
which identifies specific epistemic gaps, and the summary confidence
label, which reflects only the agent's assessment of its relative ranking
given available metrics, reveals a structural limitation: the agent
possesses no mechanism to translate articulated caveats into a
quantitative discount on its overall confidence. For practical
deployment, users cannot rely on the stated confidence level as a
reliability indicator; an external calibration mechanism (such as
ensemble-of-runs variance, metric-separation thresholds computed before
the reasoning step, or a dedicated confidence-prediction model) would be
needed to provide actionable confidence estimates, a limitation examined
further in Section~\ref{sec:disc-limitations}.

\paragraph{Limitations of the LLM reasoning layer.}
The reasoning capabilities documented in the preceding paragraphs must be
weighed against four inherent limitations of the LLM decision layer.
First, the agent's behaviour is governed by a ${\sim}7{,}800$-word system
prompt (Section~\ref{sec:agent-architecture}) that was iteratively
refined during development; a dedicated anti-bias section, for instance,
was added after early testing revealed a ``more interactions $=$ better''
failure mode. This coupling between prompt phrasing and decision
behaviour means that small modifications, such as the threshold at which
conformational strain overrides interaction quality, could alter rankings
in ways that have not been systematically characterised, because no
formal ablation varying prompt components was conducted. Second, although
the tool-mediated architecture grounds reasoning in pre-computed,
deterministic outputs rather than parametric memory
(Section~\ref{sec:bg-agentic}), the hallucination risk is reduced but
not eliminated: the model could in principle fabricate metric values or
misattribute properties across poses. Architectural safeguards (mandatory
identity verification, structured output sections, and
character-for-character code copying
(Section~\ref{sec:agent-architecture})) mitigate this risk, but no
automated post-hoc mechanism currently confirms that every number in the
reasoning trace matches the source analysis files. Third, all results
were obtained with a single model (GPT-5, OpenAI Agents SDK
version~0.5.0); performance may differ across providers or model
versions, and silent weight updates by the provider could alter reasoning
characteristics without warning, a reproducibility challenge common to
API-dependent scientific workflows. Fourth, LLM inference is inherently
stochastic, yet the evaluation employed a single run per system;
inter-run variance in pose selection has not been quantified. These
constraints are not unique to the present application but are shared by
all LLM-based scientific
agents~\citep{bran2024chemcrow, boiko2023coscientist}; they underscore
the importance of treating the agent's output as a structured
recommendation for expert audit rather than as a definitive answer. Each
limitation is examined in detail, together with partial mitigations
afforded by the modular architecture, in
Section~\ref{sec:disc-limitations}.

\subsection{Pose Anonymisation as a Design Choice Motivated by the Reasoning-Bias Literature}
\label{sec:disc-anonymisation}

\paragraph{Design rationale.}
The positional and label-based biases that systematically distort LLM
ranking behaviour (Section~\ref{sec:bg-agentic}) are especially
consequential in molecular docking, where pose identifiers assigned by
the docking engine conventionally encode the scoring-function ranking:
\texttt{pose\_01} is the top-ranked conformation, so exposing these
filenames to the model would conflate positional and content-based
signals with the independent physicochemical evaluation. The
deterministic SHA-256 anonymisation protocol described in
Section~\ref{sec:anonymisation} is a design-time intervention
introduced to pre-empt this risk, motivated by the reasoning-bias
literature cited above~\citep{liu2024lost, zheng2024large,
tang2024found}. Three properties guided the design.
Determinism ensures that the same system-pose pair always yields the
same eight-character code, enabling reproducible audits across
independent runs. SHA-256's avalanche property guarantees that
sequentially numbered poses produce identifiers with no discernible
ordinal pattern, so no ordinal cue from the docking engine's naming
convention reaches the model. The eight-character code length was
empirically calibrated: full 36-character UUIDs increased the rate of
transcription errors in early testing, whereas shorter codes preserved
the character-for-character accuracy required for automatic
de-anonymisation
(Section~\ref{sec:anonymisation}). We do not claim an established
methodological contribution. The protocol was motivated by qualitative
observations during development (in pre-anonymisation runs, the model
was observed referencing pose names as factors in its reasoning) but
its magnitude of effect on ranking accuracy, faithfulness, or
attribution has not been quantified, because no ablation comparing
anonymised and non-anonymised runs was performed on the benchmark; we
therefore present the protocol as a design choice whose effect size
remains to be measured (Section~\ref{sec:disc-future-directions}).
Moreover, while the SHA-256 hashing eliminates label-based cues, the
current implementation presents analysis files in a fixed sorted order
without shuffling, so positional bias remains an uncontrolled confound.
The underlying principle, decoupling item identity from any
quality-correlated naming signal before an LLM performs comparative
evaluation, is straightforward to transpose to other LLM-based
comparative-evaluation tasks (for example, scientific peer review or
compound triaging), but we defer any such generalisation claim until
the within-task effect size has been measured on this benchmark.

\subsection{Comparison to Related Work}
\label{sec:disc-related-work}

\paragraph{Classical rescoring and consensus approaches.}
The three scoring paradigms reviewed in
Section~\ref{sec:bg-scoring} (empirical or physics-based rescoring,
knowledge-based potentials, and consensus scoring) differ from the
agentic framework in scope, computational cost, and output format.
Physics-based rescoring with MM-GBSA or
MM-PBSA~\citep{greenidge2014mmgbsa} provides the most rigorous
individual comparison, incorporating explicit solvation modelling and
energy minimisation that the present tool suite does not replicate;
however, the result remains a single composite free-energy estimate, and
the additional computation, hours rather than minutes per system
(Supplementary Section~\ref{sec:si-scoring-benchmarks}), limits routine application to entire
docking ensembles. Knowledge-based potentials such as DrugScore and
PMF~\citep{gohlke2000knowledge, muegge1999general} achieve speed
comparable to the agentic pipeline but collapse all atom-pair preferences
into an opaque aggregate~\citep{su2019comparative}, offering no mechanism
to reveal which physicochemical dimension drove the ranking. Classical
consensus methods~\citep{charifson1999consensus, wang2003comparative,
yang2005consensus} partially address the single-function limitation by
fusing rankings from multiple independent scorers, raising
pose-prediction success rates from 66--76\,\% individually to
approximately 80\,\% or higher (Section~\ref{sec:bg-scoring}), yet
their aggregation rules are fixed at design time and applied uniformly
across all targets. As documented in
Section~\ref{sec:disc-multi-metric}, the agentic pipeline varies the
relative emphasis on each physicochemical axis from system to system
(CV~ranging from 10.9\,\% for interaction quality to 33.8--44.8\,\% for
penalty tools; Figure~\ref{fig:weight_variability}), a context-adaptive
behaviour that fixed-rule consensus cannot reproduce. The practical
consequence extends beyond numerical accuracy: for every ranking
decision, the agent produces a structured rationale identifying the
metrics that favoured or penalised each candidate
(Section~\ref{sec:disc-llm-reasoning}), enabling domain-expert audit and
iterative refinement, properties that
Section~\ref{sec:bg-scoring} characterised as absent from all existing
scoring paradigms. The comparison is not uniformly in the agent's
favour: MM-GBSA captures solvation thermodynamics that the current tool
suite omits (Section~\ref{sec:disc-failure-mode}), and consensus methods
achieve their reported accuracy gains on benchmarks of hundreds to
thousands of complexes, a scale at which the present evaluation
($m = 10$) cannot provide a statistically powered comparison.

\paragraph{ML-based scoring and pose prediction.}
Machine-learning scoring functions and generative docking models represent
the principal accuracy frontier against which the agentic approach must
be positioned. As reviewed in Section~\ref{sec:bg-scoring}, Gnina's CNN
ensemble achieves 73\,\% top-1 redocking success on CrossDocked2020,
compared with 58\,\% for AutoDock
Vina~\citep{mcnutt2021gnina, francoeur2020crossdock}, and DiffDock
reports 38\,\% top-1 success in a blind-docking setting where
conventional samplers reach
23\,\%~\citep{corso2023diffdock}. The agent's 50\,\% accuracy on ten
systems (95\,\% Wilson CI $[23.7, 76.3]$\,\%;
Table~\ref{tab:accuracy-summary}) cannot be compared to these figures on
equal statistical footing: the benchmark scales differ by two to three
orders of magnitude, the tasks are not identical (re-ranking a
pre-generated ensemble versus redocking or blind docking), and
Gnina's training corpus overlaps with the PDBbind~2016 data from which
the benchmark is drawn, an advantage the training-data-free agentic
pipeline does not share. We therefore do not claim competitive numerical
accuracy; the contribution lies in a qualitatively different output
class (a structured, metric-by-metric rationale accompanying each
ranking decision) rather than in numerical
superiority. The PoseBusters validation
framework~\citep{buttenschoen2024posebusters} offers an instructive
counterpoint: it evaluates the same physicochemical properties that
the agent's tool suite
computes (clashes, strain, interaction geometry;
Sections~\ref{sec:strain}--\ref{sec:clashes}) and demonstrated that physics-based
docking programs outperform deep-learning methods once such validity
checks are enforced. The agent's architecture can be viewed as extending
this principle from binary validity assessment to ordinal ranking: rather
than flagging a pose as physically valid or invalid, the tools
quantify each criterion on a continuous scale and the reasoning layer
integrates these scales into an explicit comparative
judgement (Section~\ref{sec:disc-llm-reasoning}). This
design also carries a practical asymmetry: ML methods require
large curated training sets, are sensitive to domain shift when targets
fall outside the training distribution
(Supplementary Section~\ref{sec:si-scoring-benchmarks}), and must be retrained as new
structural data accumulate; the agentic pipeline requires no training
data and generalises through the LLM's domain knowledge and the
physics-based tool outputs, though at the cost of the LLM-specific
limitations discussed in
Section~\ref{sec:disc-llm-reasoning}.

\paragraph{Agentic AI and LLMs for scientific reasoning.}
The agentic AI landscape reviewed in
Section~\ref{sec:bg-agentic} provides the most direct comparator for the
present framework. ChemCrow and Coscientist demonstrated that
tool-augmented LLM agents can orchestrate complex chemistry
workflows (synthesis planning, property prediction, and robotic
experimentation) but both systems were evaluated through expert judgement
or experimental outcomes on open-ended tasks rather than against an
objective structural ground
truth~\citep{bran2024chemcrow, boiko2023coscientist}.
Section~\ref{sec:bg-agentic} identified the resulting gap: no
tool-augmented chemistry agent had reported quantitative accuracy,
faithfulness, or decision-attribution metrics against a crystallographic
benchmark. The present work addresses this gap by applying the agentic
paradigm to a specific, well-defined evaluation task (ranking
pre-generated docking poses against each other on physicochemical
merit, validated against a crystallographic ground truth
(Section~\ref{sec:results-best-pose-accuracy})) whose binary
correctness criterion enables the full suite of diagnostic analyses that
open-ended tasks preclude. The narrowness of the task is not a
limitation but a design requirement: because ground-truth correctness is
computable for every decision, it becomes possible to measure not only
\emph{whether} the agent selects the correct pose (50\,\%;
Table~\ref{tab:accuracy-summary}) but also \emph{why} it succeeds or
fails, through metric separation
(Figure~\ref{fig:metric-separation-heatmap}), reasoning faithfulness
($\rho = {+}0.83$; Table~\ref{tab:faithfulness}), outcome-stratified
weight attribution (Table~\ref{tab:outcome-weights}), and per-system
failure-mode diagnosis
(Section~\ref{sec:results-representative}). This combination constitutes
a methodological template transferable to other scientific domains in
which agentic AI performs comparative evaluation: wherever an objective
ground truth exists, the same decomposition (accuracy, faithfulness,
attribution, and failure-mode analysis) can characterise not just the
agent's performance but the alignment between its stated reasoning and
the evidence it was given.

\subsection{The Retention-Recovery Asymmetry and Practical Implications}
\label{sec:disc-retention-recovery}

\paragraph{Interpreting the 80\%/20\% accuracy contrast.}
The partition-stratified accuracy analysis
(Section~\ref{sec:results-partition-stratified}) reveals an asymmetric
operational profile. On the five Smina-success systems, the agent
preserved the correct identification in four cases
($\mathrm{Acc}^{+} = 80.0\,\%$, 4/5;
Table~\ref{tab:partition-category}); the sole regression (3HS4,
Category~B) arose because the ground-truth pose was metric-inferior to a
competing conformation on every dimension the tool suite evaluates
(Section~\ref{sec:disc-failure-mode}). On the five Smina-failure
systems, only 3OXC was recovered
($\mathrm{Acc}^{-} = 20.0\,\%$, 1/5), where a
${\sim}$20~kcal/mol conformational-strain differential provided a
physics-based signal strong enough for the reasoning layer to override
the scoring-function ranking
(Section~\ref{sec:results-representative}). The four unrecovered
systems (2HA6, 4JFL, 4MLT, 4ZLS (Category~D)) are precisely those in
which the ground-truth pose offers no metric advantage over its
competitors (Section~\ref{sec:disc-failure-mode}), confirming that the
tool-suite coverage ceiling diagnosed in that section is the proximate
cause of the low recovery rate. The resulting profile (high retention,
low recovery) characterises the framework as a reliable curator of
existing correct rankings rather than a powerful recovery engine for
scoring-function failures, a pattern the results themselves described as
conservative
(Section~\ref{sec:results-partition-stratified}). A sample-size caveat
is essential: each partition contains only five systems, so a single
additional recovery would shift $\mathrm{Acc}^{-}$ from 20\,\% to
40\,\%, and the current values should be read as descriptive point
estimates rather than definitive performance bounds.

\paragraph{Practical implications for structure-based drug design.}
The RMSD distribution of the agent's selections
(Table~\ref{tab:rmsd-selected-poses}) shapes the practical deployment
model. None of the five misidentifications falls in the mild category
($\Delta_{\mathrm{RMSD}} < 1.0$~\AA{}); instead, errors span the
moderate-to-severe range (1.51--4.32~\AA{}, mean 2.92~\AA{}), indicating
that the agent either identifies a near-native conformation or selects one
that departs substantially from the crystallographic reference
(Section~\ref{sec:results-rmsd}). This absence of a ``close miss''
regime argues against treating the top-ranked pose as a final answer and
favours a workflow in which the agent's selection is regarded as a
hypothesis for visual inspection by a domain expert. The top-$k$
relaxed accuracy analysis (Figure~\ref{fig:topk_accuracy}) supports this
recommendation at matched ranks $k \leq 3$: accuracy rises from
50\,\% at $k = 1$ to 60\,\% at $k = 2$ (recovering 4MLT) and
70\,\% at $k = 3$ (recovering 4JFL). The plateau beyond $k = 3$ is a
configuration artefact of the prompt's truncated-emission convention,
not a property of the agent's underlying ranking ability, so this
figure is directly comparable to Smina only at $k \leq 3$
(Section~\ref{sec:supplementary_analyses}). Within that matched range
the three remaining failures (2HA6, 3HS4,
4ZLS) were never ranked favourably by the agent. A shortlist of three
poses (each accompanied by the structured, metric-by-metric rationale
documented in Section~\ref{sec:disc-llm-reasoning}) would therefore
capture the correct conformation in seven of ten benchmark systems while
retaining the structured per-pose rationale for each
candidate. Crucially, even when the agent's top selection is
incorrect, the reasoning trace retains diagnostic value. The 4JFL case
study (Table~\ref{tab:case-4jfl}) illustrates this: the agent
identified limitations in its own metric coverage (unmodelled water
bridges, threshold-sensitive burial ratios) and proposed concrete
remediation, including re-docking with alternative pocket definitions
and explicit water placement
(Section~\ref{sec:results-representative}). Finally, two ensemble-level descriptors covary with outcome in the
present sample: mean ensemble size (12.4~poses for correct vs.\ 20.0
for incorrect systems; point-biserial $r = -0.668$) and TQS
coefficient of variation (20.1\,\% vs.\ 35.7\,\%; $r = -0.739$;
Section~\ref{sec:supplementary_analyses}). The nominal $p$-values
($p = 0.035$ and $p = 0.015$, uncorrected for multiple comparisons)
are reported descriptively: with $n = 10$ and several candidate
descriptors examined, these correlations are exploratory and do not
support an inferential claim on their own. Because both quantities can be
computed from the tool outputs before the reasoning step, they could serve
as pre-screening indicators that flag systems warranting additional manual
review or tool-suite expansion, a prospect examined further in
Section~\ref{sec:disc-future-directions}.

\subsection{Limitations}
\label{sec:disc-limitations}

\paragraph{Benchmark scale and statistical power.}
The most consequential limitation is the benchmark's scale. Ten systems
comprising 162~docked poses
(Section~\ref{sec:results-benchmark-summary}) yield a 95\,\% Wilson
confidence interval of $[23.7, 76.3]$\,\% around the observed 50\,\%
accuracy (Table~\ref{tab:accuracy-summary};
Equation~\ref{eq:wilson}), a 53-percentage-point range that precludes
definitive comparison with any single-function baseline, including the
Smina reference that achieves 50\,\% by construction on the balanced
design (Section~\ref{sec:benchmark-construction}). The
partition-stratified estimates are more fragile still: with $n = 5$
systems per group, a single additional recovery would shift
$\mathrm{Acc}^{-}$ from 20.0\,\% to 40.0\,\%
(Section~\ref{sec:disc-retention-recovery}), and no formal hypothesis
test comparing agent and scoring-function accuracy is feasible at this
sample size. As discussed in Section~\ref{sec:bg-benchmarks}, this
constraint reflects a deliberate design trade-off. The quantitative
evaluation framework (accuracy, faithfulness quantification, metric
decomposition, and automated failure-mode
diagnosis) scales to hundreds of systems, provided the agent emits
structured machine-readable output alongside the reasoning trace. What
the small benchmark uniquely enables is the qualitative case-study
examination of individual reasoning traces with verbatim excerpts
(Section~\ref{sec:results-representative}), a narrative analysis that
would be prohibitively time-consuming at the scale of CASF-2016's
285~complexes~\citep{su2019comparative}. Several partial mitigations temper
the statistical limitation. The ten systems span five protein
families (Table~\ref{tab:selected-systems}), and the balanced
Smina-success/failure partition enables controlled comparison across
outcome categories
(Table~\ref{tab:partition-category}). The exact one-sided binomial
test against the random baseline ($p < 0.001$;
Table~\ref{tab:accuracy-summary}) confirms that the agent's
discriminative ability exceeds chance, even if the magnitude of its
advantage over Smina cannot be resolved at $m = 10$. Nevertheless,
generalisability to the broader PDBbind landscape
(${\sim}$4{,}800~refined-set
complexes~\citep{wang2004pdbbind}) or to clinical drug discovery
targets remains an open question that only a scaled evaluation, such
as an extension to the full CASF-2016 core set, can address
(Section~\ref{sec:disc-future-directions}).

\paragraph{LLM dependency and reproducibility.}
Section~\ref{sec:disc-llm-reasoning} flagged model dependency and
inference stochasticity as inherent constraints of the LLM reasoning
layer; here we examine each in detail. First, all results were obtained
through the OpenAI API
(Section~\ref{sec:system-overview}), which provides no mechanism to pin
a specific model checkpoint. Unlike open-weight models whose parameters
can be archived alongside the analysis code, API-accessed models may be
updated silently by the provider, so an identical pipeline invocation at
a later date could yield different rankings without any change to the
system prompt, tools, or input data, a version-opacity risk shared by
all API-dependent scientific
workflows~\citep{bran2024chemcrow, boiko2023coscientist}. Second, only
GPT-5 was tested
(Section~\ref{sec:agent-architecture}); whether the same system prompt
and tool outputs would elicit comparable reasoning quality from
alternative architectures (Claude, Gemini, or open-source models such
as LLaMA) remains unknown. Because the system prompt encodes
domain-specific evaluation heuristics in natural language rather than in
model-specific tokens, it is in principle architecture-agnostic, but
empirical confirmation across providers has not been attempted. Third,
LLM inference is inherently non-deterministic; the evaluation employed a
single run per system, so each ranking represents one draw from an
uncharacterised distribution. Inter-run variance (in pose selection,
stated confidence, and weight attribution) has not been quantified, and
it is therefore unknown whether the observed 50\,\% accuracy
(Table~\ref{tab:accuracy-summary}) reflects a stable operating point or
a fortunate sample. A structural mitigation partially bounds the
reproducibility risk: the analysis tools that produce the numerical
inputs to the reasoning step are fully deterministic
(Section~\ref{sec:system-overview}), so stochasticity is confined to the
LLM layer and repeated runs would operate on identical metric values.
Furthermore, the modular architecture separates the LLM backend from the
tool suite, system prompt, and evaluation protocol, enabling alternative
models to be substituted without modifying the analysis
pipeline, a flexibility leveraged in the multi-LLM evaluation proposed
in Section~\ref{sec:disc-future-directions}.

\paragraph{Computational cost and latency.}
The cost structure of the agentic pipeline differs qualitatively from
that of conventional scoring approaches. Empirical scoring functions
such as Smina evaluate a complete ensemble in seconds per
ligand~\citep{koes2013smina}, and ML-based rescoring with Gnina adds
negligible inference time once the CNN has been
trained~\citep{mcnutt2021gnina}; neither incurs per-invocation monetary
expenditure. Physics-based rescoring with MM-GBSA increases processing
time from seconds to hours per system
(Supplementary Section~\ref{sec:si-scoring-benchmarks})~\citep{greenidge2014mmgbsa}, yet still
requires only local computation. The agentic pipeline introduces two
additive cost components. First, the six primary and three derived
analysis tools (Section~\ref{sec:system-overview}) execute sequentially
for every pose; for the benchmark ensembles of 5--20~conformations
(Table~\ref{tab:selected-systems}), this computational phase completes
in minutes per system on a standard workstation but scales linearly with
ensemble size. Second, the LLM reasoning call, which concatenates all
anonymised analysis files into a single context together with the
${\sim}7{,}800$-word system prompt and processes them at reasoning
effort \texttt{"high"}
(Section~\ref{sec:agent-architecture}), adds both per-token API cost
and wall-clock latency that grow with the number of poses and the depth
of the resulting chain-of-thought ($873 \pm 101$~words across the ten
benchmark systems;
Section~\ref{sec:supplementary_analyses}). The present evaluation did
not record absolute wall-clock times or API expenditures, so the
magnitude of this overhead cannot be quantified precisely; nevertheless,
the per-system cost is orders of magnitude greater than the effectively
zero marginal cost of Smina rescoring. This cost profile renders the
agentic approach impractical for high-throughput virtual screening
campaigns in which millions of compounds must be ranked. The framework
is instead positioned for a late-stage refinement workflow in which a
docking engine has already narrowed the candidate set to a small number
of poses (5--20 per lead compound, the range represented by the
benchmark) and where the interpretability of the ranking decision
(Section~\ref{sec:disc-llm-reasoning}) justifies the additional
expense. A partial mitigation follows from the modular
architecture (Section~\ref{sec:system-overview}): because the analysis
tools are deterministic and independent of the LLM, their outputs can
be cached and re-used across multiple model invocations, confining
repeated expenditure to the reasoning call alone.

\paragraph{Force field and tool-suite limitations.}
The tool-suite coverage ceiling diagnosed in
Section~\ref{sec:disc-failure-mode} has specific technical roots in four
classes of limitation. First, the conformational strain tool employs
the MMFF94 force field~\citep{halgren1996mmff}
(Section~\ref{sec:strain}), whose absolute energies are systematically
elevated across all benchmark systems (44--97~kcal/mol;
Section~\ref{sec:disc-multi-metric}). This discrepancy with the
physically expected range (${\lesssim}\,5$~kcal/mol for the majority of
crystallographic protein--ligand
complexes~\citep{perola2004conformational}) is flagged directly in the
Methods where the strain tool is introduced
(Section~\ref{sec:strain}), so that the absolute values appearing in the
case studies and figures are read only as relative discriminators
across poses of the same ligand. The agent's system prompt adopts this
relative-only interpretation explicitly
(Section~\ref{sec:results-representative}); what the present framework
cannot recover is the absolute magnitude of the penalty, which more
accurate force fields or quantum-mechanical torsion scans could
alleviate (Section~\ref{sec:disc-future-directions}). A secondary
concern arises when MMFF94 parameterisation fails and the tool falls
back to UFF (Algorithm~\ref{alg:strain}), which lacks the atom-type
granularity needed for reliable organic-molecule energetics. Second,
water-mediated interactions are captured only through PLIP's water-bridge
detection (Section~\ref{sec:plip}), which requires an explicit
crystallographic water molecule to be present in the input structure.
Desolvation entropy, water-network reorganisation, and the energetic
cost of displacing ordered solvent molecules are not modelled by any
tool in the current suite. The 3HS4 failure exemplifies the
consequence: decomposing Smina's scoring terms
(Table~\ref{tab:smina-terms-3hs4}) revealed that pose\_01's advantage
resides entirely in the non-directional hydrogen-bond term
(0.87~kcal/mol weighted), a continuous polar complementarity
measure with no equivalent in the agent's tool suite. The agent
itself identified ``water-mediated networks and protein side-chain
repositioning'' as unmodelled confounders
(Section~\ref{sec:results-representative}), yet had no quantitative
mechanism to incorporate this insight into its
ranking, precisely the solvation gap that
Section~\ref{sec:disc-failure-mode} identified as contributing to the
tool-suite ceiling. Third, all tools operate on static, post-docking
coordinates: protein flexibility, side-chain repacking, and induced-fit
rearrangements are absent from the evidence base. The agent recognised
this constraint in the 4JFL trace, proposing re-docking with
alternative pocket definitions as a remediation
(Section~\ref{sec:results-representative}), but such proposals remain
outside the pipeline's current execution scope. Fourth, PLIP applies
fixed geometric cutoffs to detect interactions
(Section~\ref{sec:plip}): contacts marginally outside these
thresholds are invisible, while those marginally inside receive equal
weight. The interaction quality assessment tool
(\S\ref{par:interaction-quality}) partially compensates by scoring each
detected contact's geometry on a continuous scale, yet the upstream
binary detection step discards information that a continuous scoring
function would retain, an information-loss pattern that the agent
cannot mitigate through reasoning alone. Visual inspection of the
four representative systems during figure preparation corroborated
this cutoff sensitivity across multiple tools: in 3HS4, for example,
the two steric clashes penalising the correct pose (pose\_01,
Table~\ref{tab:case-3hs4}) arise against protein heavy atoms in
flexible side chains of the carbonic-anhydrase active site that would
reasonably accommodate minor rigid-body relaxation upon ligand
binding; the clash detector uses the static, unminimised complex and
therefore flags these marginal overlaps as penalties rather than as
the small, readily relievable contacts they represent. Water
molecules and metal ions are explicitly excluded from the clash
calculation by design (Section~\ref{sec:clashes}), so the 3HS4
penalty is not a solvent-inclusion artefact. In the same system, the
$\pi$-stacking contact with HIS94A that PLIP reported exclusively
for the agent's selected pose\_02 is visible at a marginally longer
distance in pose\_01 as well, placing it just beyond PLIP's detection
threshold rather than genuinely absent. These observations illustrate a broader
pattern: binary-threshold tools can systematically overstate the
metric distance between poses that are structurally more similar than
their tabulated profiles suggest, contributing to the attention-threshold
effect that excluded the correct 3HS4 pose from the agent's top-seven
ranking. Collectively, these four classes of limitation define the
physics that the reasoning layer can access and thereby constrain the
performance ceiling documented in
Section~\ref{sec:disc-failure-mode}; their remediation constitutes the
most direct path to improved accuracy
(Section~\ref{sec:disc-future-directions}).

\paragraph{Ground-truth assumptions.}
The entire evaluation rests on the assumption that the crystallographic
ligand conformation constitutes the correct binding mode and that
heavy-atom symmetry-corrected RMSD to that conformation
(Equation~\ref{eq:ground-truth}) is an appropriate measure of pose
quality~\citep{meli2020spyrmsd}. Three limitations of this assumption
warrant explicit acknowledgement. First, crystal structures are
determined under cryogenic conditions and lattice-packing constraints
that may distort the ligand geometry relative to the physiological
binding mode: differences in temperature, pH, and crystal contacts can
shift torsion angles or reposition flexible substituents, so the
deposited conformation is not necessarily the most biologically relevant
one. Second, the 2.0~\AA{} near-native threshold used throughout the
evaluation (Section~\ref{sec:pose-classification}) is a field
convention~\citep{warren2006critical, su2019comparative}, not a physical
law. A pose at 1.9~\AA{} that disrupts a critical hydrogen bond may be
less pharmacologically useful than an alternative at 2.5~\AA{} that
better recapitulates the room-temperature interaction
pattern, a disconnect between geometric proximity and energetic quality
that Section~\ref{sec:bg-benchmarks} noted and that
Greenidge et~al.\ documented
quantitatively~\citep{greenidge2014mmgbsa}. Third, the single-correct-pose
assumption implicit in Equation~\ref{eq:ground-truth} may
oversimplify reality: multiple binding modes can coexist in solution,
and a docked conformation that reproduces a secondary mode would be
penalised as incorrect even if it represents a physically accessible
state. Two features of the benchmark design partially mitigate these
concerns. All ten ground-truth RMSDs lie well within the threshold
(0.43--1.97~\AA{}; Section~\ref{sec:results-benchmark-summary}),
eliminating borderline cases where the classification would be sensitive
to small perturbations. Furthermore, the \emph{without-native}
evaluation configuration
(Section~\ref{sec:ground-truth}) excludes the crystallographic
conformation from the candidate pool, so the agent discriminates among
computationally generated docked poses rather than recognising an
experimentally determined structure, a design that tests genuine
ranking ability rather than pattern-matching to a privileged input.
Nevertheless, the reliance on a single crystallographic reference per
system means that the reported accuracy
(Table~\ref{tab:accuracy-summary}) should be interpreted as performance
against a conventional, imperfect ground truth rather than against a
definitive biological standard.

\paragraph{System prompt engineering.}
Section~\ref{sec:disc-llm-reasoning} identified the coupling between
prompt phrasing and decision behaviour as an inherent constraint; the
present paragraph examines the specific over-fitting and design-choice
risks that this coupling creates. The
${\sim}7{,}800$-word system prompt
(Section~\ref{sec:agent-architecture}) was developed iteratively
against a small pool of exploratory systems drawn from the same
PDBbind~2016 corpus as the benchmark. For transparency we state
explicitly that the ten evaluation systems listed in
Table~\ref{tab:selected-systems} were not rigorously held out from
prompt development: exploratory tuning inspected PDBbind complexes
that overlap, at least in part, with the final ten systems, so the
benchmark cannot be treated as a fully disjoint test set. The
encoded thresholds (for example, the 40\,\% burial cutoff that gates
surface-bound poses out of top-tier consideration, or the 15~kcal/mol
strain value flagged as a tiebreaker penalty) may therefore be
calibrated to the physicochemical characteristics of systems that
subsequently appeared in evaluation rather than to the broader docking
landscape, and the reported 50\% accuracy should be read accordingly.
Because no formal held-out validation set was set aside during prompt
development, the extent of this over-fitting cannot currently be
quantified; establishing disjointness via a fresh, independently
curated benchmark is identified as a priority next-round revision
(Section~\ref{sec:disc-future-directions}). A related concern is the four-level hierarchical decision
framework itself (binding mode $>$ interaction quality $>$ interaction
quantity $>$ tiebreaker penalties;
Section~\ref{sec:agent-architecture}), which encodes the designer's
domain intuitions about the relative importance of physicochemical
criteria. The anti-bias section of the system prompt states that ``surface binders should never rank first, even with high interaction counts'' (Section~\ref{sec:system-overview}), a deliberately strong directive designed to counteract the LLM's tendency to over-weight raw interaction counts. While effective for deep-pocket targets where surface binding genuinely indicates a docking artefact, this rule could penalise legitimate binding modes in targets with shallow grooves, solvent-exposed allosteric sites, or protein-protein interaction hotspots, where even the correct pose may exhibit low burial ratios. The 4JFL failure illustrates a related consequence: the agent applied the hierarchy consistently yet undervalued hydrogen-bond geometry quality relative to burial in a shallow binding site (Table~\ref{tab:case-4jfl}; Section~\ref{sec:results-representative}), a weighting limitation traceable to the specific priority ordering rather than to a reasoning error. Alternative hierarchies (for instance, elevating interaction geometry above binding-mode gating for targets with shallow or solvent-exposed pockets) have not been evaluated, and it is unknown whether the current ordering is broadly optimal or merely adequate for the benchmark. The modular prompt architecture partially mitigates these
risks: because each of the seven named sections
(Section~\ref{sec:agent-architecture}) encodes a functionally
independent aspect of the agent's behaviour, individual components can
be modified or replaced without disrupting the remainder, a property
that would enable targeted ablation studies in which, for example, the
decision hierarchy is permuted while holding all other sections
constant. Such systematic optimisation, potentially automated through
data-driven prompt-tuning frameworks, is identified as a priority
future direction (Section~\ref{sec:disc-future-directions}).

\subsection{Future Directions}
\label{sec:disc-future-directions}

\paragraph{Scaling the benchmark.}
The statistical-power limitation diagnosed in
Section~\ref{sec:disc-limitations} identifies benchmark scaling as the
most direct path to stronger conclusions. Two complementary strategies
are envisaged. First, extending the evaluation to the full CASF-2016
core set, 285~protein--ligand complexes partitioned into 57~target
clusters
(Section~\ref{sec:bg-benchmarks})~\citep{su2019comparative}, would
enable head-to-head comparison with the 25~scoring functions already
benchmarked on that corpus, placing the agent's accuracy within an
established league table rather than against the single Smina reference
available at $m = 10$. Second, a stratified sample of approximately
100~systems drawn from the broader PDBbind general set
({>}11{,}000~complexes~\citep{wang2004pdbbind}) could be designed to
preserve the balanced Smina-success/failure partition
(Section~\ref{sec:benchmark-construction}) while spanning a broader
range of protein families, ligand flexibility classes, and
scoring-function failure severities, dimensions along which the present
ten-system benchmark cannot characterise performance. The statistical
gains from either strategy are substantial: at the observed 50\,\%
accuracy, a 100-system evaluation would narrow the 95\,\% Wilson
confidence interval from the current $[23.7, 76.3]$\,\%
(Equation~\ref{eq:wilson};
Table~\ref{tab:accuracy-summary}) to approximately $[40, 60]$\,\%,
reducing the interval width from 53 to ${\sim}$20~percentage points and
for the first time permitting a formal two-proportion hypothesis test
comparing the agent with the Smina baseline. A scaled benchmark would
also provide the statistical power needed to assess the
candidate ensemble-difficulty descriptors, ensemble size and TQS
coefficient of variation, that
Section~\ref{sec:disc-retention-recovery} flagged as exploratory
covariates of outcome ($r = -0.668$ and $r = -0.739$, respectively,
uncorrected; Section~\ref{sec:supplementary_analyses}). With $m \geq 100$ systems,
multivariate logistic regression could determine whether these
descriptors jointly predict agent failure and, if so, whether they could
serve as pre-invocation screening criteria that route high-difficulty
ensembles to manual expert review or to the expanded tool suite proposed
below, thereby targeting agentic evaluation at the systems most likely
to benefit from it.

\paragraph{Multi-LLM evaluation.}
The single-model dependency acknowledged in
Section~\ref{sec:disc-limitations} can be addressed directly by
replicating the evaluation across multiple LLM backends while holding
the tool suite, system prompt, and benchmark constant. Because the
analysis tools are fully deterministic
(Section~\ref{sec:system-overview}), every model would operate on
identical metric inputs, isolating the reasoning layer as the sole
source of inter-model variation. The comparison should span three axes
already quantified for GPT-5: best-pose identification accuracy
(Table~\ref{tab:accuracy-summary}), reasoning faithfulness as measured
by the Spearman correlation between stated weights and objective metric
separations ($\rho = {+}0.83$;
Table~\ref{tab:faithfulness}), and the per-tool weight distributions
that capture each model's evaluative priorities
(Figure~\ref{fig:weight_variability}). Candidate architectures include
both proprietary frontier models, Claude and Gemini, and open-weight
alternatives such as LLaMA-3 and Mistral, whose parameters can be
archived alongside the analysis code to restore the checkpoint-pinning
capability that the current API-dependent workflow lacks
(Section~\ref{sec:disc-limitations}). Open-weight models would
additionally eliminate per-invocation API cost and enable local
deployment in environments where data confidentiality precludes
cloud-based inference. The faithfulness axis is particularly
informative: intervention-based studies have demonstrated that larger
models can exhibit \emph{lower} faithfulness than smaller
ones~\citep{lanham2023measuring}, and that biasing features in the input
can induce systematically unfaithful explanations without affecting
surface-level plausibility~\citep{turpin2023language}, findings that
caution against assuming the $\rho = {+}0.83$ concordance observed with
GPT-5 will transfer to alternative architectures. A secondary objective
is to determine whether the ${\sim}7{,}800$-word system prompt
(Section~\ref{sec:agent-architecture}) transfers intact across providers
or requires model-specific adaptation. If the same prompt elicits
comparable accuracy and faithfulness from architecturally diverse models,
the framework's claim to model-agnosticism would be empirically
grounded; if performance degrades for specific backends, prompt--model
co-optimisation, potentially automated through the data-driven tuning
frameworks discussed below, would become a prerequisite for
generalisable deployment.

\paragraph{Expanding the analysis tool suite.}
The tool-suite coverage ceiling that
Section~\ref{sec:disc-failure-mode} identified as the proximate cause of
the 20\,\% recovery rate, and whose four technical roots
Section~\ref{sec:disc-limitations} diagnosed in detail, defines the most
direct path to improved accuracy: each proposed tool addition targets a
specific physics gap that the current nine-tool suite
(Section~\ref{sec:system-overview}) leaves unmodelled. The
highest-priority expansion concerns conformational strain. MMFF94's
systematic overestimation (44--97~kcal/mol across all benchmark systems
vs.\ the ${\lesssim}$5~kcal/mol expected from crystallographic
surveys~\citep{perola2004conformational};
Section~\ref{sec:disc-limitations}) confines the agent to relative
strain comparisons
(Section~\ref{sec:results-representative}); replacing the force field
with a higher-fidelity alternative, OPLS4~\citep{lu2021opls4} or a
quantum-mechanical-accuracy neural network potential such as
ANI-2x~\citep{berenger2024ani2x}, would recover absolute physical meaning, while Cambridge
Structural Database--derived torsional strain
filters~\citep{gu2021strain} offer a complementary,
database-statistical approach that bypasses force-field parametrisation
altogether. The practical significance of improved strain fidelity is
underscored by the 3OXC result: a ${\sim}$20~kcal/mol differential was
sufficient for recovery
(Section~\ref{sec:results-representative}), suggesting that even modest
gains in strain accuracy could introduce discriminative signal in the
four Category~D systems where the current tool failed. A second
priority is explicit solvation modelling. Water-mediated interactions
are currently captured only when a crystallographic water molecule is
present in the input structure (Section~\ref{sec:plip}); desolvation
entropy, water-network reorganisation, and the energetic cost of
displacing ordered solvent remain unmodelled
(Section~\ref{sec:disc-limitations}). Incorporating a per-residue
energy decomposition via
MM-GBSA~\citep{greenidge2014mmgbsa} would simultaneously address this
solvation gap and introduce residue-level binding
energetics, enabling the agent to identify which protein residues
contribute most to binding and to assess whether key pharmacophoric
interactions are formed, a capability none of the current tools
provides. Third, pharmacophore mapping would supply
structure--activity context by comparing each pose's interaction pattern
against a reference pharmacophore model for the target, distinguishing
conformations that satisfy known pharmacophoric requirements from those
that achieve high interaction counts through non-essential
contacts, a distinction relevant to the 4JFL failure, where the
selected pose matched the ground truth on aggregate interaction metrics
yet differed in pharmacophoric alignment
(Section~\ref{sec:results-representative}). Fourth, supplementing PLIP
with a continuous interaction-profiling tool, one that scores atom-pair
contacts on a smooth potential rather than applying binary geometric
cutoffs (Section~\ref{sec:plip}), would retain the marginal
contacts that the current detection step discards, improving the
information gradient available to the reasoning layer. A guiding
principle for all four expansions is empirical validation on the existing
failure systems: inclusion should be contingent on demonstrable
improvement in metric separation
(Figure~\ref{fig:metric-separation-heatmap}), the diagnostic that
Section~\ref{sec:disc-multi-metric} established as the strongest
correlate of decision correctness.

\paragraph{Systematic prompt optimisation.}
The over-fitting and design-choice risks that
Section~\ref{sec:disc-limitations} diagnosed (thresholds calibrated to
ten systems, an untested four-level hierarchy, and the 4JFL weighting
failure (Table~\ref{tab:case-4jfl})) are amenable to data-driven
remediation once a sufficiently large benchmark is available.
Compiler-style prompt-optimisation frameworks such as DSPy~\citep{khattab2023dspy} can treat the
system prompt's tunable parameters as a search problem: given a training
partition of docking systems with known ground-truth poses, the framework
optimises natural-language modules by evaluating candidate prompts against
an objective metric (here, best-pose identification accuracy) and
retaining the variant that maximises held-out performance. The modular
seven-section architecture of the current prompt
(Section~\ref{sec:agent-architecture}) maps naturally onto this
paradigm, because each section can be varied independently while holding
the remaining six constant. Four parameter classes identified by the
present evaluation constitute the initial search space: the
binding-mode gating stringency (the burial threshold that determines
whether surface-bound poses are eligible for top-tier ranking), the
relative weight assigned to hydrogen-bond geometry quality versus raw
interaction count, the strain-override threshold above which
conformational strain triggers a tiebreaker penalty, and the overall
priority ordering among the four hierarchy levels
(Section~\ref{sec:disc-limitations}). The 4JFL case provides a concrete
validation target: a successful optimisation should learn to elevate
hydrogen-bond geometry above binding-mode gating when no pose in the
ensemble achieves deep-pocket classification
(Section~\ref{sec:results-representative}), correcting the specific
weighting limitation that Section~\ref{sec:disc-limitations} traced to
the manually encoded hierarchy. Because the outcome-stratified weight
analysis revealed nearly identical mean weight profiles between correct
and incorrect decisions (maximum difference ${\leq}$3.4~percentage
points; Table~\ref{tab:outcome-weights}), optimisation should target
\emph{conditional} weighting rules, context-dependent adjustments
triggered by ensemble properties such as burial-class homogeneity or
strain-landscape flatness, rather than global weight shifts, which the
current results suggest are already near a reasonable operating point.
A complementary experiment is a formal anonymisation ablation.
Section~\ref{sec:disc-anonymisation} introduced the SHA-256 protocol as
a preventive intervention against positional and label-based
bias~\citep{tang2024found} but noted that its magnitude of effect had
not been quantified. Comparing accuracy, faithfulness, and weight
distributions between anonymised and non-anonymised conditions, on the
same systems and with the same LLM, would isolate the contribution of
the anonymisation layer and determine whether the protocol is essential,
beneficial, or neutral for the ranking task. Together, prompt
optimisation and the anonymisation ablation would transform the current
manually engineered pipeline into an empirically calibrated system whose
design choices are justified by held-out performance rather than by
designer intuition alone.

\paragraph{Prospective validation and hybrid approaches.}
The preceding directions address limitations that are measurable on
retrospective benchmarks; the longest-term objective is to validate the
agentic framework prospectively, within active drug discovery campaigns
where experimental pose confirmation, X-ray co-crystallography of
newly synthesised compounds, provides an unambiguous ground truth that
the PDBbind-derived benchmark can only approximate
(Section~\ref{sec:disc-limitations}). In such a setting, the agent's
ranking would be generated before the crystal structure is determined,
converting each prediction into a pre-registered hypothesis testable by
structure determination. Section~\ref{sec:disc-retention-recovery}
argued that the framework's current operational profile (high retention,
low recovery) already positions it as a hypothesis generator for expert
verification rather than a standalone decision-maker; prospective
deployment would formalise this role and, crucially, accumulate labelled
systems at a pace dictated by the medicinal-chemistry programme rather
than by curator availability. A complementary strategy is to deploy the
agent alongside an ML-based rescoring method such as
Gnina~\citep{mcnutt2021gnina} and use their agreement or disagreement as
a consensus confidence signal. Because the two approaches rely on
orthogonal evidence (the agent integrates physics-based tool outputs
through natural-language reasoning, whereas Gnina learns a scoring
function from structural training data
(Section~\ref{sec:disc-related-work})) concordant top-ranked poses
would carry higher expected accuracy than either method alone, while
discordant outcomes would flag specific systems for expert review.
The ensemble-difficulty descriptors that
Section~\ref{sec:disc-retention-recovery} identified as
outcome-predictive, ensemble size and TQS coefficient of variation, could
further stratify this triage, routing high-difficulty, discordant systems
to manual analysis while accepting low-difficulty, concordant predictions
with minimal oversight. An orthogonal path to variance reduction is
internal to the agent itself. Because LLM inference is stochastic and
only a single run per system was performed
(Section~\ref{sec:disc-limitations}), the observed rankings sample one
draw from an uncharacterised distribution. Running the agent $N$ times
on identical tool outputs and aggregating via majority vote (or, more
informatively, via confidence-weighted aggregation that down-weights
low-confidence runs) would convert the single stochastic decision into
a consensus estimate whose stability can be quantified by the fraction
of runs that select the same pose. Self-consistency across
permutations has been shown to improve GPT-3.5 accuracy by
7--18~percentage points in multiple-choice
settings~\citep{tang2024found}, and an analogous gain here would begin
to close the recovery deficit that defines the framework's current
performance ceiling. Finally, the prospective setting enables a
\emph{self-improvement loop} unavailable in retrospective evaluation.
After revealing the experimentally determined ground truth, the correct
pose and its metric profile can be fed back to the agent with the
instruction to identify where its original reasoning diverged from the
evidence; the resulting self-critique would yield explicit,
natural-language hypotheses (for example, ``the strain threshold was
set too conservatively'' or ``hydrogen-bond geometry was under-weighted
relative to burial'') that can be translated into targeted prompt
modifications. A caveat from the faithfulness literature tempers
expectations for this loop: post-hoc rationalisation is well documented
in frontier models, with chain-of-thought demonstrations achieving
80--90\,\% of full-CoT performance even when the intermediate reasoning
steps are invalid~\citep{lanham2023measuring}, so reflective critiques
may reproduce surface-plausible explanations that do not identify the
true causal origin of the error. Coupling self-improvement with the
compiler-style optimisation proposed in
Section~\ref{sec:disc-future-directions}, where DSPy evaluates each
candidate prompt modification against held-out accuracy rather than
against the agent's own plausibility judgement, would provide an
objective check on the reflection loop and close the feedback circuit
between prospective performance data and prompt calibration. Together,
the four directions outlined in this section (scaled benchmarking,
multi-LLM evaluation, expanded tool suites, and systematic prompt
optimisation) converge on a single programme: transforming the current
proof-of-concept into a rigorously validated, empirically calibrated
framework whose design choices are grounded in held-out experimental
evidence rather than in retrospective analysis of ten curated systems.

\section{Conclusion}

This work introduced AgenticPosesRanker, an agentic AI framework that
couples six deterministic, physically grounded analysis tools with
large-language-model reasoning to rank protein--ligand docking poses.
By preserving each physicochemical observable as a separate input to an
explicit chain-of-thought evaluation, rather than collapsing them into
a single composite score, the framework enables context-dependent
weighting that adapts to the distributional properties of each pose
ensemble.

Evaluated on a curated benchmark of ten protein--ligand systems
(162~poses), the agent achieved 50.0\% best-pose accuracy, significantly
exceeding the 7.7\% random-selection baseline ($p < 0.001$, one-sided
binomial test) while matching the design-fixed Smina scoring baseline.
Performance was markedly asymmetric across partitions: the agent retained
80\% of Smina's correct rankings but recovered only 20\% of its failures,
characterising the framework as a reliable curator of existing correct
rankings rather than a powerful recovery engine for scoring-function
errors. Decision-attribution analysis revealed that this asymmetry
originates upstream of the reasoning layer: ground-truth poses in failure
systems are metric-indistinguishable from their ensembles, offering no
physics-based signal that any weighting scheme within the current tool
suite could exploit. Correspondingly, the alignment between the agent's stated tool weights
and the objective metric separations of the selected pose, quantified
as the per-system Spearman rank correlation, was high (median
$\rho = +0.83$) and comparable across correct and incorrect decisions,
showing that the agent applies its evaluation criteria consistently
regardless of outcome. This descriptive alignment measure does not,
on its own, distinguish faithful introspection from internally coherent
post-hoc rationalisation of the kind documented for frontier
models~\cite{turpin2023language, lanham2023measuring}.

The principal bottleneck, therefore, is not reasoning quality but
tool-suite coverage: unmodeled contributions from solvation free energy,
water-mediated hydrogen-bond networks, protein conformational
flexibility, and entropic effects define the ceiling of the current
implementation. Critically, the transparent architecture makes this
diagnosis possible, each failure can be traced to specific missing
observables rather than attributed to an opaque aggregate score. This
diagnosability distinguishes the agentic approach from conventional
scoring functions and machine-learning rescorers, which achieve higher
accuracy but offer limited mechanistic insight into individual ranking
errors.

These results establish a methodological template for evaluating agentic
AI systems against objective ground truth in the natural sciences:
accuracy alone is insufficient; faithfulness, decision attribution, and
failure-mode diagnosis are necessary to distinguish reasoning-layer
limitations from evidence-layer limitations. For structure-based drug
design, the framework is best positioned not as a replacement for
high-throughput scoring but as an interpretable curation layer for
late-stage pose refinement, where the structured, metric-by-metric
rationale accompanying each ranking decision provides actionable insight
for medicinal chemists. Expanding the tool suite to address the
identified coverage gaps, testing across multiple language models, and
scaling the benchmark to larger datasets are the immediate priorities
for advancing the approach toward prospective deployment.

\section{Code and Data Availability}

The source code and curated benchmark data will be made available upon
reasonable request to the corresponding author.
A web interface for interactive pose ranking will be available at
\url{https://sofk.ch}.

\newpage
\section{Statements and Declarations}

\paragraph{Author contributions.}
S.K.\ conceived the project, designed and implemented the agentic pipeline, curated the benchmark, conducted the evaluation, performed the analyses, produced the figures and tables, and drafted the manuscript.
A.H.M.\ contributed to methodological design, provided technical guidance on computational chemistry tooling, and reviewed the manuscript.
M.A.L.\ supervised the project, contributed to conceptual framing and methodological design, reviewed and edited the manuscript, and secured funding for the work.
All authors read and approved the final manuscript.

\paragraph{Funding.}
This work was conducted as part of S.K.'s doctoral research at the Department of Pharmaceutical Sciences, University of Basel, under the supervision of M.A.L. No specific grant from public, commercial, or not-for-profit funding agencies was received for this study.

\paragraph{Competing interests.}
S.K.\ declares no competing interests.
A.H.M.\ declares no competing interests.
M.A.L.\ declares no competing interests.

\paragraph{Use of generative AI in manuscript preparation.}
The authors used large-language-model assistants (OpenAI GPT-5 and Anthropic Claude) during manuscript preparation for language polishing and reference cross-checking. All scientific content, analyses, interpretations, and conclusions are the authors' own. All AI-assisted edits were reviewed and approved by the authors, who take full responsibility for the final manuscript. Generative AI was not used to produce research data, analyses, figures, or results. The use of GPT-5 as the reasoning backbone of the AgenticPosesRanker framework (the research object) is a methodological component described in Sections~\ref{sec:system-overview} and~\ref{sec:agent-architecture} and is distinct from this writing-assistance disclosure.

\paragraph{Author contact information.}
Sofiene Khiari: \href{mailto:research@sofk.ch}{research@sofk.ch}, ORCID \href{https://orcid.org/0000-0003-0540-2052}{0000-0003-0540-2052}.
Amr H.\ Mahmoud: \href{mailto:amr.abdallah@unibas.ch}{amr.abdallah@unibas.ch}.
Markus A.\ Lill: \href{mailto:markus.lill@unibas.ch}{markus.lill@unibas.ch}, ORCID \href{https://orcid.org/0000-0003-3023-5188}{0000-0003-3023-5188}.

\begin{appendix}
\renewcommand{\thefigure}{S\arabic{figure}}
\renewcommand{\thetable}{S\arabic{table}}
\setcounter{figure}{0}
\setcounter{table}{0}

\section{Supplementary Information}

\subsection{Detailed Scoring-Function Benchmark Performance}
\label{sec:si-scoring-benchmarks}

\paragraph{Empirical scoring functions.}
An evaluation of ten docking programs across more than 2\,000
protein--ligand complexes found that pose-reproduction success rates ranged
from roughly 60\% to 80\% depending on the program, yet binding-affinity
correlations rarely exceeded $r \approx 0.60$ even for the best-performing
methods~\cite{wang2016comprehensive}.
Similar trends emerge from the Comparative Assessment of Scoring Functions
(CASF)~\cite{su2019comparative} and from community benchmarks such as
CSAR~2014, where docking power consistently surpassed scoring power across
all participating methods and few achieved a Spearman $\rho \geq 0.5$ for
affinity ranking despite generating poses with median RMSDs below
2.0~\AA{}~\cite{carlson2016csar}.
This disparity reflects a design tension: because the weights are fitted to
reproduce absolute binding free energies, they do not necessarily produce
correct rank-orderings among the multiple poses of a single
ligand~\cite{quiroga2016vinardo}.
Physics-based rescoring with Molecular Mechanics / Generalised Born Surface
Area (MM-GBSA) or Poisson--Boltzmann Surface Area (MM-PBSA) includes
explicit solvation modelling and energy minimisation, offering a more
rigorous decomposition of the binding free
energy~\cite{greenidge2014mmgbsa}; however, the additional computation
increases processing time from seconds to hours per system and the result
remains a single composite score.

\paragraph{Knowledge-based and consensus scoring benchmarks.}
Knowledge-based scoring functions such as DrugScore and PMF capture
interaction patterns that complement the functional forms of empirical
scorers, yet they share the same fundamental limitation: all
physicochemical contributions are collapsed into a single, opaque numerical
score~\cite{su2019comparative}.
Two prerequisites for effective consensus scoring have been identified
empirically: each constituent function must achieve reasonable individual
accuracy, and the functions must be sufficiently distinctive, making
complementary rather than redundant
errors~\cite{yang2005consensus}.
Despite gains, classical consensus methods apply their aggregation rule
uniformly across every target and every pose ensemble; the weight of each
scoring function is fixed at design time and cannot adapt to the chemical
context of a given binding site.

\paragraph{Machine-learning scoring function benchmarks.}
Early machine-learning approaches trained random forests on atom-pair
distance features, demonstrating that purely data-driven models could rival
classical scorers on standardised
benchmarks~\cite{ballester2010machine}.
Performance, however, is tightly coupled to training-data coverage;
Gnina's redocking success drops from 73\,\% to 68\,\% on targets absent
from its training distribution, and analogous domain-shift effects have
been observed across deep-learning
methods~\cite{mcnutt2021gnina, francoeur2020crossdock}.
High RMSD-based accuracy does not guarantee physical plausibility: the
PoseBusters validation framework showed that many deep-learning methods
generate poses with steric clashes, incorrect stereochemistry, or
non-standard bond geometries, causing classical force-field-based programs
to outperform them once such validity checks are
applied~\cite{buttenschoen2024posebusters}.

\subsection{Pose Anonymisation Protocol Details}
\label{sec:si-anonymisation}

The following paragraphs expand on the pose anonymisation protocol
summarised in Section~\ref{sec:anonymisation} of the main text.

\paragraph{Anonymous code generation.}
Each pose is assigned a unique eight-character alphanumeric identifier
drawn from a base-36 alphabet (digits 0-9 and uppercase letters A-Z).
The identifier is derived by concatenating the system identifier
(e.g.\ \texttt{1ERR}) and the original pose stem
(e.g.\ \texttt{pose\_01}) into a single string, computing its SHA-256
hash, and converting the first eight bytes of the digest to base-36
(Equation~\ref{eq:anon-code}), where $s$ is the system identifier,
$p$ the pose name, and $\|$ denotes string concatenation (with an
underscore separator in the implementation). The function is
deterministic: the same system-pose pair always produces the same
code. The eight-character length was chosen as a practical compromise:
it provides $36^{8} \approx 2.8 \times 10^{12}$ possible codes, far
more than needed for collision resistance within a single system's
pose set, while remaining short enough for the LLM to copy accurately
into its output. By contrast, full 36-character UUIDs were found in
early testing to increase the rate of transcription errors (truncated
or mistyped identifiers) in the model's responses, complicating
automatic de-anonymisation.

\paragraph{Content sanitisation.}
Anonymous codes are used as filenames for the analysis files
(e.g.\ \texttt{K7M9N2P4.analysis}). Before each section of tool
output is appended to an analysis file, the content is passed through
a sanitisation function that replaces every occurrence of the original
pose name with a generic placeholder and strips file-path fragments
that could leak identity information. Specifically, all strings
matching \texttt{pose\_\textbackslash d+} (case-insensitive) are
replaced with a neutral token, preventing residual references from
appearing in the text consumed by the LLM. The sanitisation is applied
to every tool's output, PLIP interaction reports, SASA summaries,
strain energies, clash lists, polar-penalty tables, SMILES strings,
and all three derived assessment metrics, before the analysis file is
finalised.

The mapping between original pose names and anonymous codes is
persisted to a JSON file (\texttt{pose\_mapping.json}) stored in the
working directory alongside the analysis files, but this mapping file
is never included in the context sent to the model.

\paragraph{Context presentation.}
When the analysis files are assembled into the LLM's input context,
each file is introduced only by its anonymous code
(e.g.\ ``\texttt{Analysis File: K7M9N2P4}''). The preamble explicitly
instructs the model that pose identifiers are anonymised and that
evaluation must be based solely on the computational metrics contained
within each file. The system prompt further reinforces this with
code-handling rules requiring the LLM to copy codes
character-for-character when referencing poses in its reasoning, to
never truncate or abbreviate them, and to verify each code against the
analysis file headers before finalising its ranking.

\paragraph{Post-inference de-anonymisation.}
After the LLM returns its ranking and chain-of-thought reasoning, both
outputs are passed through a de-anonymisation step that replaces every
occurrence of an anonymous code with the corresponding original pose
name using word-boundary-aware regular-expression matching. The
resulting human-readable output allows domain experts to interpret the
ranking in terms of the familiar pose identifiers, while the original
anonymous codes are preserved in the analysis log for auditability.
The complete mapping (pose name $\to$ anonymous code) is also recorded
in the ranking log alongside the system prompt and the raw model
output, enabling full reconstruction of the anonymisation state for
any given run.

\subsection{User Interface Implementation Details}
\label{sec:si-user-interface}

\begin{figure}[htbp]
    \centering
    \includegraphics[width=\textwidth]{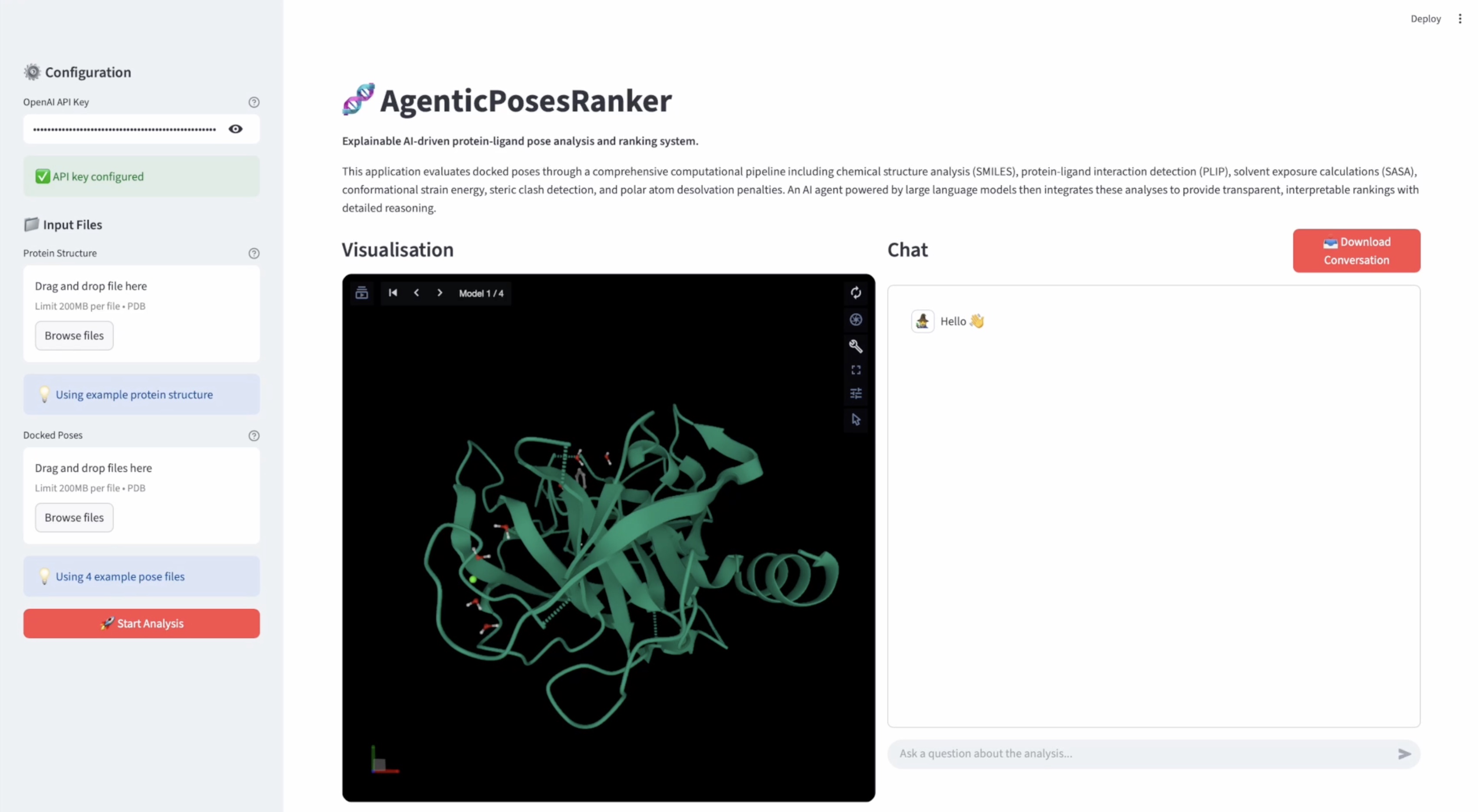}
    \caption{
        \textbf{Screenshot of the AgenticPosesRanker web interface.}
        The left sidebar contains the configuration panel with API key entry, input file upload (supporting both user-provided files and pre-loaded example systems), and the ``Start Analysis'' button. The main content area is divided into two columns: (left)~an interactive 3D molecular viewer powered by Mol*~\cite{sehnal2021molstar}, displaying the protein-ligand complex with individual docked poses navigable via the model selector; and (right)~a conversational chat interface through which the agent delivers its analysis results and users can ask follow-up questions. A ``Download Conversation'' button in the chat header enables export of the full analysis transcript.
    }
    \label{fig:streamlit_ui}
\end{figure}

The following paragraphs expand on the user-interface description
given in Section~\ref{sec:user-interface} of the main text.
Figure~\ref{fig:streamlit_ui} shows a screenshot of the deployed interface.

\paragraph{Page configuration and header.}
The application uses Streamlit's wide layout mode to maximise
horizontal space. A header at the top of the page displays the
application title and a brief description of the system's
capabilities, orienting first-time users before any files are
uploaded.

\paragraph{Sidebar and configuration.}
A persistent sidebar on the left-hand side of the page provides the
three configuration controls required before an analysis can begin:

\begin{enumerate}
    \item \textbf{API Key.}  A dialog for entering or updating the
    OpenAI API key. If a \texttt{.env} file containing the key is
    present in the working directory, the key is loaded automatically
    at startup; otherwise, the user supplies it through a
    password-masked text input. A colour-coded status badge in the
    sidebar indicates whether a valid key has been configured.

    \item \textbf{Input Files.}  A dialog that supports two
    complementary modes of data entry:
    \begin{itemize}
        \item \emph{Example systems.}  A dropdown lists pre-loaded
        protein-ligand systems bundled with the application (the
        curated benchmark systems described in
        Section~\ref{sec:benchmark-construction}). Selecting an
        example system automatically sets the protein structure and
        pose files without requiring any file upload.
        \item \emph{User uploads.}  When no example system is
        selected, two file-upload widgets appear. The first accepts a
        single protein structure in PDB format; the second accepts
        one or more docked poses in PDB format. Uploaded files are
        saved to a UUID-based workspace directory under the
        application's \texttt{data/} folder, ensuring that concurrent
        sessions do not interfere with one another.
    \end{itemize}
    A second colour-coded status badge indicates whether both the
    protein and at least one pose file have been provided.

    \item \textbf{Start Analysis.}  A primary action button that
    launches the full analysis pipeline. The button is disabled
    (greyed out) until both the API key and input files have been
    configured, preventing premature execution. Once clicked, the
    button triggers the computational pipeline described in
    Section~\ref{sec:agent-architecture}: protein filtering, complex
    construction, parallel tool execution, anonymisation, agent
    inference, and de-anonymisation.
\end{enumerate}

\noindent A legend at the bottom of the sidebar explains the
colour-coded status badges: green (``Ready'') indicates a successfully
configured component, and red (``Missing'') indicates a component that
still requires user action.

\paragraph{Two-column layout.}
The main content area is divided into two equal-width columns that are
displayed side by side, providing simultaneous access to structural
and textual information:

\begin{itemize}
    \item \textbf{Left column, 3D Molecular Visualisation.}  An
    interactive molecular viewer powered by the
    Mol*~\cite{sehnal2021molstar} plugin, embedded as an HTML
    component. Before any files are uploaded, the viewer displays a
    placeholder message; once a protein structure and docked poses are
    provided, it renders the protein-ligand complex in three
    dimensions. Each docked pose is represented as a separate model
    within a combined PDB file: the protein coordinates are replicated
    in every model frame, and the ligand coordinates for each pose are
    converted from SDF to PDB format (via Open
    Babel~\cite{oboyle2011openbabel}) and appended as \texttt{HETATM}
    records. The viewer supports standard molecular graphics
    interactions, rotation, translation, zoom, and residue selection,
    allowing users to visually inspect the poses while the agent's
    analysis proceeds. The viewer uses a black background to maximise
    contrast with protein ribbon and ligand stick representations.
    Updates to the uploaded files (e.g.\ switching between example
    systems) immediately refresh the viewer.

    \item \textbf{Right column, Chat Interface.}  A conversational
    interface that serves as the primary channel for both analysis
    output and user interaction. The chat panel has three
    sub-components:
    \begin{enumerate}
        \item \emph{Message history.}  A scrollable container (fixed
        height of 600\,px) that displays the full conversation between
        the user and the agent. Assistant messages are rendered with a
        wizard avatar and support expandable chain-of-thought (CoT)
        reasoning sections; user messages use a developer avatar.
        Custom CSS reduces inter-message spacing to display more
        content within the visible area.
        \item \emph{Text input.}  A chat input field at the bottom of
        the column, disabled until the analysis has been started at
        least once. After the initial ranking, users may type
        follow-up questions (e.g.\ ``Why was pose X ranked below pose
        Y despite having more interactions?''), and the agent responds
        with access to the full analysis context and conversation
        history (Section~\ref{sec:agent-architecture}).
        \item \emph{Download button.}  A button in the chat header
        that exports the complete conversation, including all CoT
        reasoning sections, as a Markdown file timestamped with the
        current date and time, enabling offline review and archival.
    \end{enumerate}
\end{itemize}

\paragraph{Analysis execution and progress feedback.}
When the user clicks ``Start Analysis'', the agent's introductory
message is posted to the chat, listing the six analysis categories
that will be evaluated (SMILES extraction, PLIP interaction analysis,
SASA calculations, conformational strain, steric clashes, polar atom
satisfaction). A Streamlit status widget then tracks the progress of
each tool in real time, displaying the name of the currently executing
step. The steps correspond to the full pipeline described in
Section~\ref{sec:agent-architecture}: complex PDB construction, SMILES
extraction, PLIP interaction profiling, SASA calculation,
conformational strain estimation, steric clash detection, unsatisfied
polar atom penalty computation, finalisation of analysis log files,
and generation of visualisation plots. Upon completion of the
computational tools, the status widget transitions to a ``complete''
state.

Visualisation plots generated during the analysis (metric distribution
charts and interaction network diagrams) are displayed within the chat
as an assistant message, providing immediate graphical feedback before
the agent begins its ranking inference. The agent then receives the
anonymised analysis files and produces its ranking and chain-of-thought
justification, which are streamed into the chat panel as a final
assistant message. On subsequent page loads, completed analyses are
restored from session state: the status widget is rendered in its
``complete'' state and all prior chat messages, including reasoning
sections, are faithfully reconstructed.

\paragraph{Follow-up conversation.}
After the initial ranking, the chat input field is enabled, and users
may pose clarifying or comparative questions. Each follow-up question
is processed by the same GPT-5 agent with the
\texttt{skip\_tools} flag set to true: the pre-computed analysis files
are re-loaded into the context alongside the full conversation
history, ensuring that the agent can reference specific metrics or
pose identifiers from any prior turn without re-executing the
computational tools. This design keeps follow-up responses fast
(seconds rather than minutes) while maintaining full analytical
context. A spinner (``Thinking\ldots'') provides visual feedback
during inference.

\paragraph{Error handling.}
The interface implements comprehensive error handling at every stage of
the pipeline. Missing dependencies are detected at startup and
displayed as blocking error messages with installation instructions.
Failures during tool execution (e.g.\ an SDF file that cannot be
parsed, a PLIP analysis that returns no interactions) are caught,
classified by a centralised error handler, and presented as structured
error panels within the chat, identifying the error type, affected
component, and suggested remediation. Critical errors halt the
pipeline and update the status widget to an error state; the user is
not left with a silently incomplete analysis. Network and API errors
during GPT-5 inference are similarly intercepted and displayed, with
the conversation history preserved so that users can retry after
resolving the underlying issue.

\subsection{Data File Format and Native Pose Generation}
\label{sec:si-data-files}

\paragraph{File format.}
For each system, the docking run produced a multi-conformer SDF file
containing the generated poses together with their
\texttt{minimizedAffinity} scores, which represent the
Smina-minimised binding free energy estimates (kcal/mol). Each system
directory provides the protein structure
(\texttt{\textit{id}\_protein.pdb}), the crystallographic ligand
(\texttt{\textit{id}\_ligand.sdf}), and the multi-pose docking output
(\texttt{\textit{id}\_docked.sdf}) with per-pose
\texttt{minimizedAffinity} properties.

\paragraph{Native pose generation.}
\label{sec:si-native-pose}
To enable root-mean-square deviation (RMSD) comparison between docked
poses and the experimentally determined binding mode, all structures
must share a consistent atom representation, identical atom ordering,
identical hydrogen treatment, and identical coordinate frame. Direct
comparison of the crystallographic ligand SDF against the
Smina-produced docked SDF is unreliable because the two files may
differ in atom ordering, hydrogen placement, and internal coordinate
conventions.

A \emph{native pose} is therefore regenerated for each system by
passing the crystallographic ligand through Smina in scoring-only
mode:

\medskip
\noindent\texttt{smina -r \textit{id}\_protein.pdb -l
\textit{id}\_ligand.sdf --score\_only -o \textit{id}\_native.sdf}
\medskip

\noindent The \texttt{--score\_only} flag instructs Smina to read the
input ligand coordinates without performing any conformational
sampling or minimisation; it merely re-scores the existing pose and
writes the result in the same internal format used for the docked
poses. The native SDF thus inherits the same atom ordering and
hydrogen convention as the docked conformations, enabling atom-by-atom
RMSD calculation. The Smina binary used for this step corresponds to
the October~2019 build, based on AutoDock
Vina~1.1.2~\cite{trott2010vina}.

\subsection{RMSD Calculation Details}
\label{sec:si-rmsd-details}

The symmetry-corrected RMSD was computed using the
\texttt{symmrmsd} function of
\texttt{spyrmsd}~\cite{meli2020spyrmsd}. The function receives the
three-dimensional coordinates, atomic numbers, and adjacency matrices
of both the native and docked heavy-atom substructures. Because both
structures originate from the same protein coordinate frame (via Smina
processing), the RMSD was calculated directly on the input coordinates
without prior superimposition or centering.

\subsection{SMILES Extraction and Ligand-Diversity Check}
\label{sec:si-smiles}

The following paragraphs expand on the SMILES extraction tool
summarised in Section~\ref{sec:smiles} of the main text.

\paragraph{Per-pose canonicalisation.}
The tool reads the raw pose file with RDKit~\cite{rdkit2025},
retaining explicit hydrogens during parsing, and then generates a
canonical SMILES string with stereochemistry preserved. Canonicalisation
guarantees that the same molecule always yields an identical string
regardless of atom ordering in the input file, while the isomeric flag
encodes chirality and double-bond geometry, properties that can
critically influence binding.

\paragraph{Ligand-diversity check.}
Before any pose is analysed, a diversity check is performed across all
input poses. Each pose's SMILES is normalised by removing explicit
hydrogens and regenerating the canonical form, and the resulting set
of unique normalised SMILES is inspected. If all poses share the same
normalised SMILES, the system concludes that the poses are
conformational variants of a single ligand, and the SMILES block is
omitted from the anonymised analysis files passed to the LLM, since
chemical identity is uninformative in this scenario. If two or more
distinct SMILES are detected, the SMILES block is included for every
pose so that the agent can identify structural differences between the
molecules under comparison. The agent's system prompt instructs the
LLM that the absence of a SMILES block signals identical ligands
across all poses and that, when present, SMILES should be used for
structural interpretation rather than as a ranking criterion.

\paragraph{Status in the present benchmark.}
All ten benchmark systems involve a single ligand docked in multiple
conformations, so the diversity check consistently finds identical
SMILES and the SMILES block is omitted from every analysis file.
The multi-ligand path was not exercised in the present evaluation.

\subsection{Software Environment and Tool Versions}
\label{sec:si-environment}

Table~\ref{tab:si-environment} collects the versions of all
scientific-computing tools and binaries used to produce the results
reported in this preprint. Where a snapshot identifier or build hash
is available from the provider, it is stated in the note column;
otherwise the released version label is reported and the corresponding
caveat is noted.

\begin{table}[htbp]
\centering
\small
\begin{tabular}{@{}p{0.28\textwidth}p{0.22\textwidth}p{0.42\textwidth}@{}}
\toprule
\textbf{Component} & \textbf{Version / identifier} & \textbf{Notes} \\
\midrule
OpenAI GPT-5 (reasoning backbone) & \texttt{gpt-5} & API-hosted model;
no public snapshot identifier was exposed by the provider at the time
of evaluation. Silent provider-side updates cannot be ruled out; see
Section~\ref{sec:disc-limitations}. \\
OpenAI Agents SDK & 0.5.0 & Python SDK providing the agent runner and
reasoning-event streaming.  \\
RDKit~\cite{rdkit2025} & 2024.03.x & Used for SMILES extraction,
polar-atom SMARTS matching, MMFF94 energy calculations, and SDF/PDB
conversions. Minor version fixed within the project's locked
environment. \\
PLIP~\cite{salentin2015plip, adasme2021plip, schake2025plip} & 3.0.0
(2025 release) & Used via the Python API
(\texttt{characterize\_complex}). Deterministic, no trainable
parameters. \\
BioPython~\cite{cock2009biopython} & 1.86 & Used for Shrake--Rupley
SASA and PDB parsing. \\
Open Babel~\cite{oboyle2011openbabel} & 3.1.1 & Used for SDF-to-PDB
conversion in the Streamlit front-end (Mol* viewer only). \\
\texttt{spyrmsd}~\cite{meli2020spyrmsd} & 0.5.x & Symmetry-corrected
heavy-atom RMSD. \\
Smina~\cite{koes2013smina} & October 2019 build (based on AutoDock
Vina~1.1.2~\cite{trott2010vina}) & Precompiled binary used for both the
Francoeur et al.\ cross-docked ensembles~\cite{francoeur2020crossdock}
and for native-pose generation in scoring-only mode. A stable public
build hash is not published; the release date is reported in lieu of a
hash. \\
Maestro~\cite{maestro} & Schr\"odinger release at time of figure
preparation & Used only for the structural overlays in
Figure~\ref{fig:pymol-representative} and the case-study insets; not
part of the ranking pipeline. \\
\texttt{latexmk} + pdfTeX & TeX~Live distribution (current release) &
Used to build the manuscript PDF. \\
\bottomrule
\end{tabular}
\caption{Software environment. Versions and identifiers of the tools,
libraries, and binaries used for the agentic pipeline and the
benchmark evaluation. Where a vendor does not expose a stable snapshot
identifier (for example, an API-hosted model), this is noted explicitly
and discussed as a reproducibility limitation in
Section~\ref{sec:disc-limitations}.}
\label{tab:si-environment}
\end{table}

\end{appendix}

\bibliography{manuscript}
\bibliographystyle{unsrt}

\end{document}